\documentclass[review,3p]{elsarticle}

\UseRawInputEncoding
\usepackage{hyperref}

\journal{Elsevier Engineering Applications in Artificial Intelligence}
\usepackage{amsmath,amssymb}

\RequirePackage{filecontents}
\usepackage{mathtools}
\usepackage{commath}
\usepackage{multirow}
\usepackage[linesnumbered,ruled,vlined]{algorithm2e}
\usepackage{makecell}
\usepackage{epstopdf}
\usepackage{graphicx}
\usepackage[normalem]{ulem}
\useunder{\uline}{\ul}{}

\DeclarePairedDelimiter{\ceil}{\lceil}{\rceil}
\DeclarePairedDelimiter\floor{\lfloor}{\rfloor}

\usepackage{booktabs} 
\usepackage{algorithm2e}
\usepackage{amsmath}  
\usepackage{graphicx}      
\usepackage{subcaption}    

\begin{document}

\begin{frontmatter}

\title{Advances in Small-Footprint Keyword Spotting: A Comprehensive Review of Efficient Models and Algorithms}

\author{Soumen Garai}
\ead{sg.22ec1102@phd.nitdgp.ac.in}

\author{Suman Samui\corref{mycorrespondingauthor}}
\cortext[mycorrespondingauthor]{Corresponding author}
\ead{ssamui.ece@nitdgp.ac.in}

\address{Department of Electronics and Communication Engineering, National Institute of Technology, Durgapur,713209, India}

\begin{abstract}
Small-Footprint Keyword Spotting (SF-KWS) has gained popularity in today's landscape of smart voice-activated devices, smartphones, and Internet of Things (IoT) applications. This surge is attributed to the advancements in Deep Learning, enabling the identification of predefined words or keywords from a continuous stream of words. To implement the SF-KWS model on edge devices with low power and limited memory in real-world scenarios, a efficient  Tiny Machine Learning (TinyML) framework is essential. In this study, we explore seven distinct categories of techniques namely, Model Architecture, Learning Techniques, Model Compression, Attention Awareness Architecture, Feature Optimization, Neural Network Search, and Hybrid Approaches, which are suitable for developing an SF-KWS system. This comprehensive overview will serve as a valuable resource for those looking to understand, utilize, or contribute to the field of SF-KWS. The analysis conducted in this work enables the identification of numerous potential research directions, encompassing insights from automatic speech recognition research and those specifically pertinent to the realm of spoken SF-KWS.
\end{abstract}

\begin{keyword}
Keyword spotting; Wake word detection; Speech processing; Embedded deep learning
\end{keyword}

\end{frontmatter}


\section{Introduction}
\begin{figure}[htbp]
\centerline{\includegraphics[width=9cm, height=4cm]{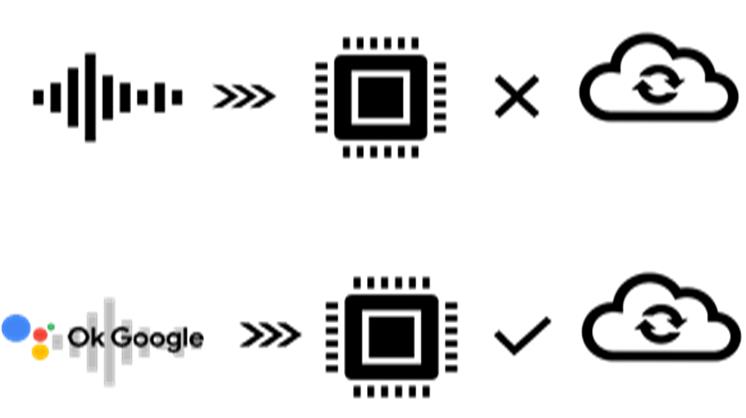}}
\caption{System overview: on edge KWS vs. cloud service based ASR}
\label{fig1}
\end{figure}

Speech recognition and voice assistants, has become a prominent and integral part of modern society. Voice assistants like Google's Home Assistant, Amazon's Alexa, Apple's Siri, and Microsoft's Cortana have demonstrated the widespread adoption of speech technologies~\cite{Hoy2018}.
In the context of the operational principle of voice assistants as illustrated in Figure \ref{fig1}, a user can initiate the system by uttering an activation keyword (wake-up words), such as ``Ok Google'' to Google Home Device \cite{Tristan2020}. Subsequently, the system establishes a connection to cloud services. Processing the trigger word ``Ok Google'' on the server is impractical, as it would involve continuously sending speech to the cloud. This approach not only compromises user privacy but also consumes more power. Instead, a low-power keyword spotter must operate at the edge, utilizing the device's on-board computational and storage capabilities to detect the activation keyword. Therefore, the task of keyword spotting (KWS) differs from the task of Automatic Speech Recognition (ASR) typically handled in a server \cite{michaely2017keyword}. Consequently, it suggests two key aspects: KWS tools should have reduced size and computational requirements, necessitating operation in a highly energy-efficient mode. Additionally, the anticipated input is a singular word or a brief phrase, distinct from the continuous sentences processed by voice recognition cloud services.
KWS is the process of identifying keywords within spoken audio streams and has numerous applications beyond voice assistant activation, including speech data mining, audio indexing, and more~\cite{LopezEspejo2021}.
Throughout the years, numerous methodologies have been explored for KWS. Initial approaches encompassed large-vocabulary continuous speech recognition (LVCSR) systems, which decoded speech signals and searched for keywords within generated lattices~\cite{rose1990hidden,wilpon1991improvements}. These methods utilized the Hidden Markov Model (HMM) to represent both the keyword and the background, which consists of non-keyword speech or non-speech noise. In some literature, this background model is referred to as the filler model. It can range from simple loops over speech and non-speech phones to more complex representations involving full phone sets or confusing word sets. Viterbi decoding is employed to identify the optimal path within the decoding graph~\cite{Motlicek2012}, and the KWS decision is made by comparing the likelihoods of the keyword and background models.

Historically, Gaussian Mixture Models (GMMs) were widely used to model the observed acoustic features. However, with the rise of Deep Neural Networks (DNNs) for acoustic modeling~\cite{LopezEspejo2021,chen2014}, this approach has evolved into a hybrid DNN-HMM framework, allowing for the incorporation of discriminative information to improve KWS accuracy. In this paradigm, DNNs directly process word posterior probabilities to detect keywords, eliminating the need for complex sequence search algorithms like Viterbi decoding. This approach offers greater flexibility in adjusting the DNN's complexity to fit resource constraints, making it particularly well-suited for deployment on microcontrollers (MCU) and edge AI platforms \cite{Saha2022}. Given the limited computational power and memory of embedded systems, optimizations such as quantization and model pruning are often employed to enable real-time keyword spotting with minimal latency.

The emergence of KWS has driven research in this field, leading to further advancements and applications in consumer electronics with limited resources, such as earphones, smartphones, and smart speakers~\cite{kwsprabhakar015,scott2016suspending}. Additionally, microcontroller-based KWS systems have gained traction in low-power IoT devices, enabling hands-free interaction in battery-operated applications such as smart home automation and wearable electronics. Despite the progress made, it's anticipated that research on KWS will continue to be a significant area of focus due to its practicality, effectiveness, and ongoing relevance~\cite{rybakov20_interspeech}.

Deep learning (DL)-based KWS \cite{LopezEspejo2021}  has attracted a lot of attention lately because of its three benefits: 
1. It employs a considerably more straightforward posterior processing technique in place of a complex sequence search algorithm (such as Viterbi decoding). 2. It is simple to modify the complexity of the DNN that generates posteriors (the acoustic model) to accommodate constraints in computational resources.
3. Under both clean and noisy environments, it regularly yields notable performance improvements over the keyword/filler HMM technique, especially in small-footprint applications with constrained memory and compute capability.

This paper primarily focuses on the implementation of small-footprint keyword spotting (SF-KWS ), which is crucial for the on-device realization of KWS. The key motivation for this implementation stems from concerns such as (a) network latency, (b) significant power consumption due to energy-intensive network communication, and (c) security, privacy, and regulatory issues. Spoofing and voice cloning are particularly critical concerns related to the security and integrity of voice-based systems. In the context of SF-KWS, the primary challenges arise because KWS models with high accuracy often have considerable complexity and resource requirements. This complexity makes it challenging to deploy models (even small models such MobileNetV2 \cite{Sandler2018}, Squeezenet \cite{Iandola2016}, etc.) on devices with limited memory (typically having less than 2 MB flash memory), including smartphones, wearables (such as smartwatches), and Internet of Things (IoT) devices. These devices demand efficient and lightweight KWS models. Therefore, SF-KWS focuses on developing lightweight models tailored for recognizing specific keywords or phrases in audio, with a primary emphasis on minimizing model size and computational requirements. This approach renders the models suitable for deployment on resource-constrained devices, leveraging the principles of Tiny Machine Learning (TinyML) \cite{Warden2019}\cite{Lin2023}. TinyML is a broader concept encompassing the deployment of machine learning (ML) models on low-power, resource-constrained devices, employing techniques such as model quantization, pruning, and compression \cite{Shafique2021}.

This article offers a concise overview of SF-KWS technology and details different TinyML frameworks, representing one of the initial efforts to comprehensively assess significant scientific contributions in SF-KWS research to the best of our knowledge. 
The existing literature provides only a limited number of comprehensive overview articles on KWS \cite{LopezEspejo2021, Giraldo2021, Tabibian2020}. This article builds upon a recently published short overview paper on KWS \cite{Garai2024}, which primarily outlined various approaches for developing SF-KWS. However, Most existing works only superficially address the state-of-the-art (SOTA) general approaches to discriminative KWS, focusing mainly on acoustic modeling, feature extraction, and model training. While \cite{Giraldo2021} reviews SOTA advancements in software/hardware co-design for embedded KWS and analyzes recent hardware architectures, SF-KWS introduces additional inherent challenges compared to general-purpose KWS. These challenges include designing model architectures, optimizing features, developing advanced learning techniques, and adapting to user-defined keywords. Potential solutions, such as effective Neural Architecture Search (NAS) \cite{ChittyVenkata2022}, Knowledge Distillation (KD) \cite{Menghani2023}, and Few-Shot Learning (FSL), are critical for addressing these challenges. Furthermore, incorporating insights from TinyML frameworks into diverse SF-KWS strategies is essential to facilitate seamless deployment on embedded edge devices, such as Microcontroller-Class Hardware \cite{Saha2022}.

Consequently, the current overview articles (survey papers) on KWS provide only limited coverage of recent advancements. This article aims to fill that gap by offering researchers and practitioners a comprehensive and up-to-date overview of SF-KWS, addressing the intricacies of the technology. The primary contributions of this study are as follows:
\begin{itemize}
\item On the basis of diverse algorithms, architecture or models, and other parameters, we grouped SF-KWS's works into seven groups. These include model architecture, learning techniques, model compression, attention awareness, feature optimization, NAS, and Hybrid Approach. These approaches aim to adapt the KWS to IoT or Edge devices while maintaining accuracy and resource efficiency.
\item  Put a detailed discussion of various TinyML frameworks \cite{Lin2023}. This TinyML Framework aims to present a comprehensive solution for building and deploying ML models on low-power devices, making it easier for developers to design edge computing applications.
\item Comprehensive evaluation of various SF-KWS architectures across different model sizes using the Google Speech Commands Dataset (GSCD) \cite{warden2017launching}. By applying TinyML frameworks such as TensorFlow Lite and EdgeImpulse, the study optimizes these architectures for IoT edge devices, demonstrating that quantization to Int8 format reduces model size by up to 69\% while maintaining accuracy and improving inference efficiency. Moreover, to address the inherent trade-offs between model performance and resource constraints, the study employs multi-objective optimization (MOO) techniques such as Simulated Annealing (SA), Bayesian Optimization (BO), and NSGA-II revealing that these Sequential Model-Based Optimization methods have the potential of producing the best Pareto-optimal models \cite{Liang2022}. These optimized models cater to two deployment goals: high accuracy for performance-critical applications and low memory usage for embedded systems with strict resource constraints. The findings provide actionable insights into balancing model complexity, computational efficiency, and memory footprint for real-world SF-KWS deployment, with an open-source implementation available on GitHub\footnote{https://github.com/sumansamui/Small-footprint-keyword-spotting.git}  for further research and application. 
\end{itemize}
\noindent
\textbf{Paper Organization:} The remainder of this article is organized as follows: \textbf{Section~\ref{sec:trends}} provides a comprehensive overview of recent trends in keyword spotting research. It covers the rise in research activity, the architectural evolution from classical models to deep learning frameworks, the emergence of TinyML for on-device inference, and recent directions such as multilingual/multimodal KWS, privacy-preserving learning, and expanding application areas. \textbf{Section~\ref{sec:task}} offers an overview of the KWS task and its challenges. It discusses the acoustic pipeline for KWS implementation, highlights key factors influencing performance, reviews available datasets and evaluation metrics, and addresses advanced aspects such as latency control, robustness against noise and variability, and user customization techniques. \textbf{Section~\ref{sec:sfkws}} introduces a set of efficient approaches to KWS development, emphasizing the transition from traditional systems to SF-KWS. It categorizes related research efforts and presents a comprehensive review of efficient algorithms and model architectures tailored for lightweight, resource-aware KWS solutions suitable for deployment on IoT edge devices. \textbf{Section~\ref{sec:tinyml}} explores the integration of SF-KWS within the TinyML ecosystem. It examines various TinyML frameworks, the limitations of microcontroller-based deployment, and the broader relevance of SF-KWS in power- and memory-constrained environments. \textbf{Section~\ref{sec:exp}} presents an experimental case study that evaluates model architectures designed for SF-KWS across multiple TinyML frameworks. This section also investigates multi-objective optimization to identify Pareto-optimal models that balance accuracy and F1-score with model size and RAM usage. \textbf{Section~\ref{sec:conclusion}} concludes the paper and discusses future research directions in the field of KWS and TinyML-based deployments. A list of abbreviations used throughout the manuscript is provided in Table 1.
\begin{table}[]
\caption{List of Abbreviations}
\centering
\scriptsize
\begin{tabular}{cc}
\hline
\textbf{Abbreviations} & \textbf{Full meaning}                                 \\
\hline
ACR                    & Absolute Cosine Regularization                         \\
ASR                    & Automatic Speech Recognition                          \\
AUC                    & Area Under the Curve                                  \\
CNN                    & Convolutional Neural Network                          \\
CMSIS-NN               & Cortex Microcontroller Software Interface Standard-NN \\
CRNN                   & Convolutional Recurrent Neural Network                \\
DAG                    & Directed Acyclic Graph                                \\
DARTS                  & Differentiable Architecture Search                    \\
DL                     & Deep Learning                                         \\
DNN                    & Deep Neural Network                                   \\
DQ                     & Dynamic Quantization                                  \\
DS-CNN                 & Depth-wise Separable Convolution Neural Network       \\
FFT                    & Fast Fourier Transform                                \\
FLOPs                  & Floating-Point Operations                             \\
FSL                    & Few-Shot Learning                                     \\
GCN                    & Graph Convolutional Network                           \\
GRU                    & Gated Recurrent Unit                                  \\
GSCD                   & Google Speech Commands Dataset                        \\
HMM                    & Hidden Markov Model                                   \\
IoT                    & Internet of Things                                    \\
KD                     & Knowledge Distillation                                \\
KWS                    & Keyword Spotting                                      \\
LSTM                   & Long Short-Term Memory                                \\
LVCSR                  & Large-Vocabulary Continuous Speech Recognition        \\
MCU                    & Microcontroller Unit                                  \\
MFCC                   & Mel-Frequency Cepstral Coefficients                   \\
MFSTS                  & Multi-Frame Shifted Time Similarity                   \\
MTConv                 & Multi-branch Temporal Convolution Module              \\
NAS                    & Neural Architecture Search                            \\
NN                     & Neural Network                                        \\
PTQ                    & Post-Training Quantization                            \\
QAT                    & Quantization-Aware Training                           \\
QNN                    & Quaternion Neural Networks                            \\ 
RNN                    & Recurrent Neural Network                              \\
SF-KWS                 & Small-Footprint Keyword Spotting                      \\
SNN                    & Spiking Neural Network                                \\
SOTA                   & State-of-The-Art                                      \\
SSL                    & Semi-Supervised Learning                              \\
SSRL                   & Self-Supervised Representation Learning               \\
SVD                    & Singular Value Decomposition                          \\
SWSA                   & Shared Weight Self-Attention                          \\
TCN                    & Temporal Convolutional Network                        \\
TDNN                   & Time-Delay Neural Network                             \\ 
TC-ResNet              & Temporal Convolutional ResNet                         \\
TENet                  & Temporal Efficient Neural Network                     \\
TinyML                 & Tiny Machine Learning                                 \\
\hline                       
\end{tabular}
\end{table}

\section{Trends in Keyword Spotting Research \label{sec:trends}}
Over the past decade, Keyword Spotting has evolved from basic acoustic pattern matching to a sophisticated subdomain of speech recognition, driven by advancements in deep learning, embedded AI, and edge computing. This section summarizes key developments and emerging trends.
\subsection{Steady Rise in Research Activity}
Keyword spotting research has seen an exponential increase in publications since 2015. In this survey, we collect and analyze the most recent surveys on SF-KWS that have been published in refereed journals and
conference. To understand the overall research landscape, we reviewed almost 250 papers published between 2010 and 2025.

\begin{figure}[htbp]
	\centerline{\includegraphics[scale=0.5]{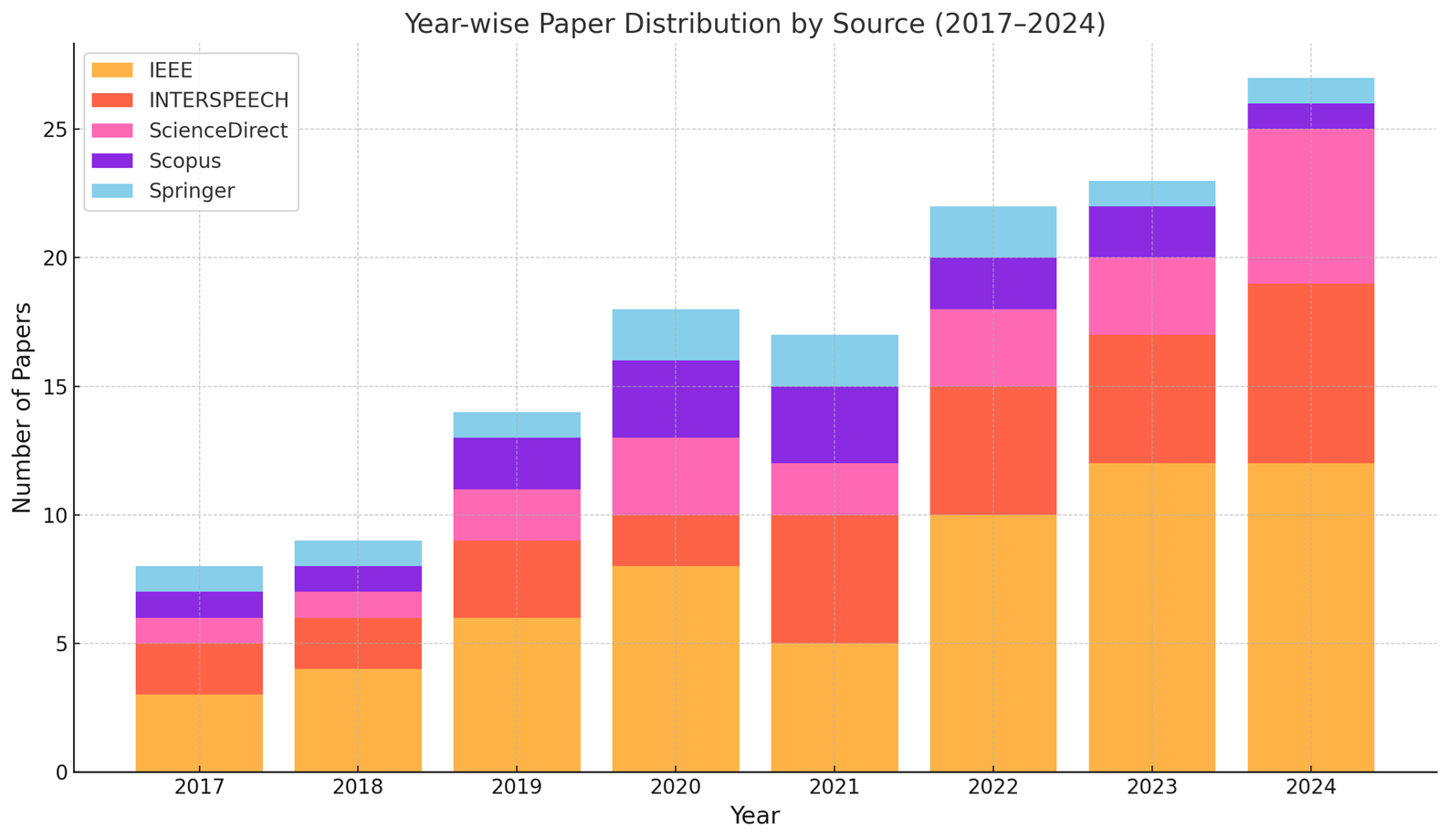}}
	\caption{Year-wise distribution of keyword spotting publications (2017-2024) categorized by publication source. The chart highlights contributions from IEEE, INTERSPEECH, ScienceDirect, Scopus, and Springer.}
	\label{dist_paper}
\end{figure}

To curate the relevant literature, we systematically evaluated the initially selected papers and excluded those considered outside the scope of our study. Our research draws upon publications from several prominent repositories, including AAAI, ACM, ICML, ICLR, NeurIPS, IEEE, INTERSPEECH (ISCA), SCOPUS, MDPI, ScienceDirect, and the arXiv repository hosted by Cornell University.
 As shown in Figure \ref{dist_paper}, a sharp growth occurred after 2017, which coincides with the proliferation of voice assistants like Amazon Alexa, Google Assistant, and Apple's Siri. The introduction of standardized datasets like GSCD \cite{warden2017launching} and the surge in edge-AI platforms further catalyzed research in this domain.	
\subsection{Architectural Evolution: Classical to Deep Models}
\begin{figure}[h]
	\centerline{\includegraphics[scale=0.8]{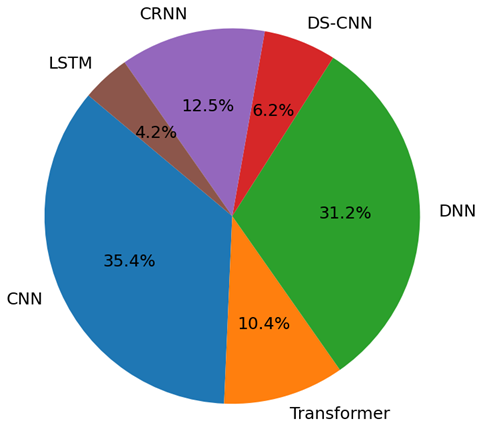}}
	\caption{Distribution of deep learning architectures used in recent KWS literature.}
	\label{fig:arch_dist}
\end{figure}

Initially, KWS was developed using dynamic time warping (DTW) and Gaussian mixture model-hidden Markov model (GMM-HMM) approaches \cite{wilpon1991improvements}. These traditional methods were later replaced by deep learning-based architectures. Convolutional Neural Network (CNN)-based models were introduced for their compactness and ability to perform efficient parallel computations \cite{Sainath2015conv}. Recurrent Neural Network (RNN)-based models, such as long short-term memory (LSTM) and gated recurrent unit (GRU) networks, were employed to effectively model temporal dependencies in speech signals \cite{Li2022}. More recently, Transformer-based models, such as Audio Spectrogram Transformers (AST) \cite{Gong2021} \cite{Berg2021} have gained popularity for their ability to capture long-range dependencies using self-attention mechanisms \cite{Samui2023}.
Figure~\ref{fig:arch_dist} shows the relative distribution of architectures used across the surveyed literature.  CNNs are the most frequently employed architecture, accounting for 43.9\% of the total, followed by DNNs at 36.6\%. Transformer-based models have gained notable attention with a usage share of 12.2\%, whereas CRNNs and LSTMs were comparatively less frequent, constituting 4.9\% and 2.4\% respectively. This trend reflects a strong preference toward computationally efficient and spatially-aware models like CNNs for edge device deployments in tasks such as keyword spotting.

\subsection{TinyML and On-Device Inference}
A key research direction involves optimizing models for low-resource edge devices through TinyML. Frameworks such as TensorFlow Lite Micro~\cite{David2021}, STM32Cube.AI, and Edge Impulse enable real-time inference on microcontrollers with less than 1MB of memory. Model optimization strategies such as quantization-aware training (QAT) \cite{Menghani2023}, network pruning and compression, and knowledge distillation~\cite{hinton2015distilling} allow these models to run continuously on ultra-low-power hardware, making KWS viable in embedded systems.

\subsection{Multilingual and Multimodal KWS}
The research landscape is also evolving to address multilingual and multimodal requirements. Techniques like transfer learning and shared phoneme embeddings~\cite{Lei2023} have facilitated recognition in multilingual and code-switched environments. Concurrently, multimodal KWS systems incorporating EEG~\cite{Chakravarthi2022}, EMG, or visual cues have demonstrated improved robustness in noisy or unconstrained conditions.
\subsection{Expanding Applications}
Applications of KWS have expanded beyond traditional smart assistant use cases. In healthcare, KWS systems enable hands-free interaction for elderly care and assistive technologies. In the automotive sector, they support voice-based control for enhanced driver safety, and in industrial IoT scenarios, they facilitate voice-triggered machine interactions \cite{LopezEspejo2021}.
\subsection{Privacy and Federated Learning}
Due to data privacy concerns, researchers are exploring federated learning approaches to train KWS models locally without uploading raw data~\cite{Leroy2019}. Google's Gboard project~\cite{Esch2019} demonstrated real-world on-device personalization with privacy-preserving learning techniques.	
\section{Fundamentals of KWS: overview and challenges \label{sec:task}}
Spoken KWS is commonly regarded as a subproblem within the domain of ASR shown in Figure \ref{fig1}. A typical KWS identifies specific words or phrases of interest within spoken utterances or an audio stream. Smart voice assistant devices like Amazon Echo, Google Home, and similar systems incorporate on-device KWS functionality. These devices first detect predefined keywords such as ``Alexa'', ``OK Google'' or ``Hey Siri''. Upon detecting a keyword, the device is activated and only then streams the audio to the cloud for LVCSR.  In such applications, accurate on-device KWS is essential, as it must operate with minimal CPU and memory usage. The system must achieve high recall to ensure ease of use while maintaining a low false acceptance rate to address privacy concerns. Additionally, the system must maintain low latency for real-time responsiveness.  KWS has many uses beyond only enabling on voice assistants, as listed below:
\begin{itemize}
	\item \textbf{Home automation systems} \\
	KWS is used in smart home devices to control appliances and systems via voice commands. For example, it allows users to turn on lights, adjust thermostats, or activate security systems with predefined keywords.  
	
    \item \textbf{Audio indexing} \\
     It also is used to index audio recordings so that they can be searched by keyword. This is useful for applications such as call centres, where it is important to be able to find recordings that contain specific keywords quickly.
	
	\item \textbf{Wearable devices} \\
	In wearable technology, such as fitness trackers and smartwatches, KWS enables hands-free interaction by recognizing voice commands, making these devices more user-friendly and accessible.  
	
	\item \textbf{Telecommunications and customer support} \\
	KWS is used in call centers and automated customer support systems to detect specific keywords or phrases, helping route calls or trigger automated responses, thereby improving efficiency.  
	
	\item \textbf{Healthcare monitoring systems} \\
	KWS is implemented in assistive devices for healthcare applications, such as voice-controlled medical monitoring systems or emergency response devices, enabling patients to call for help or interact with medical devices using voice commands.

\end{itemize}

\subsection{System requirements and key features of KWS on TinyML at the edge}
KWS on TinyML at the edge represents a cutting-edge approach to enabling intelligent and efficient processing on resource-constrained devices. This technology leverages compact ML models tailored for small-scale hardware, making it well-suited for applications where power consumption, memory limitations, and real-time responsiveness are critical factors. The essence of TinyML lies in its ability to bring ML capabilities directly to the edge, eliminating the need for constant connectivity to centralized servers. Here are some essential points that encapsulate the key features and advantages of implementing KWS using TinyML at the edge:
\begin{itemize}
\item \textbf{Low power requirements:} TinyML models for KWS are designed to operate efficiently on low-power edge devices, making them suitable for battery-powered applications.
\item  \textbf{Low latency:} KWS can respond more quickly than traditional speech recognition systems because they do not need to wait for the entire utterance to be processed.
\item  \textbf{Compact model size:} The models used for KWS on TinyML are optimized for small memory footprints, enabling deployment on devices with limited storage capacity.
\item  \textbf{Accuracy in real-time processing:} TinyML implementations for KWS aim to achieve real-time processing, ensuring prompt detection and response to specified keywords or phrases.
\item  \textbf{Edge computing capability:} KWS on TinyML leverages the capabilities of edge computing, allowing for on-device processing without the need for constant connectivity to a centralized server. Hence, the Locality of computation also resolves the privacy-related issues.
\item  \textbf{Customizable keywords:} TinyML offers an on-board training facility. By this, Users can often customize the set of keywords or commands based on their specific application requirements, providing flexibility for different use cases.
\end{itemize}
\begin{figure}[htbp]
	\centerline{\includegraphics[scale=0.5]{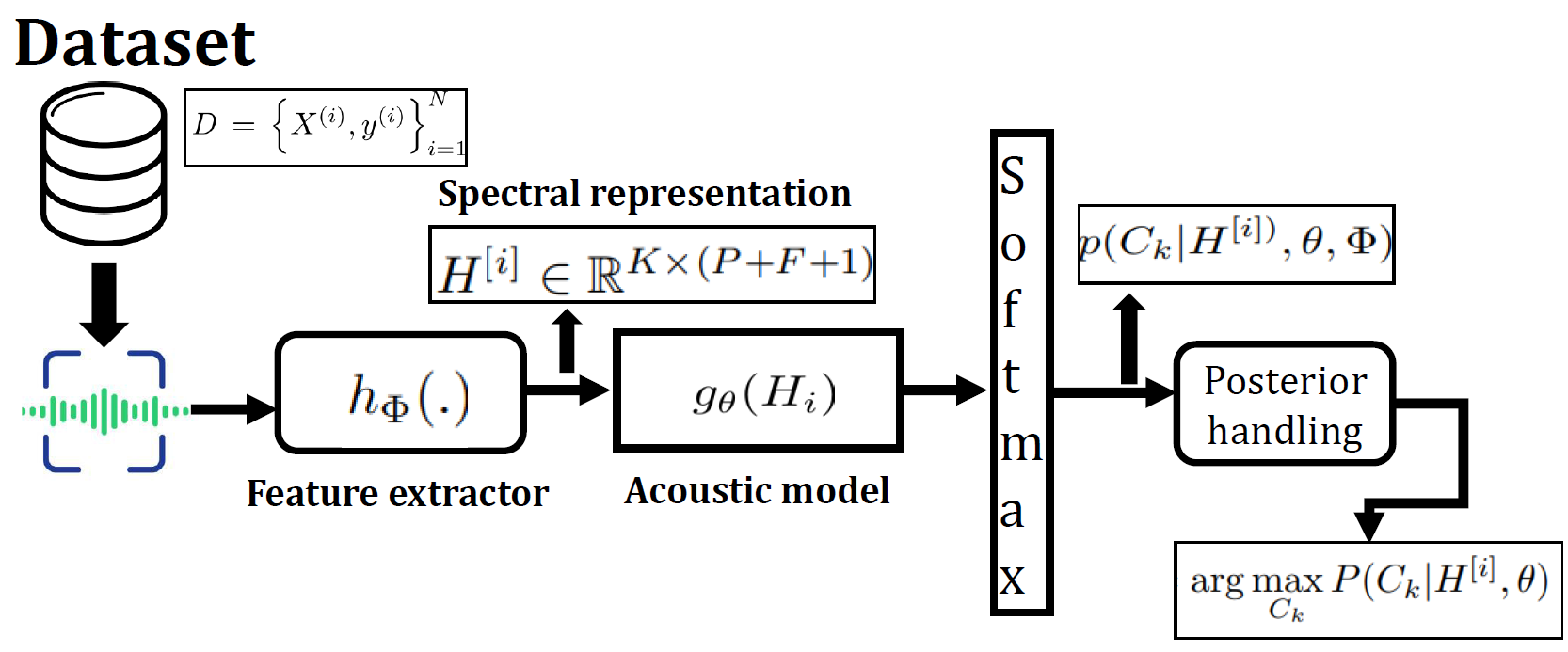}}
	\caption{The typical workflow of a deep learning-based spoken keyword spotting system involves three main steps: (i) extracting features from the speech signal, (ii) utilizing a deep neural network (DNN) acoustic model to generate posterior probabilities for different keyword and non-keyword (filler) classes, and (iii) Posterior Handling for detecting the presence of potential keywords \cite{LopezEspejo2021}.}
	\label{fig2}
\end{figure}
\subsection{Supervised learning based discriminative approach for KWS} 
In general, for supervised learning, KWS requires a labeled data set $D=\set{X^{(i)},y^{(i)}}^{N}_{i=1}$, where $X^{i}$ denotes an audio segment containing a specific keyword and $y^{(i)}$ represents the corresponding label of the keyword.  
A DNN-based predictive model can be built from this data set as follows: $\hat{y}^{(i)}=g_{\theta}(h_{\Phi}(x_{i}))$ where $h_{\Phi}(.)$ is a representation extractor function, and  $g_{\theta}(.)$ denotes the classifier function.  Each of this $x^{(i)}$  is a time series data $s(n)$ of specific length (higher in dimension) of specific sampling frequency i.e. a function of discrete time index. From $s(n)$, a more compact spectral representation $H$ can be obtained by using sort of $h_{\Phi}(.)$, and it can be represented as a two-dimensional matrix with a temporal series of $K$-dimensional feature vectors:
\begin{equation}
	H=(h_0 ,h_1, . . , h_t , . . , h_{T-1} )\in {\Bbb R}^{T \times K} \label{eq1}
\end{equation}
Here, $T$ represents the total number of feature vectors extracted from the signal $s(n)$. Using $H$ as input, the DNN-based acoustic model generates a series of posterior probabilities for different keyword and non-keyword classifications. In particular, until the complete feature sequence is examined, the acoustic model sequentially processes temporal segments of $H$.
\begin{equation}
	H^{[i]} =(H_{is-P}, . . , H_{is}, . ., H_{is+F})\label{eq2}
\end{equation}
where $ i= \ceil[\big]{\dfrac{P}{s}}, . . . . . ., \floor[\big]{\dfrac{(T-1-F)}{s}}$. Here, the time frameshift is represented by $s$, and an integer segment index $i$. Additionally, $F$ and $P$ represent the quantity of following and previous frames (temporal background) in each segment, respectively, in $H^{[i]}\in {\Bbb R}^{K \times (P+F+1)}$.

The DL-based acoustic model: $g_\theta(.)$  maps each $H^{[i]}$ to an N-dimensional vector $P^{N \times 1}$ which is the probability distribution corresponding to $N$ different keyword classes. $\theta$ denotes the parameters of the acoustic model.  We can train this predictive model by minimizing a loss function $\mathcal{L}$, such as the
negative log-likelihood or cross-entropy loss:
\begin{equation}
\arg\min_{\theta} \sum_{{H^{[i]},y^{(i)} \in D}} \mathcal{L}(g_{\theta}(H^{[i]}),y^{(i)})
\end{equation}
the acoustic model produces for input segment $H^{[i]}$
\begin{equation}
	y^{[i]}_k = g_{\theta}(H_{i})=p(C_k|H^{[i])},\theta,\Phi), k= 1, 2, . . , N. \label{eq3}
\end{equation}
where $y^{[i]}_k$ is the posterior of the $k$-th class $C_k$. Hence, the DL model would have a dense (fully- -connected) output layer with softmax activation to ensure that $\sum^N_{k=1} y^{[i]}_k = 1 \forall i $ KWS. The model's parameters are typically calculated by discriminative training  by backpropagation using annotated diverse voice samples representing the various $N$ classes.

KWS task is dynamic rather than static, requiring the system to continuously monitor the input signal $s(n)$ to produce a sequence of posterior probabilities $y^{[i]}$, where $ i= \ceil[\big]{\dfrac{P}{s}}, . . . . . ., \floor[\big]{\dfrac{(T-1-F)}{s}}$, for detecting keywords in real-time.  a straightforward approach to achieve this is to select the class $\hat{C}^{[i]} $ with the highest posterior probability. This can be expressed as:
\begin{equation}
	\hat{C}^{[i]} = \arg\max_{C_k} y^{[i]}_n = \arg\max_{C_k} P(C_k | H^{[i]}, \theta) \label{eq4}
\end{equation}
where $ C_k $ represents the candidate classes, and $ P(C_k| H^{[i]}, \theta,\Phi)$ is the posterior probability for each class given the input features $H^{[i]}$ and model parameters $\theta$. The entire KWS pipeline is illustrated in Figure \ref{fig2}.
\subsubsection{Posterior handling}
The series of posterior probabilities is used to determine whether a keyword is present in an audio stream, $p^{[i]}$, generated by the acoustic model must be processed. There are two primary approaches for handling these posteriors: non-streaming (static) and streaming (dynamic) modes. 
\begin{itemize}
	\item \textbf{Non-Streaming Mode: }In non-streaming mode, each input segment containing a single word is classified independently (i.e., classification of isolated words). To ensure the full duration of each word is captured, input segments must be sufficiently long, typically around one-second \cite{warden2018speech}. For a given input $H^{[i]}$, the class with the highest posterior probability is used for classification, as expressed in \eqref{eq4}. This approach is generally preferred over selecting classes based on posteriors exceeding a sensitivity threshold, as non-streaming KWS typically generate highly peaked posterior distributions \cite{LopezEspejo2021}.

	\item \textbf{Streaming Mode:} Keywords are not separated or segregated in this mode of operation, which processes an audio stream continuously and in real time. Any given segment may contain all or a portion of a keyword in this form. The acoustic model, therefore, generates a time sequence of raw posterior probabilities $.., y^{[i-1]}, y^{[i]}, y^{[i+1]}, ..,$ which are strongly correlated over time. Because these raw posteriors are inherently noisy, they are typically smoothed using methods like a moving average across each class \cite{kwsprabhakar015, chen2014small} Let $p^{[i]}$ denote the smoothed posterior values. In many KWS, each class represents a distinct word. The smoothed posteriors $p^{[i]}$ can then be directly used to detect keywords, either by comparing them against a sensitivity threshold \cite{sun2016max} or by identifying the highest posterior within a sliding time window \cite{kumar2018convolutional}. However, consecutive segments, $  .., H{[i-1]}, H{[i]}, H{[i+1]}, .., $ may contain parts of the same keyword, which could lead to false alarms identifying the same term more than once. To prevent this, a simple process is introduced after a keyword detection, during which the system is temporarily prevented from triggering again\cite{sun2016max,  sorensen2020depthwise}.
\end{itemize}
Both non-streaming and streaming modes are used in KWS research. While streaming mode is more realistic for continuous audio streams, non-streaming performance has been found to be highly correlated with streaming performance, making it relevant for practical experimentation.
\subsubsection{Speech Feature Extraction} 
\begin{figure}[]
	\centering
	\begin{subfigure}[b]{0.8\textwidth}
		\centering
		\includegraphics[width=\textwidth]{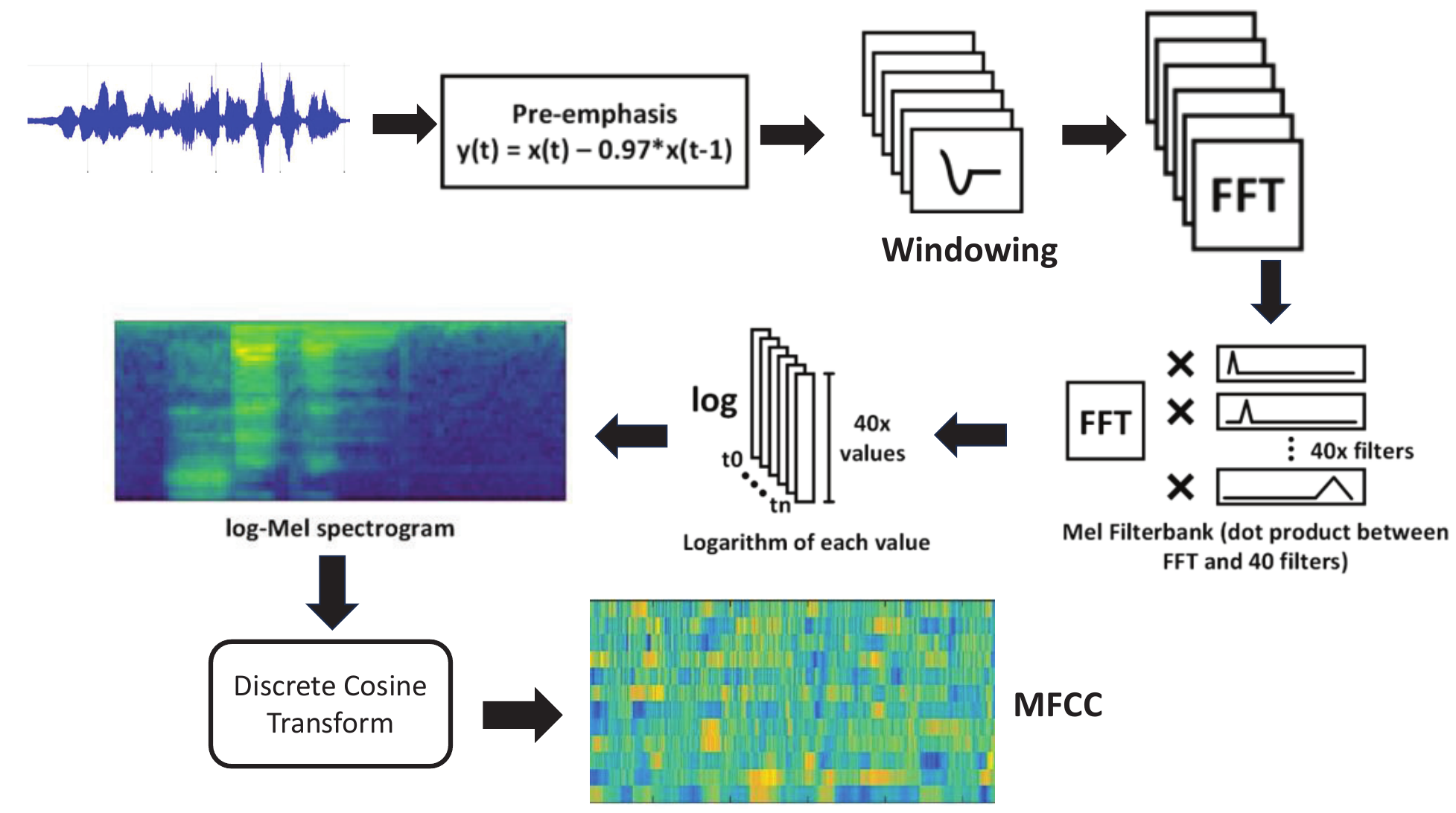}
		\caption{}
		\label{fig3:image1}
	\end{subfigure}
	
	\begin{subfigure}[b]{0.8\textwidth}
		\centering
		\includegraphics[width=\textwidth]{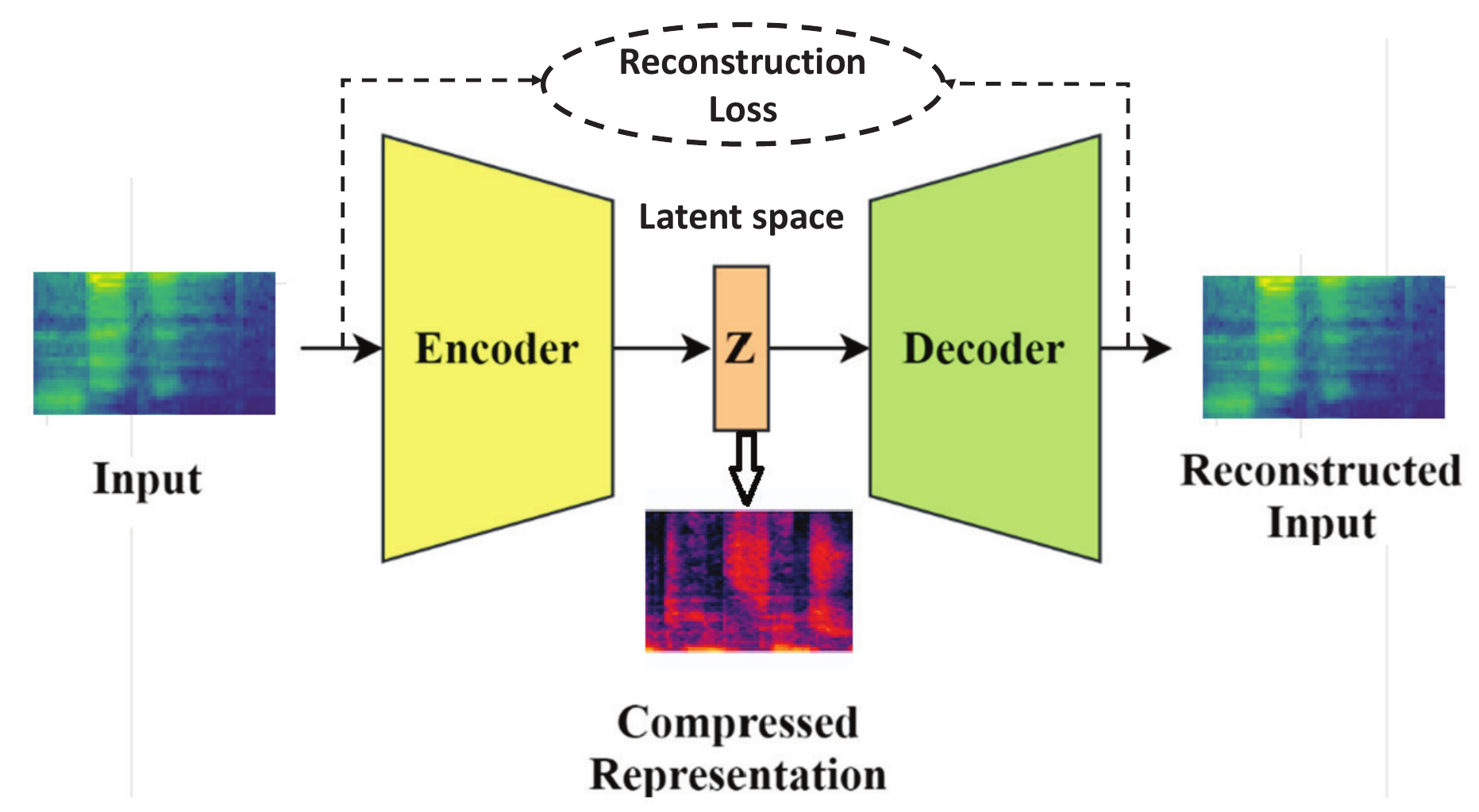}
		\caption{}
		\label{fig3:image2}
	\end{subfigure}
	
	\caption{(a) The traditional procedure for obtaining Mel-frequency cepstral and log-Mel spectral speech features involves employing FFT on each short time frame of the audio signal.Overall figure caption describing both images (b) Automatic feature representation learning using Autoencoder \cite{Vitolo2023}.}
	\label{fig3}
\end{figure}

Most popular KWS extraction features techniques are the Mel-scale-based feature. The Fast Fourier transform (FFT) is a traditional method for obtaining log-mel spectral and Mel-frequency cepstral speech characteristics. Mel scale filter banks, such as log-Mel spectral coefficients and Mel-frequency cepstral coefficients (MFCCs) \cite{davis1980comparison}, have been prevalent in ASR and KWS for a long time. Figure \ref{fig3:image1} shows the steps involved for log-mel spectrogram and MFCCs. These features are considered robust and competitive choices even with attempts to explore alternative representations. In KWS, log-Mel spectral coefficients and MFCCs are typically normalized before being used in the KWS model, enhancing training stability and model generalization. MFCCs with temporal context are widely used in KWS, and the log-Mel spectrogram is also preferred over MFCCs due to its ability to capture spectro-temporal correlations effectively. A lower number of filterbank channels can be used without significantly affecting KWS performance, as long as the computing complexity is kept to a minimum by having a very good Mel-frequency resolution.
Furthermore, to lower memory and energy utilization in deep KWS on devices with limited resources, quantization is applied by lowering the precision of model parameters
Furthermore, some work \cite{Vitolo2023} introduced an innovative data-driven approach that utilizes an autoencoder to automate audio feature extraction (as shown in Figure \ref{fig3:image2}) while ensuring low computational complexity and achieving accuracy comparable to SOTA KWS. 
\subsubsection{Neural network based acoustic modelling}
Once the FE phase is completed, the processed features are fed into a ML classifier that determines the probabilities of different output classes. In contrast, end-to-end ML models bypass FE and directly process raw audio samples. Each output class typically represents a specific keyword, background noise, or speech without target keywords. For example, a neural network (NN) designed for detecting 10 keywords may have 12 outputs, with 10 representing keyword probabilities and the remaining two accounting for background noise and non-target speech.

Historically, HMMs combined with the Viterbi Algorithm were widely used for keyword recognition. However, DL methods have now largely replaced them, offering higher accuracy, more deterministic inference, and greater robustness to noise. Early DL-based KWS models relied on fully connected DNNs \cite{LopezEspejo2021}, which, despite their effectiveness, struggled with capturing temporal relationships. CNNs and RNNs were introduced to address this. CNNs leverage convolution operations across the temporal dimension, while RNNs, LSTM and GRU, exploit hidden states to retain temporal dependencies \cite{LeCun2015}, thereby achieving high accuracy with reduced computational overhead.

\begin{figure}[htbp]
	\centerline{\includegraphics[scale=0.5]{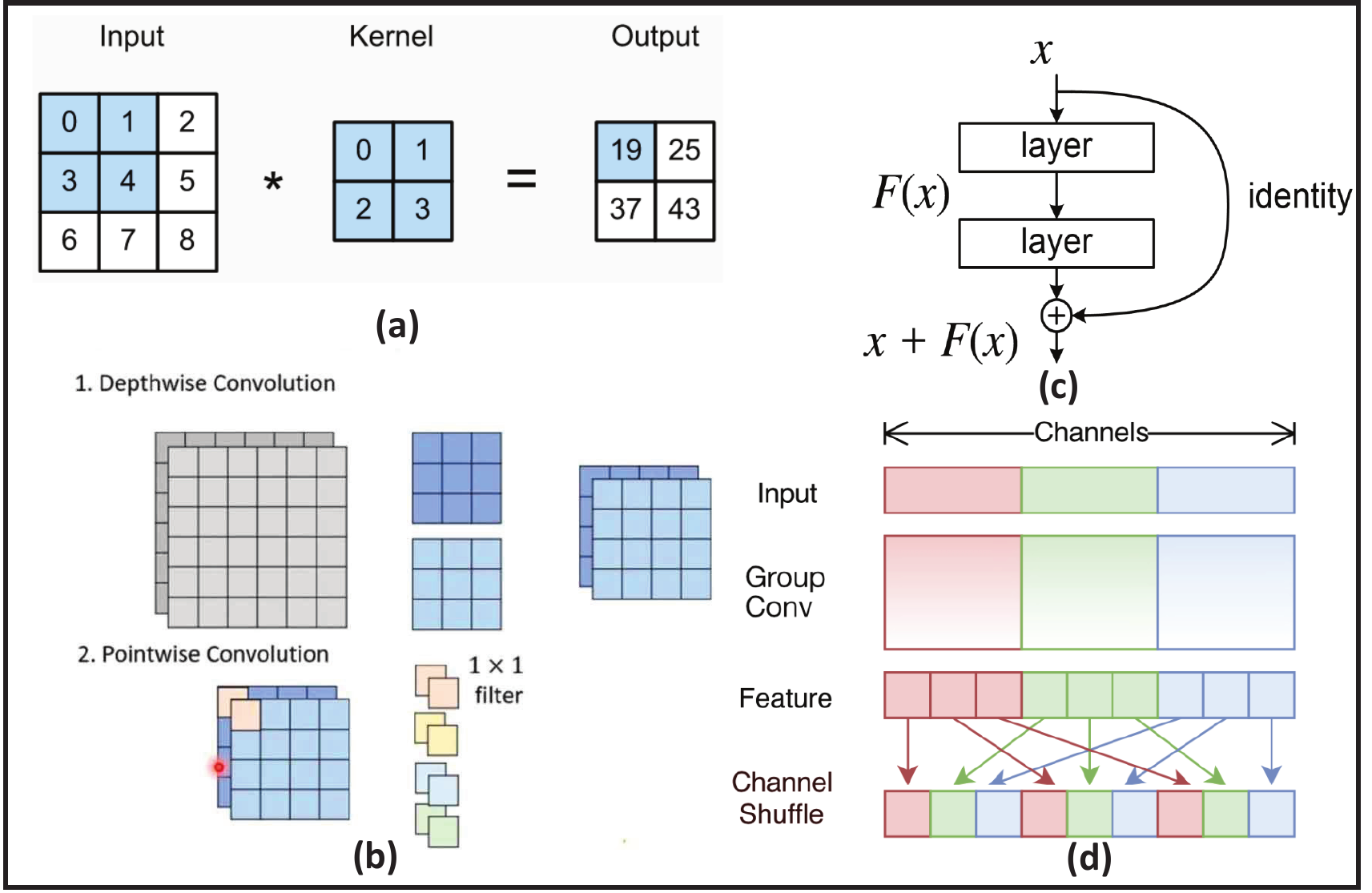}}
	\caption{Different types of the convolution operation. (a) Basic
		convolution. (b) Depthwise separable convolution.
		(c) Residual connection. (d) Channel shuffle \cite{Shuvo2022}}
	\label{fig9}
\end{figure}

Compact network design focuses on optimizing convolution operations in CNNs to enhance efficiency while maintaining accuracy, particularly for resource-constrained scenarios. Instead of using standard convolutions, which process all input channels together and require high computational power, alternative approaches help reduce the number of operations while maintaining good performance. One such method is depth-wise separable convolution (DS-CNN) as shown in Figure \ref{fig9}, which is used in MobileNet \cite{Chen2024} \cite{howard2017mobilenets}.  DS-CNNs enhance efficiency by decomposing standard convolution layers into depthwise and pointwise operations, maintaining accuracy while reducing computational complexity. On the other hand, temporal convolutional networks (TCNs) employ dilated convolutions, which allow for an expanded receptive field while keeping the computation low. Unlike standard convolutions that rely on adjacent values, dilated convolutions skip certain inputs in the temporal dimension, progressively expanding their receptive field across network layers \cite{Goodfellow2016a}. Moreover, trained network filters can be decomposed using tensor decomposition techniques such as \textit{Canonical Polyadic } and \textit{Tucker decomposition} \cite{Lebedev2015}. These methods approximate large filters with smaller ones, enabling compression and computational speed improvements. The more details discuss on efficient network architecture for SF-KWS will be found on Section 4.1.
\subsection{Latency control}
As highlighted in Section 2, smart conversational agents often rely on KWS to initiate voice interactions with users. To ensure a seamless user experience and address privacy concerns, most existing approaches in the literature have traditionally emphasized novel architectures and learning techniques to enhance accuracy. Because waiting for additional context can result in improved accuracy, this concentration can occasionally result in increased delay. However, developing a SF-KWS  that operates with minimal CPU and memory usage while maintaining low latency is essential.   This multi-objective scenario inherently demands a balance between accuracy and latency. Insufficient accuracy can lead to missed responses or unintended actions. In contrast, excessive latency not only delays interactions but may also alter user behaviour, such as prompting users to repeat the keyword while waiting for a response. It has often been observed that incorporating additional audio context following a detected trigger phrase (wake-up word) enhances decision accuracy. S. Sigtia et al. \cite{Sigtia2021} experimentally demonstrated that detection accuracy improves as more audio is added after the trigger phrase, enabling the model to refine its estimation progressively. 
They suggested a two-stage design in which the first stage, a low-power detector that is always on, produces an early score and then re-scores the keyword segment that the first stage found using a broader verification model. However, inference of the second stage requires more device resources. Their results show that the early score suffices for most test set examples, while the late score enables more accurate decisions in challenging or borderline cases. This way, the two-stage design effectively balances accuracy and latency. 
Geng-shen Fu et al. \cite{Fu2022} introduced a Convolution Recurrent Neural Network (CRNN)-based unified model for keyword detection that integrates speculation, detection, and verification, utilizing a latency-aware max-pooling loss. The max-pooling loss \cite{sun2016max} eliminates the need for manually selecting frames, instead automatically minimizing cross-entropy loss on specific frames. For positive examples, it selects the frame with the highest class score, and for negative examples, it minimizes the highest positive score or maximizes the lowest negative score. This encourages consistently low positive scores across all frames. However, the method may delay detection, as the model leverages more contextual information for improved accuracy. The latency-aware max-pooling loss is an improved version of the standard max-pooling loss, designed to balance accuracy and latency in KWS models. Unlike conventional loss functions that optimize solely for accuracy, this method incorporates a latency constraint to ensure that detections occur within a predefined time window. The latency-aware max-pooling loss is defined as \cite{wang2023wekws}:

\begin{equation}
	L(t_l) = -\log(p_{y_t})
\end{equation}

where:

\begin{equation}
	t =
	\begin{cases}
		\arg\min\limits_i (p_{yi}), & \text{if } y = 0 \quad (\text{negative class}) \\
		\arg\max\limits_{i \in (i - e \leq t_l)} (p_{yi}), & \text{if } y \neq 0 \quad (\text{positive class})
	\end{cases}
\end{equation}
where,  \( t \) and \( i \) represent the frame indices, while \( y \) denotes the class label, where \( y = 0 \) corresponds to the negative class. The parameter \( e \) marks the end of the keyword, and \( p_{yi} \) represents the output score of class \( y \) at frame \( i \). Finally, \( t_l \) defines the target latency, which constrains the detection time within the specified limit. Empirical results demonstrate that this approach effectively trains the model to optimize accuracy while adhering to latency constraints.

In \cite{Park2024}, a set of RepTor models were utilized by applying the structural re-parameterization technique to temporal convolutions for efficient KWS models. Structural re-parameterization \cite{howard2017mobilenets}\cite{Ding2021} introduces a novel method to improve CNN efficiency. This technique separates the model structure for training and inference, allowing for a multi-branch model with enhanced training capacity during the training phase. The model is re-parameterized at inference time into a single-branch architecture for faster speed. RepTor-k models achieved a better accuracy-latency trade-off through the re-parameterisation technique than previous KWS models. Another interesting approach to obtaining a faster model involves designing hardware-efficient networks. Linear operators are typically highly optimized on processors to fully leverage computational capacity. Inspired by the efficient use of linear operators and the modeling efficacy of convolutions, the Linearized Convolution Network (LiCo-Net) was proposed in \cite{Yang2022}. This dual-phase system utilizes computation-efficient 8-bit linear operators during inference to optimize hardware utilization, while streaming convolutions are applied during training to maintain high model capacity.
\begin{figure}
	\centerline{\includegraphics[scale=0.4]{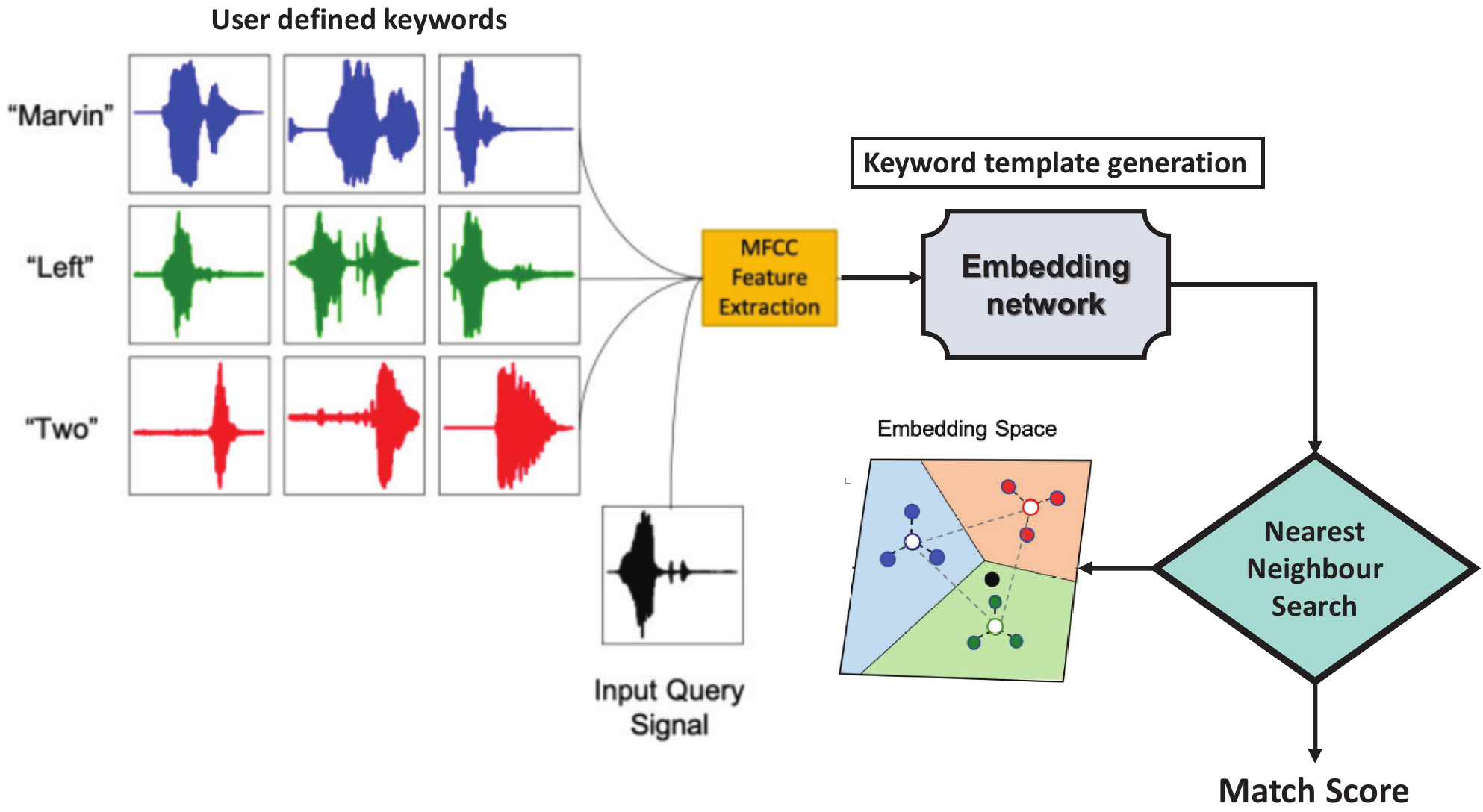}}
	\caption{Framework of a typical Query-by-Example based KWS}
	\label{fig5}
\end{figure}
\subsection{User customization}
When designing a DL-based KWS, a model is usually trained using a dataset of pre-specified keywords \cite{LopezEspejo2021}, enabling deployment across various IoT devices. Smartwatches and house robots are examples of consumer IoT gadgets that are intrinsically private and personal. On such devices, using a keyword spotter with preset keywords decreases personalization because other speakers can readily activate them. While these systems may perform effectively with specific predefined keywords, they pose an inconvenience for users who must memorize these keywords, and retraining the NN becomes necessary when the keywords change. Supporting customisable keywords that may be changed without retraining is more appropriate because many consumer IoT devices, such pet robots and smart speakers, are made for individual use. Therefore, the difficulty in creating a KWS for low-power IoT devices is to provide accurate detection of user-customized keywords while also decreasing memory and processing needs.

As per the findings in the study by Huang et al.~\cite{huang2022}, the DL model's performance in KWS can suffer a significant degradation when applied to unseen keywords in the target domain during runtime, despite its efficiency in terms of lower computation and a smaller footprint when trained on source-domain data. This challenge can be addressed through FSL, which offers a workable way to address the lack of plentiful user-defined keyword data. The Prototypical Network (ProtoNet) idea is the foundation of a number of new FSL techniques~\cite{snell2017}. In the intended scenario, users are requested to supply a few enrollment samples for every keyword during system setup. After that, a trained feature encoder processes these reference voice samples to create a set of feature vectors. Next, for each user-defined keyword, the mean of the feature vectors is calculated to create a class prototype. In inference, the distances are computed between the class prototypes and the test sample's output feature vector, or embedding \cite{huang2022}. 

Another widely adopted approach for detecting user-specified keywords is QbyE. A typical QbyE-KWS consists of two key components: FE and keyword search. The FE module plays a crucial role in the performance of keyword searches by utilizing an NN-based model to extract frame-level features. These features are generated by processing transcribed utterances and producing frame-level phoneme posteriorgrams. For example,  \cite{Li2022} employed an LSTM network as a feature extractor, as shown in Figure \ref{fig5}.  If the score surpasses a preset threshold, the keyword is detected; otherwise, it is not recognized. An in-depth exploration of these topics is outside the purview of this article. Readers seeking more information are directed to \cite{Rusci2023}\cite{Lei2023}.

\subsection{Robustness in KWS}
To ensure good performance of a KWS in real-life conditions, robustness must be maintained against various challenges, including but not limited to background noise, far-field scenarios, and other acoustic variabilities. To address these issues, various front-end methods \cite{Wang2017} such as speech enhancement \cite{Samui2019}, gain control, beamforming and adaptive noise cancellation are commonly employed. Usually, these techniques alter the voice signal prior to sending it to the acoustic model.

Additionally, advanced techniques like Adversarial training and multi-style training \cite{kwsprabhakar015} are used in the acoustic model to improve its generalization to various acoustic circumstances. These approaches often rely on data augmentation and the Fast Gradient Sign Method \cite{Goodfellow2014} to generate synthetic audio data for KWS training. 
A detailed discussion on these topics lies beyond the scope of this article. Interested readers are encouraged to refer to \cite{Zhang2018} for further information.
\begin{table*}
	\centering
	\scriptsize
	\label{table1}
	\caption{Summarized Description of the Notable Corpora Developed for KWS} 
	
	\begin{tabular}{|cccccc|}
		\hline \hline
		\multicolumn{1}{|c|}{\textbf{Name}} &
		\multicolumn{1}{c|}{\textbf{Language}} &
		\multicolumn{1}{c|}{\textbf{Developer}} &
		\multicolumn{1}{c|}{\textbf{\begin{tabular}[c]{@{}c@{}}No.of \\ KWS\end{tabular}}} &
		\multicolumn{1}{c|}{\textbf{Noisy?}} &
		\multicolumn{1}{c|}{\textbf{\begin{tabular}[c]{@{}c@{}}Publically \\ available?\end{tabular}}} 
		\\ \hline
		\multicolumn{1}{|c|}{AISHELL 2 \cite{du2018aishell}} &
		\multicolumn{1}{c|}{MANDARIN} &
		\multicolumn{1}{c|}{AISHELL} &
		\multicolumn{1}{c|}{20} &
		\multicolumn{1}{c|}{N} &
		\multicolumn{1}{c|}{Y} 
		\\ \hline
		\multicolumn{1}{|c|}{ALEXA \cite{Mishchenko2019}} &
		\multicolumn{1}{c|}{ENGLISH} &
		\multicolumn{1}{c|}{AMAZON} &
		\multicolumn{1}{c|}{1} &
		\multicolumn{1}{c|}{Y} &
		\multicolumn{1}{c|}{N} 
		\\ \hline
		\multicolumn{1}{|c|}{GOOGLE SPEECH COMMANDS V1 \cite{warden2017launching}} &
		\multicolumn{1}{c|}{ENGLISH} &
		\multicolumn{1}{c|}{GOOGLE} &
		\multicolumn{1}{c|}{30} &
		\multicolumn{1}{c|}{Y} &
		\multicolumn{1}{c|}{Y} 
		\\ \hline
		\multicolumn{1}{|c|}{GOOGLE SPEECH COMMANDS V2 \cite{warden2018speech}} &
		\multicolumn{1}{c|}{ENGLISH} &
		\multicolumn{1}{c|}{GOOGLE} &
		\multicolumn{1}{c|}{35} &
		\multicolumn{1}{c|}{Y} &
		\multicolumn{1}{c|}{Y} 
		\\ \hline
		\multicolumn{1}{|c|}{HEY SIRI \cite{higuchi2020stacked}} &
		\multicolumn{1}{c|}{ENGLISH} &
		\multicolumn{1}{c|}{APPLE} &
		\multicolumn{1}{c|}{1} &
		\multicolumn{1}{c|}{Y} &
		\multicolumn{1}{c|}{N} 
		\\ \hline
		\multicolumn{1}{|c|}{\begin{tabular}[c]{@{}c@{}}HEY SNAPDRAGON \\ KEYWORD DATASET \cite{kim2019query}\end{tabular}} &
		\multicolumn{1}{c|}{ENGLISH} &
		\multicolumn{1}{c|}{QUALCOMM} &
		\multicolumn{1}{c|}{4} &
		\multicolumn{1}{c|}{N} &
		\multicolumn{1}{c|}{Y} 
		\\ \hline
		\multicolumn{1}{|c|}{MOBVOIHOTWORDS \cite{hou2019}} &
		\multicolumn{1}{c|}{MANDARIN} &
		\multicolumn{1}{c|}{MOBVOI} &
		\multicolumn{1}{c|}{2} &
		\multicolumn{1}{c|}{Y} &
		\multicolumn{1}{c|}{Y} 
		\\ \hline
		\multicolumn{1}{|c|}{MULTILINGUAL SPOKEN WORDS CORPUS \cite{mazumder2021multilingual}} &
		\multicolumn{1}{c|}{\begin{tabular}[c]{@{}c@{}}50+\\ LANGUAGES\end{tabular}} &
		\multicolumn{1}{c|}{MLCOMMONS} &
		\multicolumn{1}{c|}{344,286} &
		\multicolumn{1}{c|}{Y} &
		\multicolumn{1}{c|}{Y}
		\\ \hline
		\multicolumn{1}{|c|}{Hi-MIA \cite{Qin2019}} &
		\multicolumn{1}{c|}{\begin{tabular}[c]{@{}c@{}}MANDARIN\end{tabular}} &
		\multicolumn{1}{c|}{AISHELL} &
		\multicolumn{1}{c|}{Wake words} &
		\multicolumn{1}{c|}{Y} &
		\multicolumn{1}{c|}{Y}
		\\ \hline
		\multicolumn{1}{|c|}{ARABIC SPEECH COMMANDS \cite{Ghandoura2021}} &
		\multicolumn{1}{c|}{ARABIC} &
		\multicolumn{1}{c|}{\begin{tabular}[c]{@{}c@{}}HIAST, SVU\\Innopolis University \end{tabular}} &
		\multicolumn{1}{c|}{40} &
		\multicolumn{1}{c|}{Y} &
		\multicolumn{1}{c|}{Y} 
		\\ \hline \hline
	\end{tabular}
\end{table*}
\subsection{Review of major corpora available for KWS}
KWS relies on well-structured speech datasets to enable efficient model training and evaluation. The GSCD is one of the most widely used corpora, providing labelled utterances of short words like ``yes'', ``no'',``stop'', and ``go'' in English, with two versions (V1 and V2) supporting DL models. For multilingual applications, the Common Voice (Mozilla) dataset offers diverse, crowdsourced speech data in over 100 languages, making it suitable for KWS tasks beyond English. The Multilingual Spoken Words Corpus (MSWC) expands this scope by providing keyword-labeled data in 50+ languages, enabling robust multilingual keyword recognition. Datasets such as the Arabic Speech Commands and HiMIA (Chinese Wake Word Dataset) cater specifically to non-English KWS applications. Additionally, Fluent Speech Commands focuses on natural phrase-based keyword detection for intent recognition, while VoxForge provides open-source multilingual speech data. For wake-word detection, specialized datasets like Hey Snips and Porcupine Wake Word Dataset offer real-world wake-word recordings. These corpora collectively drive advancements in KWS, supporting applications in voice assistants, IoT devices, and edge AI. The list of popular Publicly available datasets  for KWS is shown in Table 2.
\subsection{Evaluation metrics} 
\begin{figure}[htbp]
	\centerline{\includegraphics[scale=0.6]{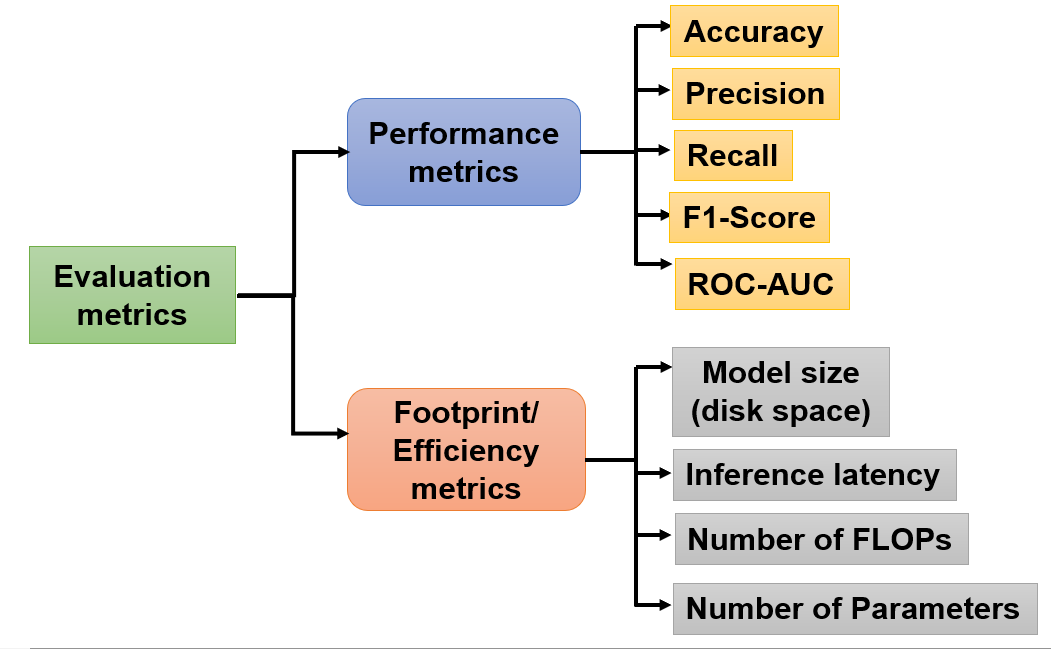}}
	\caption{A broad classification of efficiency metrics. When deploying a model, its feasibility is typically assessed based on performance (quality metrics such as accuracy, precision, and recall) and footprint metrics like model size, latency, and the number of FLOPs required for convergence. To effectively compare the efficiency of two models, both quality and footprint metrics must be considered \cite{Menghani2023}.}
	\label{fig6}
\end{figure}

In the era of DL-based KWS, two primary categories of metrics are of interest: performance metrics and model footprint metrics (refer to the efficiency metrics taxonomy in Figure~\ref{fig6}).

\begin{itemize}
	\item \textbf{Performance Metrics:} These metrics evaluate the model's effectiveness, including accuracy, F1-score, recall, and other indicators. High accuracy, true positive rate (TPR), and precision are desirable, while a low false positive rate (FPR) is preferred. Metrics such as the Receiver Operating Characteristic (ROC) curve and Area Under the Curve (AUC) provide insights into performance across various decision thresholds, while the F1-score balances precision and recall. 
	
To evaluate KWS  processing continuous audio streams, additional metrics such as False Alarm Rate per Hour (FAR), Miss Rate, Equal Error Rate (EER), and $C_\text{avg}$ are critical. The KWS task, often framed as a classification problem, can be binary (keyword/non-keyword distinction) or multiclass (with multiple keywords). For multiclass scenarios, performance metrics are typically computed for each keyword and then averaged.
	
	\item \textbf{Footprint/Efficiency Metrics:} These metrics quantify the costs associated with training and TinyML deployment of the model, including model size, latency, the number of parameters, number of floating-point operations (FLOPs), and RAM consumption during inference.
\end{itemize}
When two models achieve similar performance metrics for a given KWS task, the one with a lower footprint cost during training, inference, or both should be preferred, depending on the use case. Deploying models on resource-constrained devices (e.g., mobile and embedded systems) \cite{Shuvo2022} or costly computational platforms (e.g., cloud servers) further emphasizes the importance of inference efficiency.

\begin{figure}[htbp]
	\centerline{\includegraphics[scale=0.6]{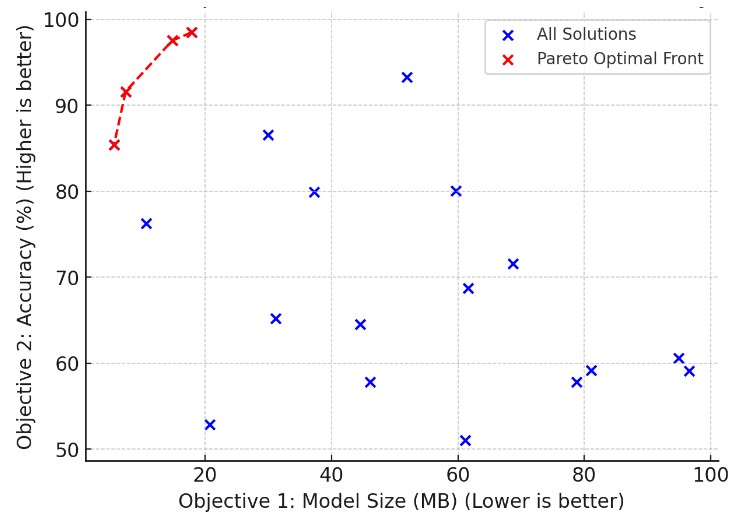}}
	\caption{Pareto optimal front for accuracy vs. model size. The red points (x) represent the optimal trade-offs, where accuracy is maximized while model size is minimized. The blue points (x) show all possible solutions. }
	\label{fig7}
\end{figure}

Regardless of the optimization goal, achieving Pareto-optimality is critical. A Pareto-optimal model ensures the best possible trade-offs for the metrics we prioritize. In TinyML-based KWS applications, achieving Pareto optimality involves balancing competing objectives such as model accuracy, inference latency, memory footprint, and power consumption. Traditional DL models often prioritize accuracy at the cost of increased computational complexity and model size, making them unsuitable for edge devices with constrained resources \cite{Shuvo2022}.  As Figure \ref{fig7} illustrates, the Pareto front (red points) represents the best trade-offs between accuracy and model size. Models along this front provide the highest possible accuracy for a given model size, ensuring efficient TinyML deployment. By analyzing the Pareto front, developers can select an optimal TinyML KWS model that meets specific deployment constraints without unnecessary overhead. This method guarantees that the chosen model is efficient and effective in real-world applications such as wake word detection on microcontrollers, wearables, and IoT devices.
\section{Review on efficient algorithms and models for SF-KWS \label{sec:sfkws}}
\begin{figure}[h]
	\centering
	\includegraphics[scale=0.7]{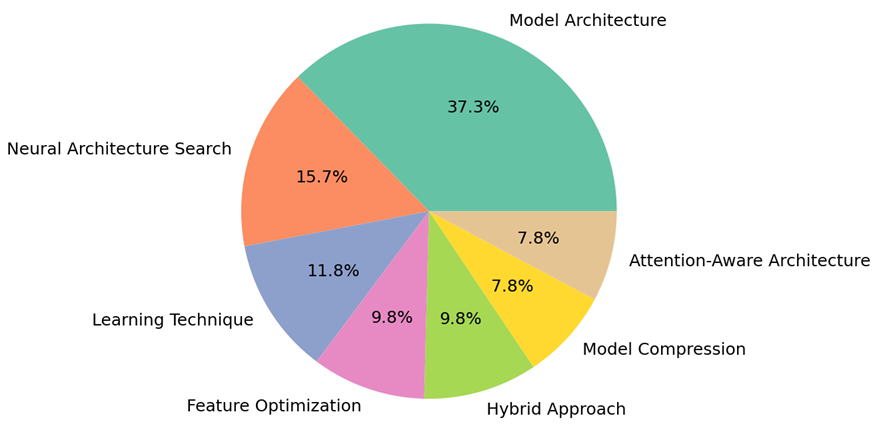}
	\caption{Distribution of SF-KWS research works based on algorithmic categories.}
	\label{fig:research_dist}
\end{figure}

This section presents notable research studies on SF-KWS that are specifically designed for resource-constrained devices based on the methods described in Section 3. These works are categorized following the same approach outlined in Section 3 and shown in the Table~\ref{table4}.  To quantify the distribution of research across these categories, we analyzed the surveyed literature and compiled a pie chart, as shown in Figure.~\ref{fig:research_dist}. The chart reveals that network architecture innovations dominate the field (37.3\%), followed by neural architecture search (15.7\%) and learning techniques (11.8\%). Other areas like hybrid approaches, feature optimization, model compression, and attention-aware architectures each contribute significantly, indicating a well-rounded exploration of approaches in recent years.
\begin{table}[htp]
\scriptsize
\renewcommand{\arraystretch}{1.2}
\centering
\caption{Summary of Methodologies Followed in the Major Research Contributions for Small-Footprint Keyword Spotting}
\label{table4}
\begin{tabular}{|c|c|c|c|}
\hline
\textbf{Category}                                                                                 & \textbf{Proposed Approach}                                                                                                                          & \textbf{Year} & \textbf{Authors}                                                       \\ \hline
\multirow{13}{*}{\textbf{\begin{tabular}[c]{@{}c@{}}Network \\ Architecture\end{tabular}}}          & DNN with posterior handling method  & 2014          & Guoguo Chen et al. \cite{chen2014}                     \\ \cline{2-4} 
                                                                                                  & CNN with pooling and striding                                                                                                                       & 2015          & Tara N. Sainath et al. \cite{Sainath2015conv}          \\ \cline{2-4} 
                                                                                                  & CRNN                                                                                                                                                & 2017          & Sercan Arik et al. \cite{Arik2017a}                 \\ \cline{2-4} 
                                                                                                
                                                                                                  & \begin{tabular}[c]{@{}c@{}}Use the residual learning techniques and dilated \\ convolutions,  and tune the depth and width of networks\end{tabular} & 2018          & Raphael Tang et al. \cite{tang2018}                    \\ \cline{2-4}
                                                                                                   &
                                                                                                   \begin{tabular}[c]{@{}c@{}}TC-ResNet\end{tabular} & 2019          & S. Choi et al. \cite{choi2019temporal}
                                                                                                   
                                                                                                  \\ \cline{2-4} 
                                                                                                  & Graph convolutional network & 2019          & Xi Chen et al. \cite{chen2019}                         \\ \cline{2-4} 
                                                                                                  & Multi-branch temporal convolution module (MTConv)                                                                                                   & 2020          & Ximin Li et al. \cite{Li2020}                          \\ \cline{2-4} 
                                                                                                  &  Matchbox Architecture                                                                                                                & 2020          & Somshubra Majumdar et al. \cite{majumdar2020}          \\ \cline{2-4} 
                                                                                                   & Broadcasted residual learning                          & 2021          & B. Kim et al. \cite{kim2021broadcasted}          \\ \cline{2-4} 
           
                                                                                                  & Quaternion neural models                                                                                                               & 2023          & Aryan Chaudhary et al. \cite{chaudhary2023towards}     \\ \cline{2-4} 
                                                                                                  & Slimming CNNs and Slimming Transformers Models                                                                                                      & 2023          & Zuhaib Akhtar et al. \cite{akhtar2023}                 \\ \hline \hline
\multirow{4}{*}{\textbf{\begin{tabular}[c]{@{}c@{}}Learning \\ Technique\end{tabular}}} & Low-rank weight matrices and knowledge distillation mechanism &   2016          & George Tucker et al. \cite{tucker2016}  \\ \cline{2-4} & Anchor-based region proposal network & 2019 & Jingyong Hou et al. \cite{hou2019}  \\ \cline{2-4}  & SSL based KWS using Transformer architecture    & 2019          & C. Gao et al. \cite{Gao2023}                      \\ \cline{2-4}   & SSL + KD                                                                                                   & 2023          & G. P. Yang et al. \cite{Yang2023}    \\ \hline \hline
\multirow{4}{*}{\textbf{\begin{tabular}[c]{@{}c@{}}Model \\ Compression\end{tabular}}}            & Post training Dynamic quantization                                                                                     & 2019          & Yusuf Goren et al. \cite{Mishchenko2019}               \\ \cline{2-4} 
																								   & \begin{tabular}[c]{@{}c@{}} Fixed point convolutional KWS \\ by combining two quantization-aware-training (QAT) techniques.\end{tabular}    & 2023          & Sashank Macha et al. \cite{macha2023}                 \\ \cline{2-4}  	
																								     & TDNN + SVD based  model complexity reduction          & 2017          & Ming Sun et al. \cite{Sun2017a}                         \\ \cline{2-4}
                                                                                                  & Error diffusion based speech feature quantization                                                                                                   & 2022          & Mengjie Luo et al. \cite{luo2022} 
                                                                                                  \\ \cline{2-4}
                                                                                                  & SSL+ QAT & 2024 & G. P. Yang \cite{Yang2024} 	                     \\ \hline \hline 
\multirow{3}{*}{\textbf{\begin{tabular}[c]{@{}c@{}}Attention-Aware \\ Architecture\end{tabular}}} & Average attention and soft attention                                                                                                                & 2018          & Changhao Shan et al. \cite{Shan2018}             \\ \cline{2-4} 
                                                                                                  & TDNN with a Shared Weight Self-Attention (SWSA) module                                                                                              & 2019          & Ye Bai et al. \cite{bai2019}                           \\ \cline{2-4} 
                                                                                                  & Keyword Transformer model                                                                                                                           & 2021          & Axel Berg et al. \cite{Berg2021}                       \\ \hline \hline
\multirow{4}{*}{\textbf{\begin{tabular}[c]{@{}c@{}}Feature\\ Optimization\end{tabular}}}          & Multi-Frame Shifted Time Similarity (MFSTS)                                                                                                         & 2019          & E. A. Ibrahim et al. \cite{mfslbrahim2019}            \\ \cline{2-4} 
                                                                                                  & SincConv layer extracts features                                                                                                                    & 2020          & Simon Mittermaier et  al. \cite{mittermaier2020}       \\ \cline{2-4} 
                                                                                                  & Binary speech features                                                                                                                   & 2019          & Alexandre Riviello et  al. \cite{riviello2019binary}       \\ \cline{2-4} 
                                                                                                  
& Autoencoder based data-driven approach & 2023          & P. Vitolo et  al. \cite{Vitolo2023}       \\ \hline \hline

\multirow{6}{*}{\textbf{\begin{tabular}[c]{@{}c@{}}Neural Architecture \\ Search\end{tabular}}}   
                                                                                                 & Performance-Oriented Neural Architecture Search                                                                                                     & 2019          & Andrew Anderson et al. \cite{anderson2019performance} \\ \cline{2-4} 
                                                                                                  & Stochastic Adaptive Neural Architecture Search                                                                                                      & 2019          & Tom Veniat et al. \cite{veniat2019stochastic}         \\ \cline{2-4} 
                                                                                                  & DARTS, a gradient-based differentiable NAS technique                                                                                                & 2020          & Tong Mo et al. \cite{mo20_interspeech}                \\ \cline{2-4} 
                                                                                                  & Search space on top of TC-ResNet with squeeze-and-excitation module                                                                                 & 2021          & Bo Zhang et al. \cite{Zhang2021autokws}                       \\ \cline{2-4} 
                                                                                                  & Micronets                                                                                                                                           & 2021          & Colby Banbury et al. \cite{Banbury2021}               \\ \cline{2-4} 
                                                                                                  & Target-Aware Neural Architecture Search                                                                                                             & 2022          & Paola Busia et al. \cite{Busia2022}                   \\ \hline \hline
\multirow{2}{*}{\textbf{\begin{tabular}[c]{@{}c@{}}Hybrid \\ Approach\end{tabular}}}   
                                                                        & Target-aware NAS + Fixed point quantization                                                                                        & 2018          & Yundong Zhang et al. \cite{zhang2017}                  \\ \cline{2-4}  
                                                                        & DARTS + SincConv + weight and activation quantization                                                                                               & 2022          & David Peter et al. \cite{Peter2022}                   \\ \hline
\end{tabular}
\end{table}
\subsection{Model architecture}
Developing compact and lightweight deep learning architectures is a crucial aspect of the SF-KWS approach. In the pre-deep learning era, Hidden Markov Models (HMMs) were widely used for KWS. However, their computational expense, particularly due to Viterbi decoding, a resource-intensive sequence search algorithm posed significant challenges. In 2014, Chen et al. \cite{chen2014} proposed a DNN-based architecture for SF-KWS. By comparing it with HMMs, they showed that the HMM system used an input window of 10 left frames and 5 right frames, with 2,002 output states, resulting in approximately 373K parameters. In contrast, the DNN-based KWS used 30 left frames and 10 right frames with only 3 or 4 output labels, resulting in a significantly smaller model ($\sim$ 244K parameters) and no need for Viterbi decoding or a separate decoder. This reduced runtime complexity and simplified implementation, making it more suitable for real-time applications. The DNN also made predictions every 10 ms, which minimized detection latency a critical factor for real-time KWS on resource-constrained devices.

Subsequently, convolutional neural networks (CNNs) were found to be more effective than fully connected DNNs for several reasons. First, CNNs operate on localized receptive fields, enabling them to detect relevant local patterns and achieve translation invariance via pooling. Second, DNNs are not inherently designed to handle translational variance in speech signals, often caused by variations in speaking style~\cite{Toth2014}. These variations shift formant positions in the frequency domain, typically requiring speaker adaptation techniques. While large DNNs can theoretically model this variation, they demand significantly more data and parameters. In contrast, CNNs efficiently model such variation with fewer parameters by averaging outputs over local time-frequency regions. CNNs are particularly well-suited to spectrogram-based speech features due to the strong correlations in both time and frequency dimensions. However, since frequency and time shifts have different implications in speech (unlike 2D image data), convolution along the frequency axis is more impactful~\cite{Toth2014}. Abdel-Hamid et al.\cite{AbdelHamid2012} and Sainath et al.\cite{Sainath2015conv} observed minimal benefits from time-axis convolution alone.

Nevertheless, standard CNN architectures such as \texttt{cnn-trad-fpool3} were computationally demanding due to the high number of multiplications, particularly in the second layer handling 3D inputs across time, frequency, and channels. These models were not ideal for low-power SF-KWS applications. To address this, Sainath et al.~\cite{Sainath2015conv} proposed more efficient CNN architectures by increasing the number of feature maps while keeping the model size under 250K parameters. They explored sub-sampling in both time and frequency domains (via striding or pooling) to expand feature map capacity without increasing model complexity. Their approach using time-striding filters followed by non-overlapping pooling yielded over 41\% relative performance improvement compared to DNNs under both clean and noisy conditions.

Following this line of research, numerous models have since been proposed for small-footprint KWS. Table~\ref{tab:kws_comparison} provides a comprehensive comparison of these recent models, highlighting their accuracy, parameter count, and architectural innovations.
\begin{table*}[ht]
	\centering
	\caption{Comparison of recent model architectures for SF-KWS}
	\scriptsize
	\begin{tabular}{||l|c|c|c||}
		\hline
		\textbf{Model} & \textbf{Accuracy (GSCD v2)} & \textbf{Params} & \textbf{Key Features / Novelty}\\
		\hline \hline
		\textbf{BC-ResNet-1 \cite{kim2021broadcasted}} & 96.9\% & 9.2K & Broadcasted residual learning; combines 1D/2D convolutions.\\
		\textbf{CENet-GCN \cite{chen2019}} & 96.8\% & 72K & Graph-based residual CNN; captures long-range temporal context. \\
		\textbf{CRNN \cite{Arik2017a} } & 97.71\% & 230K & CNN+RNN hybrid.\\
		\textbf{ED-sKWS \cite{song2024ed}} & 93.15\% / 90.14\% & 306K / 28K & Early-decision SNN; cumulative temporal loss.\\
		\textbf{MatchboxNet \cite{majumdar2020}} & 97.3--97.6\% & 77--140K & Scalable 1D depthwise-separable convolution.\\
		\textbf{QMatchboxNet \cite{chaudhary2023towards}} & 97.4\% & 140K & Quaternion convolutions; multi-view acoustic compression.\\
		\textbf{Res15 \cite{tang2018}} & 95.8\% & 238K & Deep residual CNN with dilated convolutions.\\
		\textbf{Slimmable CNN \cite{akhtar2023}} & 90.5\%  & 243K & Super-network supporting multiple widths.\\
		\textbf{SNN-KWS  \cite{wang2024global} } & 94.4\% & 70K & Global-local spike convolutions; PLIF neurons.\\
		\textbf{TC-ResNet14 \cite{choi2019temporal} } & $\sim$96.6\% & 305K & 1D temporal convolution for utilization of MFCC features. \\
		\textbf{TENet + MTConv \cite{Li2020}} & 96.84\% & 100K & Multi-scale temporal convolution.\\
		\textbf{WaveNet KWS \cite{coucke2019efficient}} & $\sim$98\% & 222K & Gated dilated convolutions; end-of-keyword supervision. \\
		\hline \hline
	\end{tabular}
	\label{tab:kws_comparison}
\end{table*}

\begin{figure}[htbp]
	\centerline{\includegraphics[scale=0.3]{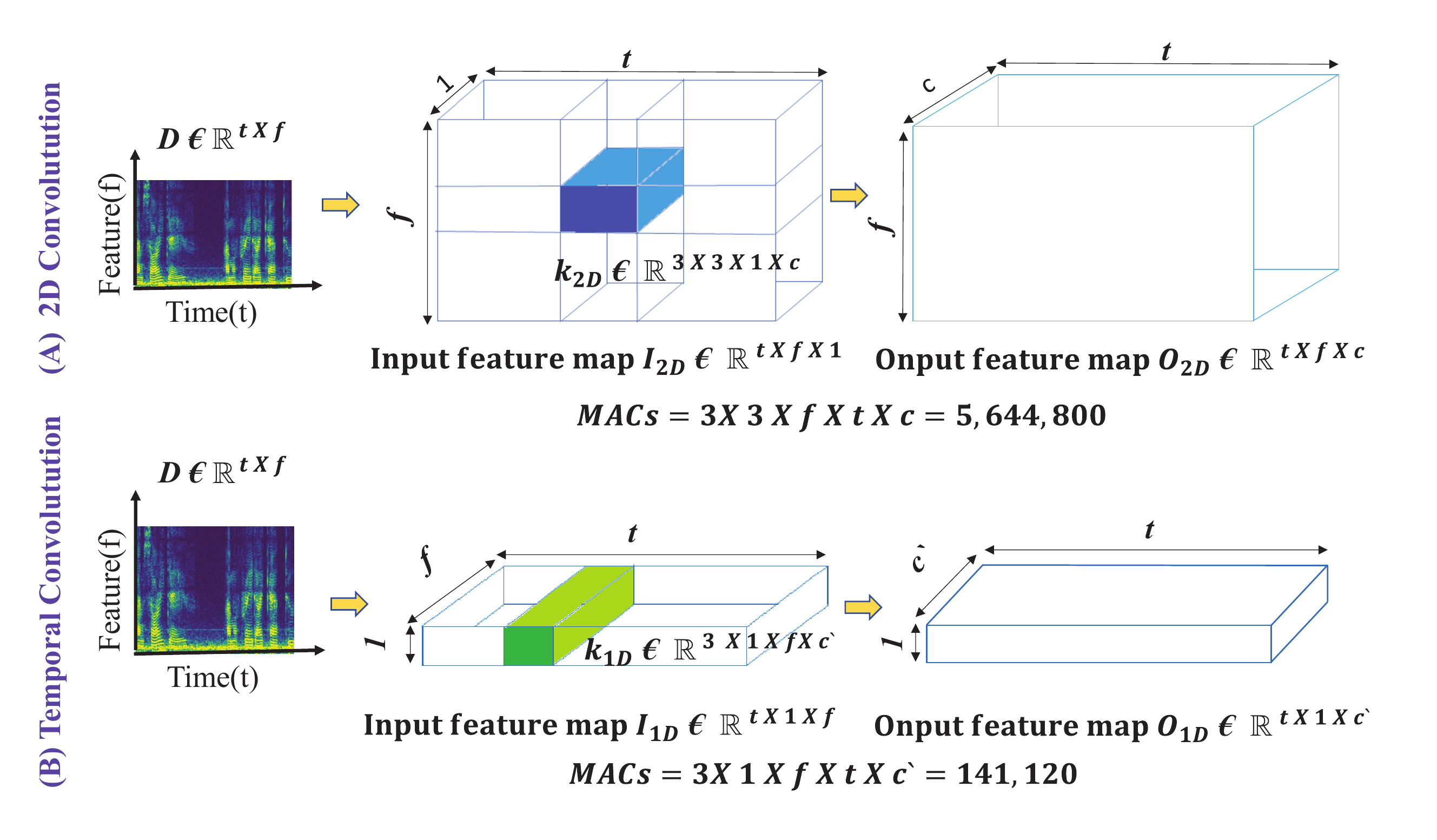}}
	\caption{A simplified example highlighting the difference between 2D convolution and temporal convolution: (A) \textbf{2D Convolution:} Typically used in conventional CNN-based approaches. (B) \textbf{Temporal Convolution:} Focuses on processing sequential data over time. In this example, both convolution types have the same parameter dimensions \cite{choi2019temporal} , set as follows: time steps (t) = 98, frequency bins (f) = 40, input channels (c) = 160, and output channels (c0) = 12.}
	\label{fig18}
\end{figure}

An alternative approach to achieving a computationally efficient model using a traditional 2D-CNN architecture is to employ Temporal Convolutional ResNet (TC-ResNet)\cite{choi2019temporal} which performs temporal convolution by reshaping 2D audio features (e.g., MFCC, log-Mel spectrogram) into 1D feature maps. This is done by treating the frequency axis as the channel direction and setting the height to 1. While CNNs are effective at transforming low-level features into higher-level concepts, capturing informative features across both low and high frequencies with a shallow network and small kernel size can be challenging. Author S. Choi et al. proposed reshaping 2D MFCC features from the input format $X_{2D}\in {\Bbb R}^{t \times f \times 1}$  (as shown in Figure \ref{fig18}) to $X_{1D}\in {\Bbb R}^{t \times 1 \times f}$ to create a fast and accurate model for real-time KWS. Their main idea was to treat per-frame MFCCs as time-series data rather than as intensity or grayscale images, which aligns more naturally with the representation of audio signals.  As illustrated by the colored box in Figure \ref{fig18}, this method leverages informative features across the entire frequency range in the lower layers, eliminating the need to stack multiple layers to extract higher-level features. This enables better performance with fewer layers, resulting in a smaller model size. Furthermore, applying this method shrinks the size of the two-dimensional feature map while keeping the number of parameters constant, as shown in Figure \ref{fig18}. Additionally, the output feature map (used as input for the next layer) of the temporal convolution is smaller than that of a standard 2D convolution. This reduction in feature map size significantly decreases the computational load and memory footprint in subsequent layers, making it ideal for fast and efficient KWS implementations.
While CNN-based KWS approaches, including 1D temporal convolution and 2D convolution, are widely used, they come with certain limitations. Temporal convolution requires fewer computations compared to 2D convolution; however, it cannot achieve translation equivariance along the frequency dimension due to its inherent biases. Conversely, 2D convolution-based methods address translation equivariance but are computationally more demanding than their 1D counterparts. To overcome these challenges, Author B. Kim et. al. proposed broadcasted residual learning \cite{kim2021broadcasted}, a technique that combines the strengths of both 1D and 2D convolution while minimizing computational overhead. Building on this innovation, they developed a family of networks called BC-ResNets, which achieved SOTA performance on the GSCD v1 and v2.

CNNs are effective at capturing local features, such as spatial details in images or short-term patterns in audio. However, they struggle to understand long-range dependencies across an entire input sequence without either large filter sizes or deep, computationally heavy architectures. So, Arik et al.\cite{Arik2017a} proposed a CRNN architecture, combining convolutional and recurrent layers to exploit the local structure and long-range context efficiently. They utilized GRU for their CRNN architecture, achieving a six-fold reduction in size compared to CRNN with Connectionist Temporal Classification (CTC) loss. This CRNN model leverages CNNs to extract local features from the audio spectrograms, which reduces the input dimensionality before feeding it into the RNNs. 
\begin{figure}[htbp]
	\centerline{\includegraphics[scale=0.5]{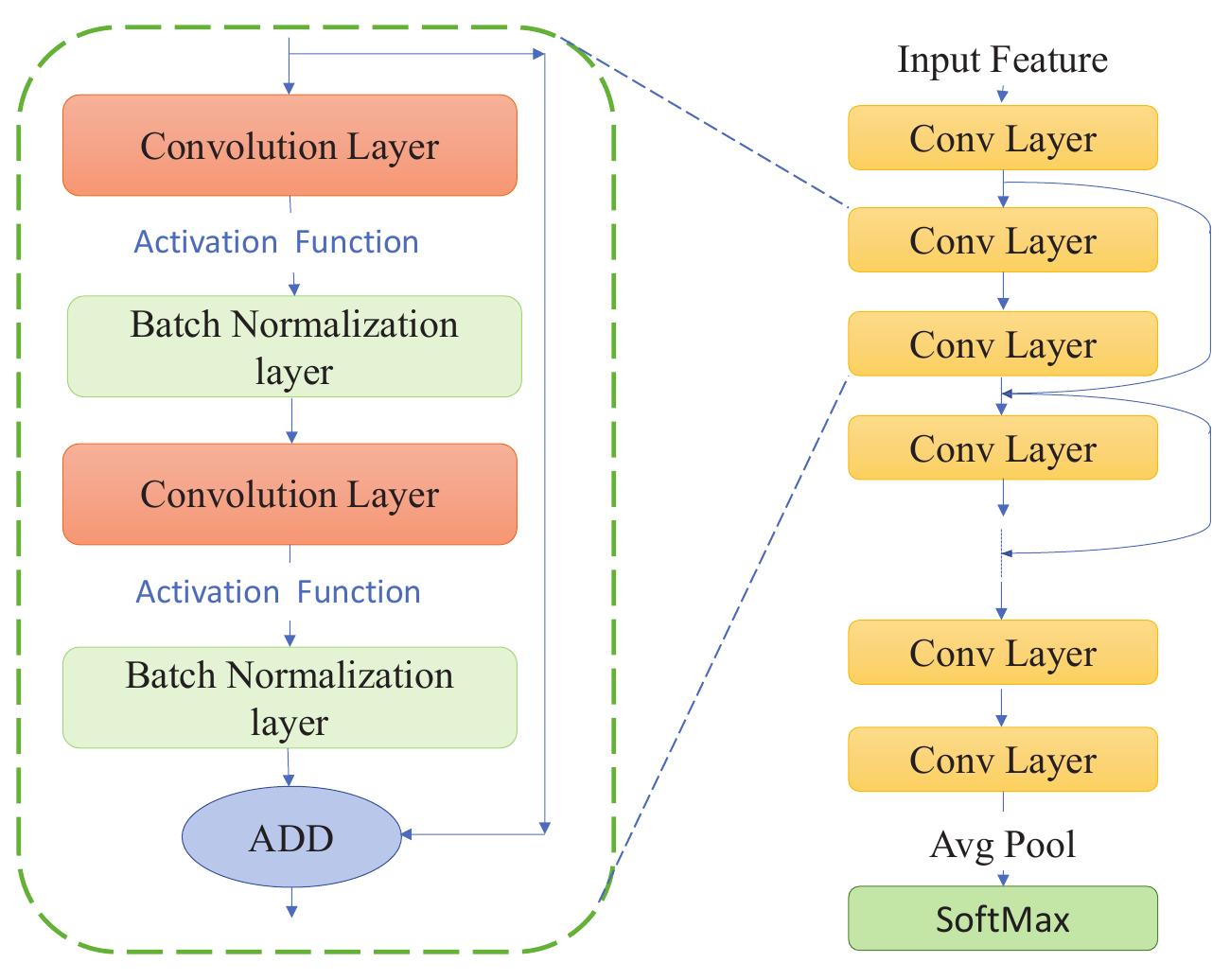}}
	\caption{ResNet architecture, with a magnified residual block.}
	\label{fig19}
\end{figure}
As we know, CNNs for KWS may need deep or wide architectures to achieve high accuracy, leading to larger models. However, with the increase in network depth, the number of required parameters also increases. This causes issues like vanishing gradients, which makes the training difficult and less efficient. The problem of vanishing gradient has been addressed by Tang et al.\cite{tang2018}. They used residual learning to resolve this issue and also utilized dilated convolutions to increase the receptive field without increasing the number of parameters. They used ResNet architecture, which directly connects the input of a residual block to its output, bypassing a few layers in between as visualized in Figure \ref{fig19}. Their best-performing model, Res15, achieves 95.8\% accuracy, significantly outperforming Google's prior CNN baselines. They also introduce compact versions like Res8 and Res8-narrow, which retain high accuracy (up to 94.1\%) while dramatically reducing model parameters and computational cost down to just 19.9K parameters and 5.65M multiplies for Res8-narrow. These models outperform prior compact CNNs (like Google's one-stride1) by a wide margin. The study highlights that model width impacts accuracy more than depth, and that residual learning and dilated convolutions are effective for enabling compact, accurate, and deployable KWS systems.

RNNs are indeed well-suited for handling temporal dependencies, but they struggle with state saturation in continuous input streams. State saturation in RNNs occurs when the hidden state starts to carry redundant information or loses relevant past information due to long sequences, which leads to increased computational costs and detection delays. To address these limitations, the authors in \cite{chen2019} proposed a compact and context-aware architecture for KWS that combines deep residual learning with graph convolutional networks (GCNs).  They introduced CENet, a convolutional model built with bottleneck residual blocks to minimize model size and computation. To further enhance performance, they integrate GCN modules into CENet. This hybrid model, CENet-GCN, significantly outperforms previous compact models on the GSCD. For example, CENet-GCN-6 achieves 95.2\% accuracy with only 27.6K parameters and 2.55M multiplies, outperforming ResNet-based models like Res15 in both accuracy and efficiency. 

Another notable work \cite{coucke2019efficient} proposed a WaveNet-inspired architecture for efficient, small-footprint keyword spotting (KWS) that avoids the limitations of recurrent networks. The model uses stacked dilated causal convolutions to capture long-term temporal dependencies, gated activations to control information flow, and residual skip connections for effective deep network training. A unique end-of-keyword labeling strategy is introduced, allowing the model to trigger detection only at the end of the keyword, improving robustness and reducing false alarms. Evaluated on the ``Hey Snips'' dataset, the proposed model outperforms both a CNN and a max-pooling LSTM baseline with significantly lower false rejection rates in both clean and noisy conditions achieving a 1.60\% FRR in noise, compared to 11.21\% for LSTM and 13.18\% for CNN. Despite having a similar number of parameters ($\sim$222K), it requires far fewer FLOPS than CNNs, making it highly suitable for real-time, on-device KWS applications.

In 2020, Ximin Li et al.\cite{Li2020} proposed a model for KWS called Temporal Efficient Neural Network (TENet) which utilizes depthwise temporal convolutions with an inverted bottleneck block structure inspired by MobileNetV2. To improve temporal feature extraction, the authors further introduce MTConv, a multi-branch temporal convolution module that captures both short-term and long-term dependencies using multiple convolution kernels of varying sizes (e.g., 3$\times$1, 5$\times$1, 9$\times$1). While MTConv is used during training, its fused equivalent is deployed during inference, thus incurring no additional parameter or computational cost. On the GSCD, TENet12 with MTConv achieves 96.84\% accuracy with only 100K parameters and 2.9M multiplies, outperforming several popular models in both accuracy and efficiency.

Another notable work is MatchboxNet \cite{majumdar2020}, a lightweight, end-to-end deep residual neural network . Its key innovation lies in using 1D time-channel separable convolutions, a variant of depthwise separable convolutions, which significantly reduce the model's parameter count and computation. The architecture is modular and scalable, allowing flexible tuning of depth, width, and number of residual blocks. It achieves state-of-the-art accuracy on GSCD (v1 and v2) with fewer parameters than comparable models such as ResNet.

In 2023, Aryan Chaudhary et al.\cite{chaudhary2023towards} proposed Quaternion Neural Networks (QNNs) for on-device KWS with a focus on achieving a low footprint. Quaternion models operate in a four-dimensional space (real + 3 imaginary components) and can inherently capture interdependencies in multi-view acoustic features such as Mel-spectrogram energy, velocity, acceleration, and jerk. The authors convert standard CNN-based KWS models (e.g., ResNet18, MatchboxNet, and BCResNet) into their quaternion counterparts (e.g., QResNet18, QMatchboxNet, QBCResNet). Experiments on the GSCD V2  showed that QNN-based models achieve comparable or better accuracy than real-valued models, often with significantly fewer parameters e.g., QMatchboxNet with just 30K parameters achieves 96.20\% accuracy. 

In 2023, Slimmable Networks were introduced as adaptable neural networks that dynamically adjust their width at runtime, enabling flexible trade-offs between accuracy and computational efficiency \cite{akhtar2023}. The core idea is to train a large, over-parameterized network (i.e., the super-network and Each sub-network represent a smaller variant of the super-network, defined by its width, which refers to the number of channels or neurons in each layer) that supports multiple configurations or widths. During training, the model adjusts its width dynamically to simulate different sub-network configurations. For example, a network with widths of 1.0, 0.75, 0.5, and 0.25 would be trained at each width by sharing weights among the sub-networks and using separate Batch Normalization layers for each width. When the network is slimmed down, only a subset of the weights is utilized. Slimmable Networks provide a flexible and efficient approach to small-footprint KWS by allowing a single model to adapt to different device constraints. This adaptability is achieved through dynamic slimming techniques that balance accuracy and computational efficiency, making it highly suitable for deployment in various edge devices with different performance and resource requirements.

Traditional deep learning based KWS models often require extensive computation and wait until the end of an audio sequence to make a prediction, which increases latency and energy consumption.  In contrast, several Spiking Neural Network-based KWS models have shown energy-efficient and low-latency keyword spotting for edge devices, though they tackle the problem with different approaches. For example, ED-sKWS \cite{song2024ed} emphasizes early decision-making by stopping the inference process once a confidence threshold is met, thereby considerably reducing both computation and energy usage. It enhances temporal learning through adaptive  Leaky Integrate-and-Fire  neurons and a novel Cumulative Temporal loss, achieving up to 52\% energy savings with just 27.63K parameters. In contrast, the SNN-KWS model by Wang et al. \cite{wang2024global}  introduces architectural innovations namely the Global-Local Spiking Convolution (GLSC) and Bottleneck-Parametric Leaky Integrate-and-Fire (PLTF) modules to balance fine-grained and contextual features without traditional front-end processing like MFCCs. This model emphasizes frame-by-frame inference and efficient spiking operations, yielding a 10x energy reduction while maintaining performance with only 70.1K parameters. While ED-sKWS excels in early and adaptive inference for ultra-low-power scenarios, the GLSC-based model stands out in structural efficiency and real-time, high-fidelity processing, making both highly suitable for small-footprint, on-device KWS under different deployment constraints.
\subsection{Learning techniques}
Leveraging advanced learning techniques has been instrumental in improving the efficiency of KWS systems.  Various researchers have explored techniques like KD, self-supervised learning, data augmentation, low-rank weight matrices, etc. aim to train models in a way that improves generalization while maintaining a small memory footprint and low computational requirements. In general, these methods enhance accuracy and efficiency, making KWS models more suitable for real-world applications with limited resources. One key advantage of focusing on learning techniques is that they are applied exclusively during the training phase, leaving the inference process unaffected.
\begin{figure}[h]
	\centerline{\includegraphics[scale=0.5]{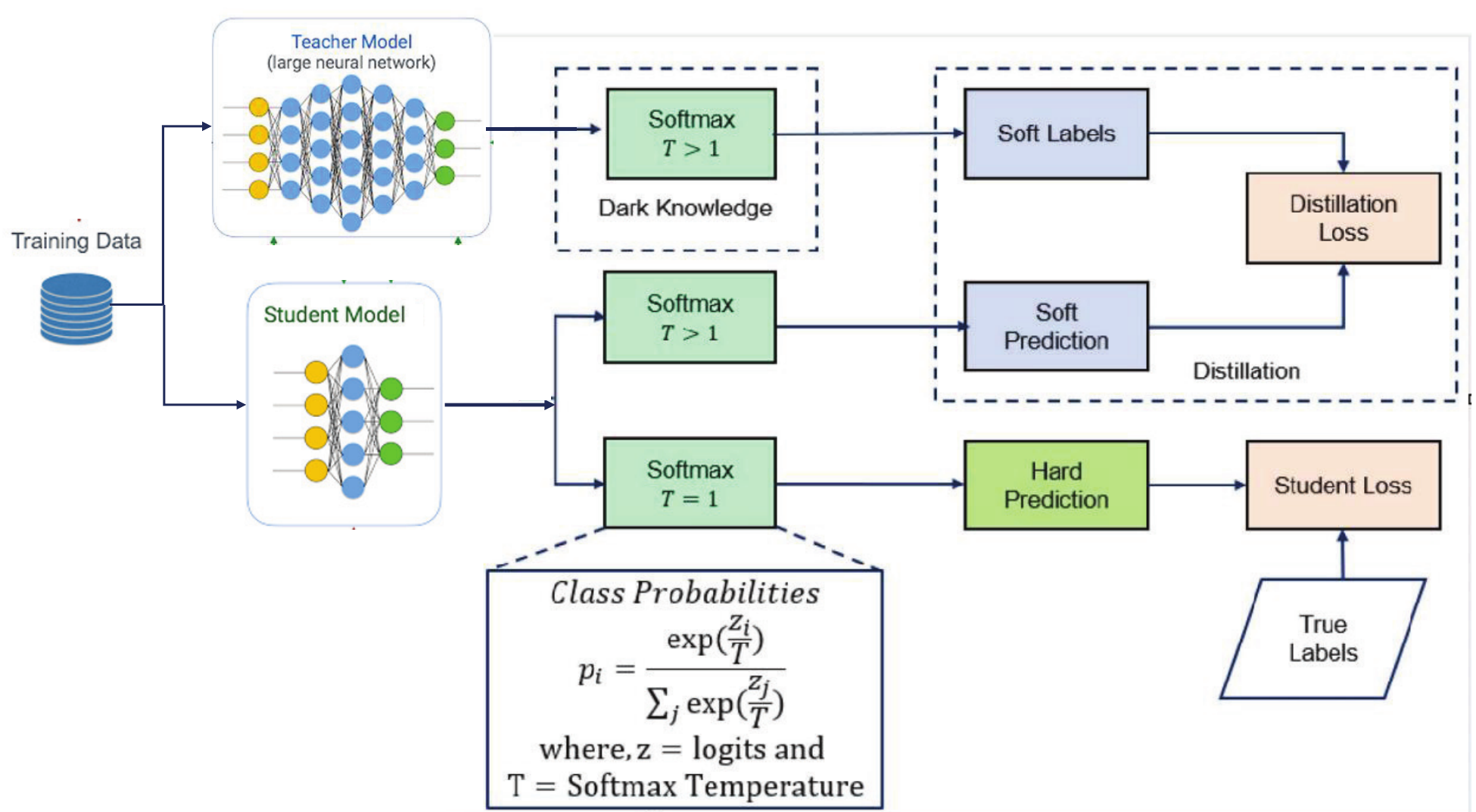}}
	\caption{Knowledge Distillation Process: A pre-trained Teacher model generates soft labels (probabilities) through a softmax layer. These soft labels guide the training of a smaller Student model, which learns to mimic the teacher’s predictions using a distillation loss function \cite{Shuvo2022}}
	\label{fig10}
\end{figure}
In KD \cite{hinton2015distilling}, a smaller, lightweight \textit{student} model $S(.)$ is trained to emulate a larger, pre-trained \textit{teacher} model $T(.)$. Given an input-label pair $(x, y)$, the teacher model predicts $\hat{y}^{t} = T(x)$, and the student model is learned to generate a comparable prediction, $\hat{y}^{s} = S(x) \approx \hat{y}^{t}$ as shown in Figure \ref{fig10}. This is achieved by optimizing the student model's parameters using a composite loss function consisting of two main components: 
\begin{enumerate}
	\item The cross-entropy loss $\mathcal{L}_{CE}$, which uses the ground-truth labels as targets, and
	\item The distillation loss $\mathcal{L}_{KD}$, designed to impart knowledge from a teacher to the student.
\end{enumerate}
Usually, the model's output is presented as a probability distribution across classes:
\begin{equation}
	p(z_j) = \sigma(z_j) = \frac{\exp(z_j)}{\sum_{k} \exp(z_k)},
\end{equation}
where $z_j$ represents the logits for class $j$. The student model parameters $\theta_s$ are optimized as follows:
\begin{equation}
	\theta_s^* = \arg\min_\theta \mathcal{L}_s = \arg\min_\theta \lambda \mathcal{L}_{CE} + (1 - \lambda) \mathcal{L}_{KD},
\end{equation}
where $\lambda$ balances the contributions of $\mathcal{L}_{CE}$ and $\mathcal{L}_{KD}$. The cross-entropy loss $\mathcal{L}_{CE}$ is defined as:
\begin{equation}
	\mathcal{L}_{CE} = -\sum_{i=1}^C y_i \log(p(z_s, T)),
\end{equation}
where $y$ is the one-hot encoded label, $z_s$ represents the student logits, and $C$ is the number of classes.

The distillation loss $\mathcal{L}_{KD}$ is generally expressed as:
\begin{equation}
	\mathcal{L}_{KD} = D[\Psi(T(.), \theta_T), \Psi(S(.), \theta_s)],
\end{equation}
where $\theta_T$ and $\theta_s$ are the trainable parameters of the teacher and student, respectively. $\Psi(.)$ represents the knowledge extracted from the models, and $D(.)$ measures the divergence between their representations. Enhancements in KD involve modifying either the type of knowledge transferred or the transfer strategy.

Building on this idea, Tucker et al. \cite{tucker2016} applied a combination of quantization, low-rank matrix factorization, and KD to compress large feedforward DNNs used for KWS tasks. They begin with a baseline model of 1.8 million parameters and systematically reduce the size using SVD and 8-bit quantization, achieving up to 93\% reduction in model size with negligible loss in performance. Furthermore, the use of KD from a large teacher model to smaller student models enhances robustness, especially under noisy conditions. The resulting compressed models maintain high accuracy while being computationally efficient and suitable for deployment on resource-constrained devices like smart speakers and IoT hardware.

On the other hand, the work described in \cite{Gao2023}  explores the application of SSL to KWS using lightweight transformers suitable for compute-constrained edge devices. Traditional SSL methods require large models with millions of parameters, making them impractical for on-device applications. This study addresses the challenge by employing transformers with only 330K parameters and introducing an utterance-wise distinction mechanism to enhance classification accuracy. The primary objective is to improve performance while reducing reliance on labelled datasets. The authors evaluate three different SSL methods: Auto-Regressive Predictive Coding (APC), Masked Predictive Coding (MPC), and Contrastive Learning (CL). APC is a generative-loss-based approach where a model predicts future frames based on past information, with a training objective given by:
\begin{equation}
L_{\text{APC}} = \sum_{i=1}^{T-n} \| x_{i+n} - y_i \|
\end{equation}
where \( x \) represents the target sequence, \( y \) is the predicted sequence, and \( n \) denotes the number of future steps predicted. MPC reconstructs masked features using bi-directional context, and its loss function is:
\begin{equation}
 L_{\text{MPC}} = \sum_{i=1}^{T} w_i \| x_i - y_i \|
\end{equation}
 where \( w \) controls the contribution of masked and unmasked regions. CL maximizes the distinction between positive and negative sample pairs using:
\begin{equation}
 L_{\text{CL}} = - \sum_{i=1}^{T} w_i \log \frac{\exp (\Phi(y_i, q_i) / \kappa)}{\sum_{q \in Q_p} \exp (\Phi(y_i, q) / \kappa)} + \beta L_D
\end{equation}
where \( Q_p \) represents acoustic unit-level centroids, \( \Phi(\cdot) \) is a cosine similarity function, and \( \kappa \) is a temperature scaling factor. To improve utterance-wise distinction, the study introduces a two-step contrastive-learning mechanism, which extracts utterance-level representations and enhances classification accuracy. The utterance-wise contrastive learning loss is given by:
\begin{equation}
 L_{\text{utt}} = - \log \frac{\exp (\Phi(u_1, Q^+) / \kappa)}{\sum_{q \in Q} \exp (\Phi(u_1, q) / \kappa)} + \beta L_D
\end{equation}
where \( u_1 \) is the extracted utterance representation and \( Q^+ \) represents positive samples. The final training objective is:
\begin{equation}
 L = \alpha L_{\text{SSRL}} + (1 - \alpha) L_{\text{utt}}
\end{equation} 
where \( \alpha = 0.9 \) maintains self-supervised learning properties. The approach was evaluated on the GSCD V2 and their in-house dataset, achieving a 1.2\% accuracy improvement on the former and a 6\% to 23.7\% FAR reduction on the latter. Among the tested SSRL methods, APC yielded the best results, improving KWS accuracy. The study confirms that SSL can be effectively applied to lightweight models for on-device KS, offering a viable alternative to supervised learning. The utterance-wise distinction mechanism further enhances performance, making SSRL-based KS models more reliable for real-world applications. 
 
 Self-supervised KD is also effective for on-device KWS. To mitigate the problem of high computational complexity issue, a teacher-student KD approach was introduced in \cite{Yang2023}, enabling knowledge transfer from a large, pre-trained model to a smaller, lightweight model. This process is facilitated by dual-view cross-correlation distillation and teacher codebook distillation. The method is evaluated on a 16.6K-hour in-house Alexa keyword dataset, demonstrating significant improvements in both normal and noisy conditions.  
 
 Within the teacher-student framework, the student model is trained to learn high-fidelity representations from the teacher by minimizing L1 distance and cosine similarity loss, formulated as:
\begin{equation}
 L = \sum_{t=1}^{T} \left[ \|h_t - o_t\|_1 - \lambda \sigma \left( \cos(h_t, o_t) \right) \right]
\end{equation} 
where \( h_t \) and \( o_t \) represent the hidden features of the teacher and student models at time \( t \), and \( \lambda \) is a weighting factor set to 1. Unlike frame-wise approaches, utterance-wise features are considered, modifying the loss function to:
\begin{equation}
 L = \|h - o\|_1 - \lambda \sigma \left( \cos(h, o) \right)
\end{equation}  
 where \( h \) and \( o \) denote averaged features over time.
 
To enhance the effectiveness of distillation, Dual-View Cross-Correlation Distillation (DVCC) is employed to regularize feature redundancy across batch-view and feature-view correlations. The feature-view loss function is defined as:
\begin{equation}
 C_{ij} = \frac{\sum_b H_{bi} O_{bj}}{\sqrt{\sum_b (H_{bi})^2} \sqrt{\sum_b (O_{bj})^2}}
\end{equation}  
where \( C \) represents a square matrix that captures feature correlations. The redundancy reduction objective is expressed as:
\begin{equation}
 L_C = \sum_i (C_{ii} - 1)^2 + \alpha \sum_{i, j \neq i} C_{ij}^2
\end{equation}

Similarly, the batch-view correlation matrix \( G \) is computed as:
\begin{equation}
 G_{ij} = \frac{\sum_d H_{id} O_{jd}}{\sqrt{\sum_d (H_{id})^2} \sqrt{\sum_d (O_{jd})^2}}
\end{equation} 
with the corresponding contrastive loss:
\begin{equation}
 L_G = \sum_i (G_{ii} - 1)^2 + \beta \sum_{i, j \neq i} G_{ij}^2
\end{equation}

By combining both views, the dual-view cross-correlation loss is formulated as:
\begin{equation}
 L_{\text{DVCC}} = \frac{L_C}{\text{sg}(L_C)} + \frac{L_G}{\text{sg}(L_G)}
\end{equation}  
where \(\text{sg}\) represents the stop gradient operation.
 
Additionally, Teacher Codebook Distillation is introduced to mitigate data bias in small student models by utilizing the pre-trained codebook from the teacher model rather than learning a new one. The student model follows the Wav2Vec 2.0 objective \cite{Baevski2020}, selecting positive and negative samples from the teacher's quantized features to reduce spurious noise and improve generalization. The teacher codebook contrastive loss is defined as:
\begin{equation}
 L_{\text{t-code}} = - \sum_t \log \frac{\exp(\cos(o_t, k_t))}{\sum_{k \sim K_t} \exp(\cos(o_t, k))}
\end{equation}   
where \( o_t \) represents the student model's output, \( k_t \) is the positive sample from the teacher codebook, and \( K_t \) includes both positive and negative samples. The final combined objective integrates both techniques:
\begin{equation}
 L_{\text{combined}} = L_{\text{DVCC}} + \gamma L_{\text{t-code}}
\end{equation}  
where \( \gamma \) controls the trade-off between the two losses.
 
Experiments conducted on the 16.6K-hour Alexa dataset demonstrate that dual-view cross-correlation distillation surpasses baseline methods by reducing the relative FAR by 14.6\% under normal conditions and 21.3\% in noisy conditions. When teacher codebook distillation is incorporated, further improvements are observed, leading to a FAR reduction of up to 23.8\%. 
\subsection{Model compression}
Two main categories can be used to group different DL model compression strategies. The first category focuses on reducing the number of trainable parameters, a process known as Network Pruning. The second category, called Network Quantization \cite{Sze2017}, involves decreasing the bitwidth of operations and operands (such as weights, activations, or both) by transitioning from floating-point to fixed-point representations \cite{DNNbook2020}.
\begin{itemize}
	\item Network Pruning: 
	Consider a neural network \( N_W(X) \), where \( X \) denotes the input and \( W \) represents the set of learnable parameters (weights). \textit{Pruning} is a model compression technique aimed at identifying a smaller subset of important parameters, denoted as \( W_p \), while setting the remaining weights in \( W \) to zero. This process typically leads to a sparse representation of the network, where most weights are zero, without causing a significant drop in performance or accuracy. The pruned network is thus characterized by sparse weight tensors. The level of sparsity is quantified as the ratio of pruned (zeroed-out) parameters to the total number of parameters in the original network. A higher sparsity ratio implies that the pruned model retains fewer non-zero weights, reflecting a more compact and efficient representation \cite{Menghani2023}.
	
	Two foundational works in this domain are \textit{Optimal Brain Damage}~\cite{LeCun1989} and \textit{Optimal Brain Surgeon}~\cite{Hassibi1993}, which utilize the Hessian matrix of the loss function to assess the importance of each weight (i.e., weight saliency). Weights with the lowest saliency are pruned, and the remaining ones are fine-tuned to recover any lost accuracy.
	
	Network pruning typically starts with a large, pre-trained, DNN and proceeds in two stages:
	\begin{enumerate}
		\item Removing specific weights from the trained DNN based on a defined criterion.
		\item Fine-tuning the network to update the values of the remaining (non-zero) weights.
	\end{enumerate}
	These two stages are typically repeated in an iterative manner to progressively enhance the sparsity of the DNN. The strategy for deciding the number of weights to prune in each iteration is referred to as \textit{scheduling}. Biases are usually not pruned, as their quantity is significantly smaller compared to the number of weights.

	\item Quantization:
	In trained NN, activations and weights are typically represented using a 32-bit floating-point format. While this retains information with minimal loss, it results in slower processing. Quantization addresses this by mapping model parameters and activations to lower-precision levels, often using 8-bit fixed-point integers, thereby avoiding expensive FLOPs. Quantization offers two key benefits: (1) reduced memory usage due to the compact representation with fewer bits and (2) faster inference times. When the primary concern is model size, weight quantization can be applied, where only the model weights are represented in reduced precision. To achieve latency improvements, activation quantization is also necessary, ensuring that all operations in the quantized graph are performed using fixed-point arithmetic \cite{Menghani2023}.
	
	A straightforward method for quantizing weights to reduce model size is as follows. Given a 32-bit floating-point weight matrix, the minimum weight value ($x_{min}$) in the matrix can be mapped to 0, while the maximum value ( $x_{max}$) is mapped to $2^b$-1, where  b represents the number of bits used for quantization and is less than 32.
	All intermediate values are then linearly scaled to an integer within the range [0,$2^b$-1]. This process effectively converts each floating-point value into a fixed-point representation, requiring fewer bits than the original floating-point format. A commonly used value for is 8, as it provides a 4$\times$ reduction in storage size ($\frac{32}{8} $)  while maintaining compatibility with widely supported uint8-t and int8-t datatypes. During inference, the process is reversed, where a lossy approximation of the original floating-point value is reconstructed through dequantization using only $x_{min}$ and $x_{max}$ \cite{DNNbook2020}.

	The process of quantizing a pre-trained model's weights to reduce its size is commonly referred to as post-training quantization (PTQ) in the literature. This approach is often sufficient for model size reduction without requiring any additional retraining of the DNN. However, PTQ may degrade model performance during inference, as noted in previous studies \cite{Jacob2018}\cite{Nahshan2021}. The primary reason for this degradation is that quantization introduces perturbations to the trained model parameters, potentially shifting the model away from the optimal state it reached during training with floating-point precision \cite{Gholami2022}. To mitigate this issue, the model can be retrained with quantized parameters, allowing it to reconverge to a more optimal state with improved loss. QAT, a widely adopted method is employed for achieving this. In QAT, training is conducted in floating-point precision, but the forward pass simulates quantization effects that occur during inference. This is done by passing both weights and activations through a function that mimics the quantization process, commonly referred to as fake quantization in the literature \cite{Menghani2023}\cite{Jacob2018}. During the backward pass, updates are performed using floating-point precision since accumulating gradients in a quantized format can lead to zero gradients or highly inaccurate gradients, particularly in low-bit precision settings.
	
\end{itemize}
Various works explored novel quantization techniques aimed at reducing the memory and computational footprint of DNNs for KWS while maintaining high accuracy. Traditional PTQ methods typically quantize model weights at the matrix level, which can lead to performance degradation in low-bit quantization. To address this, a dynamic quantization (DQ)  approach is introduced \cite{Mishchenko2019}, where column-wise quantization is applied to weight matrices, and input/output activations are dynamically quantized per frame based on actual min-max values. This technique allows for more precise quantization and reduces performance loss in 8-bit and 4-bit models. Experiments conducted on an in-house 500-hour far-field speech dataset demonstrate that 8-bit quantized models achieve near full-precision performance, while 4-bit models achieve up to 80\% memory footprint reduction with moderate accuracy loss. A hybrid 4-8 bit quantization approach further balances performance and efficiency, reducing the model size by 70\% while maintaining near full-precision accuracy.

M. Luo et al. \cite{luo2022} proposed error-diffusion-based speech feature quantization  to improve low-precision input feature representations in small-footprint KWS models. Unlike previous quantization approaches that primarily focus on reducing model weights and activations, this method applies Floyd-Steinberg error diffusion \cite{Ostromoukhov2001}, an image-processing technique, to quantize log-Mel spectrograms while preserving speech feature quality. The study demonstrates that a 3-bit error-diffused representation results in only a 0.45\% accuracy drop compared to full-precision log-Mel spectrograms, whereas conventional linear and max-min quantization suffer from over 3\% accuracy degradation. Additionally, 1-bit quantization achieves practical performance with only a 1.7\% accuracy loss, enabling deployment in binary neural networks. The effectiveness of time-direction error diffusion  over filter-direction diffusion is also analyzed, concluding that the former diffusion better preserves temporal feature structures in TCNs.

Ming Sun et. al.\cite{Sun2017a}  introduced a compressed Time Delay Neural Network (TDNN) approach for small-footprint keyword spotting (KWS), targeting devices with limited CPU, memory, and latency budgets. The authors propose training a full-rank TDNN using transfer learning and multi-task learning, and then compressing it with Singular Value Decomposition (SVD) to significantly reduce model size and computational cost without sacrificing performance. The TDNN architecture uses temporal sub-sampling to reduce redundant computations and is integrated with an HMM-based decoder for keyword detection. Experiments conducted on a large, in-house far-field dataset show that the compressed TDNN achieves up to 37.6\% reduction in Detection Error Tradeoff (DET) AUC compared to a similarly sized DNN baseline. The system outperforms both uncompressed TDNN and SVD-compressed DNN alternatives, demonstrating its effectiveness for real-time, on-device KWS.

Another interesting work \cite{Yang2024} investigated the application of QAT to optimize SSL models for kWS on resource-limited edge devices. It aimed to improve computational complexity  by reducing bit precision of model weights and activations through QAT, ensuring that high accuracy is maintained despite a 75\% model size reduction. The proposed approach is evaluated on a 16.6K-hour in-house KWS dataset, demonstrating that quantized models perform comparably to full-precision models, even with a fourfold reduction in bit size. In their proposed methodology, QAT is integrated with Self-Supervised Learning (SSRL), employing Autoregressive Predictive Coding (APC) as the self-supervised learning objective. The APC loss function is given by:
\begin{equation}
L_{\text{APC}} = \sum_{t=1}^{T-k} \| x_{t+k} - f_W(x_{1:t}) \|^2_2
\end{equation} 
where \( x_{t+k} \) represents the future frame prediction, \( x_{1:t} \) denotes past frames, and \( f_W \) is the model function parameterized by weights \( W \). 
To facilitate low-bit quantization, Absolute Cosine Regularization (ACR) is introduced to guide weights toward predefined quantized values. The ACR loss function is expressed as:
\begin{equation}
L_{\text{ACR}} = -\sum_{i} | \cos(\pi f w_i) |
\end{equation}
where \( w_i \) represents model weights, and \( f \) determines the frequency of quantized values. The total training objective, incorporating APC and ACR, is formulated as:
\begin{equation}
L_{\text{total}} = L_{\text{APC}} + \alpha L_{\text{ACR}}
\end{equation}
where \( \alpha \) is a hyperparameter controlling the impact of quantization constraints.
Beyond model weight quantization, activation quantization is explored using Moving Average Quantization and DQ.
Their experimental results on the 16.6K-hour dataset indicate that QAT models achieve performance comparable to full-precision models while significantly reducing model size. It is observed that PTQ alone results in a 32\%-41\% degradation in FAR, whereas QAT preserves accuracy. The best performance is obtained using DQ without ACR, as ACR excessively restricts weight distribution, leading to suboptimal model utilization. 
\subsection{Attention aware architecture}
The attention mechanism in SF-KWS systems plays a crucial role in creating small-footprint models by efficiently focusing on the most relevant portions of the audio input, thereby reducing unnecessary computations \cite{Ding2022}. By dynamically assigning more importance to key segments like the specific keyword and less to background noise, these models can achieve high accuracy without processing the entire input sequence in detail. This selective focus allows for fewer parameters and lower computational load, making the system more suitable for for deployment on resource-constrained devices.

 Changhao Shan et al.\cite{Shan2018} proposed an efficient, attention-based end-to-end architecture for keyword spotting on resource-constrained devices. The model combines a recurrent encoder (such as LSTM, GRU, or CRNN) with an attention mechanism to generate a fixed-length representation directly from the audio input, eliminating the need for frame-level alignments. Despite its compact size (84K parameters approx.), the model achieves competitive performance, making it suitable for real-world deployment in small-footprint applications.. GRUs are generally more parameter-efficient than LSTMs because they have fewer gates and are simpler in structure. This makes them a better choice for small-footprint models. They have also  shown that GRU models achieve lower FRR with fewer parameters than LSTM models. 
 
 The attention mechanism in keyword spotting models can be implemented in two ways: average attention, which assigns equal importance to all time steps, and soft attention, which allows the model to dynamically focus on the most relevant parts of the input \cite{Niu2021}. Soft attention calculates scores using a learned function and normalizes them with a softmax to determine how much attention to pay to each time step. This enables the model to selectively highlight important segments of the audio, such as keyword regions, while ignoring background noise. The resulting context vector is then used to predict the presence of a keyword.

\begin{figure}[htbp]
\centerline{\includegraphics[scale=0.3]{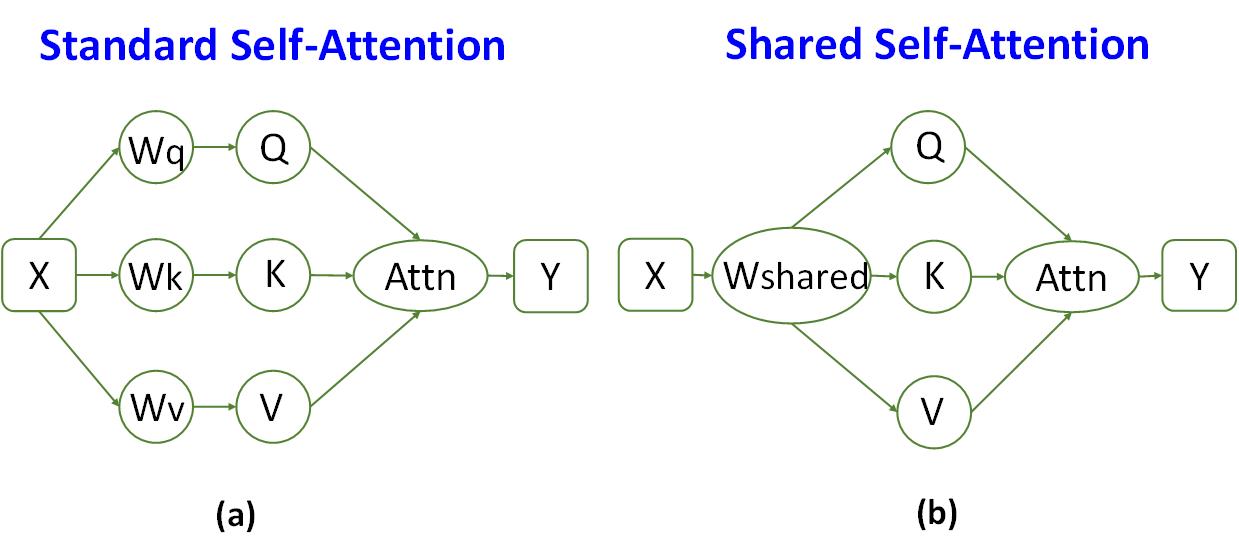}}
\caption{(a) \textbf{Standard Self-Attention} use three weight matrices, $W_q$, $W_k$, and $W_v$, each transforming the input X into queries (Q), key (K), and values (V). (b) \textbf{Shared-Weight Self-Attention} uses a single weight matrix of $W_{shared}$ for all three transformations. This reduces the number of trainable parameters by a factor of 3, leading to lower memory usage and faster computations \cite{Niu2021}.}
\label{fig22}
\end{figure}

 Ye Bai et al. \cite{bai2019} proposed integrating a  with a Shared Weight Self-Attention (SWSA) mechanism. This design captures both local and global features of audio sequences while significantly reducing the number of parameters. By sharing weights within the self-attention module, the model achieves a parameter count of just 12,000-approximately 1/20th of a comparable ResNet-based KWS model without compromising performance. Evaluated on the GSCD, the model attains an error rate of 4.19\%, closely matching the 4.12\% error rate of the much larger ResNet model. SWSA mechanism as visualized in Figure \ref{fig22} offers several advantages over traditional self-attention. By sharing the attention weights across the input sequence, SWSA significantly reduces the number of parameters, making the model more compact and efficient. This results in faster computation and lower memory usage, which is particularly beneficial for deployment on resource-constrained devices.

 In 2021, Berg et al.\cite{Berg2021} introduced the Keyword Transformer (KWT), which is a model that applies the Transformer architecture to KWS, leveraging self-attention mechanisms to process and analyze audio input sequences for keyword detection.
 The KWT relies on a pure self-attention mechanism across spectrogram patches, achieving the state-of-the-art results on the GSCD, with no need for convolutional or recurrent layers.
 KWT's architecture, inspired by the Vision Transformer, treats spectrogram patches as inputs and employs attention mechanisms in the time-frequency domain to improve performance and efficiency in keyword detection tasks.

\subsection{Feature optimization}
Feature optimisation in SF-KWS involves selecting or engineering the most informative audio features to improve accuracy and efficiency. Earlier approaches of kWS relied on MFCCs for feature extraction, followed by HMM-based keyword probability estimation using a Viterbi search, where a keyword was recognized once its probability exceeded a predefined threshold. However, the need for energy-efficient KWS on microcontrollers arose from power consumption considerations. To address this, the use of a SincConv layer \cite{mittermaier2020}  to extract features directly from raw input samples was proposed by \cite{mittermaier2020}. In this approach, log compression was applied instead of a common activation function (such as ReLU). Subsequently, five  Grouped Depth-wise Separable Convolution layers were employed. As a result, the number of parameters was reduced to 62K, significantly lowering power consumption, since memory accesses were found to contribute substantially more to power usage than computation

\begin{figure}[htbp]
\centerline{\includegraphics[scale=0.3]{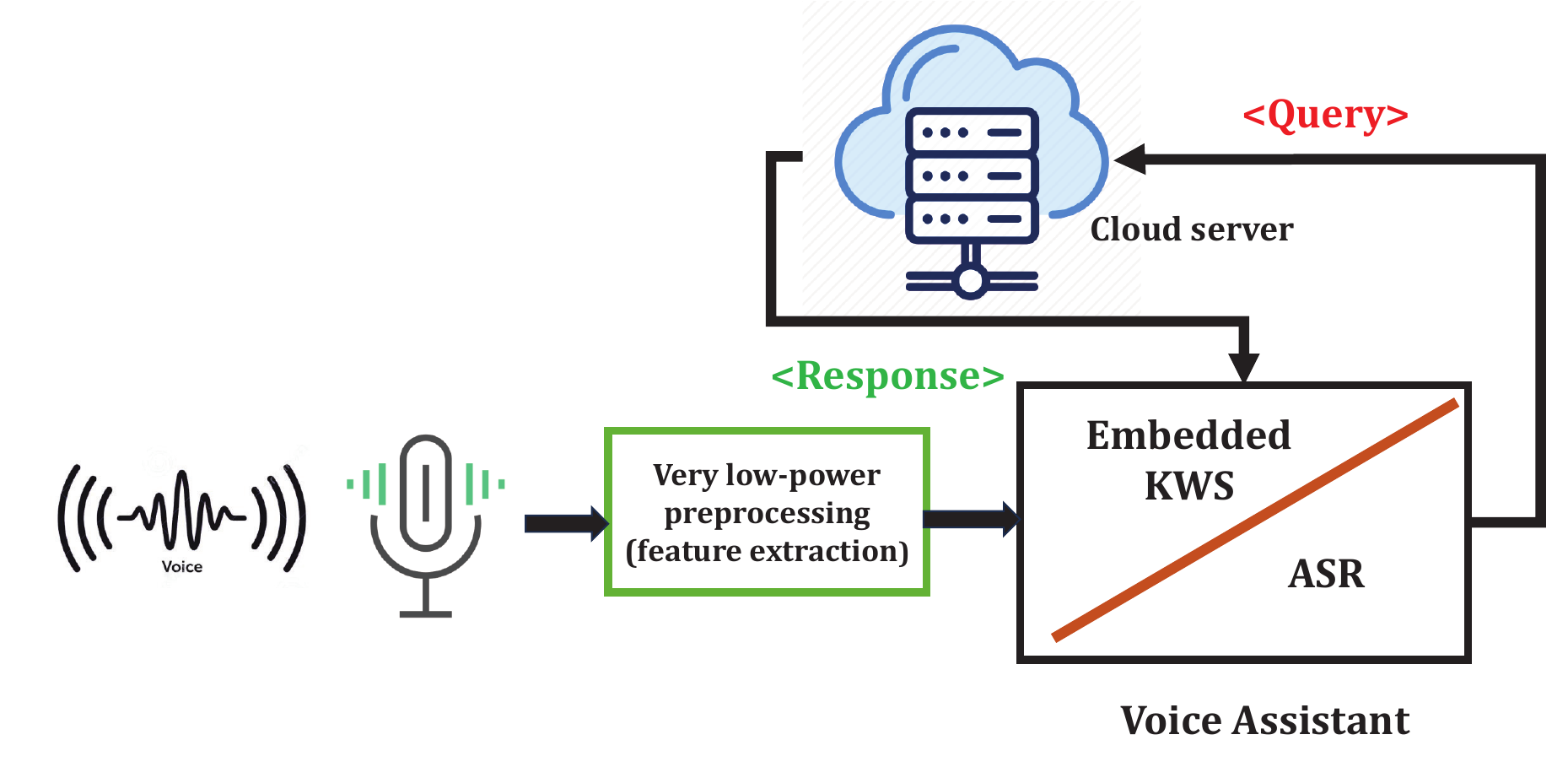}}
\caption{On-edge deployment of  Fast Keyword Spotting System}
\label{fig23}
\end{figure}

Alternatively, a more computationally efficient feature known as Multi-Frame Shifted Time Similarity (MFSTS)  was introduced in \cite{mfslbrahim2019} to decrease the computational burden during the preprocessing phase of a keyword spotter as shown in Figure \ref{fig23}. This enhancement aims to make it feasible to operate in low-power edge environments and identify the activation keyword. 
An assessment of performance comparing MFSTS and MFCC as input features for TCN-based KWS revealed similar results. This comparability stems from the fact that calculating MFSTS coefficients up to a specific lag maintains speech intelligibility. MFSTS may be a viable option for low-power/memory devices, as illustrated in the resource-aware case study  \cite{mfslbrahim2019}, particularly when dealing with the implementation of light-weight  classifiers. On the other hand, the work in \cite{Vitolo2023} presents a completely novel data-driven approach that employs an autoencoder to automate audio FE, maintain a low computational complexity, and achieve accuracy levels comparable to SOTA KWS systems. To assess its effectiveness, they have conducted a comparison with the widely-used MFCC based FE method using the publicly available GSCD, and the results reveal that their proposed audio feature extractor attains a decent classification accuracy (90\%) across 12 classes at low noise scenario, outperforming the MFCC by up to 5.2\%. Additionally, the necessary number of operations is an order of magnitude lower than that of the MFCC, leading to reduced computational complexity and processing time. Recently, I. L. Espejo et. al. \cite{LopezEspejo2022}  has conducted an experimental approach to reduce the computational complexity of a SOTA CNN acoustic model for KWS, commonly equipped with residual connections. Their focus was mainly on diminishing the spectro-temporal resolution of the speech feature matrix as an indirect means. Their experimental findings reveal a notable reduction in the size of standard feature matrices without significantly compromising KWS performance, leading to a substantial decrease in computational load.  
\subsection{Neural Architectural Search}
NAS \cite{Sze2017} can play a significant role in improving SF-KWS by automatically discovering and optimizing the architecture of NNs. However, designing accurate NNs is challenging as the vast amount of hardware platforms makes it difficult to design one globally efficient architecture.

In general, a NN's architecture can be represented as a Directed Acyclic Graph (DAG), where each node corresponds to an operator applied to its parent nodes \cite{Goodfellow2016a, Menghani2023}. Examples of such operators include convolution, pooling, activation, and self-attention. Connecting these operators in different ways leads to diverse network architectures. A crucial factor in designing an effective DNN is determining the type and number of nodes, as well as how they are structured and interconnected.  Beyond structural design, both architectural hyperparameters (e.g., stride and number of channels in a convolution) and training hyperparameters (e.g., learning rate, number of epochs, and momentum) significantly impact overall performance. 

DL models following this representation can consist of hundreds of layers and millions-or even billions of parameters. These models are typically either manually designed through iterative experimentation or adapted from existing architectures. Over time, they have grown increasingly complex and large, making manual design a challenging, time-consuming, and error-prone task that demands significant expertise and mathematical intuition. Consequently, techniques for automating NN design referred to as NAS have gained considerable attention in recent years, offering a more efficient way to discover optimal architectures tailored to specific datasets for a particular tasks.  The NAS process begins with a seed architecture (a manually designed baseline model) and explores different configurations of NN parameters

\begin{figure}[htbp]
	\centerline{\includegraphics[scale=0.8]{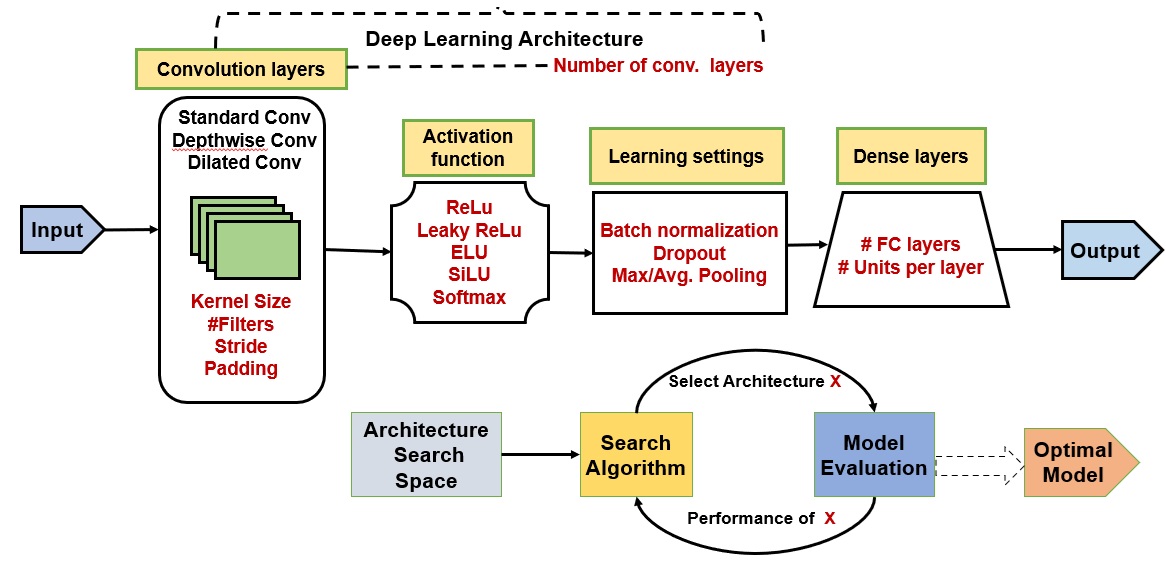}}
	\caption{Traditional NAS components: The algorithm explores the search space by generating candidate models and iteratively refines them based on evaluation feedback to improve performance \cite{Benmeziane2021}.}
	\label{fig16}
\end{figure}

A conventional NAS process requires the definition of three main components: the search space, the search strategy, and the evaluation methodology \cite{Benmeziane2021}, as illustrated in Figure \ref{fig16}. 
\begin{itemize} 
	\item \textbf{Search Space}: This defines the set of architectures that can be explored during the search. To narrow down the possibilities, constraints are applied to network architectures, operations, layers, and recurring patterns. While these constraints help streamline the search, they also introduce human bias into the NAS process.  
	
	\item \textbf{Search Algorithm and State}: This refers to the mechanism that drives the architecture search. Common algorithms used in Hyper-Parameter Optimization, such as grid search, random search, BO, and evolutionary algorithms, are also applicable to NAS.  
	
	\item \textbf{Evaluation Strategy}: This determines how a model's performance is assessed. It can be based on conventional metrics like validation loss or accuracy, or it may involve a more complex, composite metric \cite{tan2019mnasnet}. The chosen metric helps guide the search algorithm toward promising architectures within the defined search space.  
\end{itemize}

In 2019, Andrew Anderson et al.\cite{anderson2019performance} evaluated each NAS candidate architecture based on two criteria:  classification accuracy and operation count (a proxy for inference cost). For instance, fewer operations mean lower computational demand, which is crucial for edge devices. After generating a set of candidate architectures, the NAS applies  optimization method to find models that offer the best trade-offs between accuracy and inference cost. As convolutions are typically the most computationally expensive part of a DNN, the NAS focuses on optimizing the size and number of convolutional filters in the network. By reducing the kernel size or filter count, the number of operations required for inference can be significantly reduced without drastically impacting accuracy. After identifying Pareto-optimal models, the final step involves fine-tuning the top models using techniques such as pruning and quantization to further reduce the memory and computational footprint. These steps ensure that the final models are both lightweight and efficient for deployment on devices with strict resource constraints. They achieve 95.1\% accuracy with only 223.4 MFPOPS, representing a 2.6$\times$ reduction in operations compared to the seed model.

The authors in \cite{veniat2019stochastic}  introduced an efficient and flexible framework to automatically discover compact neural architectures optimized for KWS. They have used a stochastic adaptive NAS method that employs a controller network to dynamically sample architectures from a search space composed of building blocks such as depthwise separable convolutions and squeeze-and-excitation modules. Unlike traditional NAS approaches, this method adapts the sampling distribution using policy gradient reinforcement learning, which progressively focuses the search toward high-performing and efficient models. The framework achieves competitive accuracy while significantly reducing the number of parameters and operations, demonstrating its effectiveness across multiple KWS benchmarks. The study emphasizes the importance of balancing performance and computational cost, making it well-suited for real-time KWS applications on embedded systems.

On the other hand, Tong Mo et al.\cite{mo20_interspeech} proposed a Cell-Based NAS approach to automatically design compact and high-performing CNNs for KWS task using Differentiable Architecture Search (DARTS). This approach seeks optimal cell structures building blocks of neural networks by evaluating candidate operations (e.g., convolutions, pooling) within a defined search space. The resulting architectures are then scaled in depth and width to build the final model. Evaluated on the GSCD (12-class setup), their method achieves a state-of-the-art accuracy of over 97\% with a smaller model size than existing baselines like ResNet variants and SincNet-based models \cite{Ravanelli2018}. The study demonstrates that NAS, even when constrained to standard operations, significantly outperforms manually designed architectures, making it highly suitable for on-device KWS systems where memory and performance trade-offs are critical.

Another notable contribution is AutoKWS by Bo Zhang et al. \cite{Zhang2021autokws}, which leverages DARTS and its enhanced variants (FairDARTS and NoisyDARTS) to efficiently explore a well-designed search space based on TC-ResNet with squeeze-and-excitation modules. Unlike cell-based NAS systems that may be too complex for direct deployment on edge devices, AutoKWS produces streamlined, high-performing models achieving up to 97.44\% accuracy on GSCD V2 with under 110K parameters making it significantly more practical for real-time, on-device applications compared to prior handcrafted or compute-intensive NAS solutions.

A recent approach, MicroNets \cite{Banbury2021} has shown efficiency in generating highly efficient neural network architectures through differentiable neural architecture search (DNAS), specifically tailored for deployment on resource-constrained microcontrollers. By optimizing for memory usage, latency, and energy consumption, MicroNets achieve state-of-the-art performance while remaining suitable for real-time, always-on applications. In the context of KWS, MicroNet delivers compact models that can accurately detect spoken keywords from short audio clips, even under strict constraints on memory (SRAM and flash). These models are deployable using the open-source TensorFlow Lite for Microcontrollers (TFLM), ensuring broad compatibility and ease of integration \cite{Lin2023}. Notably, MicroNet-based KWS models outperform traditional architectures like DS-CNN in both speed and size while maintaining high accuracy, making them ideal for IoT devices, wearables, and smart assistants that require continuous voice input processing with minimal power consumption.

\subsection{Hybrid approaches}
Combining multiple methods and techniques mentioned above can lead to even more efficient and compact KWS models. Hybrid approaches leverage the strengths of different methods to achieve a balance between model size, accuracy, and computational requirements \cite{Shrivastava2021}. When crafting resource-efficient DNNs for KWS, numerous considerations come into play. The method proposed in \cite{Peter2022} focuses on key aspects simultaneously, namely: (i) The architecture of the deep learning model, (ii) FE scheme during preprocessing stage, and (iii) the quantization of weights and activations once the model is trained. They employed NAS to acquire streamlined end-to-end CNNs tailored for KWS. Moreover, the adopted end-to-end KWS model incorporates a SincConv  at the input layer Ravanelli2018, facilitating classification directly on raw audio waveforms.  An alternative approach is observed in \cite{zhang2017}, where the authors implemented a model on the Cortex-M7-based STM32F746G-DISCO development board utilizing Cortex Microcontroller Software Interface Standard-NN (CMSIS-NN) kernels. This implementation required a restricted memory footprint and minimal computing resources. To address the challenges, the authors introduced DS-CNN and quantized techniques. In this context, HMM and Viterbi decoding were deemed computationally expensive during inference, and RNN suffered from substantial detection latency. Meanwhile, the limitations of DNN included neglecting local temporal and spectral correlations in the input speech feature, and CNN overlooked long-term temporal dependencies. In their approach, the authors trained the KWS model with an 8-bit fixed-point instance, derived from a 32-bit floating-point representation. The DS-CNN, when subjected to 8-bit quantization, achieved an accuracy of approximately 94.9\%, outperforming DNN, CNN, Basic LSTM, GRU, and CRNN.
\begin{table}[htp]
    \centering
    \caption{Representative devices supported by TensorFlow Lite for Microcontrollers}
    \label{Microcontrollers}
    \begin{tabular}{ccccc}
        \hline \hline
         \textbf{MCU Platform} & \textbf{Processor} & \textbf{ \begin{tabular}{@{}c@{}}Frequency\\(MHz)  \end{tabular}} & \textbf{\begin{tabular}{@{}c@{}}SRAM\\ (KB) \end{tabular}} & \textbf{Flash}\\
         \hline
          \begin{tabular}{@{}c@{}}Arduino Nano \\33 BLE Sense \end{tabular} & ARM Cortex M4 & 64 & 256  & 1 MB\\ \hline
         ESP32 &  \begin{tabular}{@{}c@{}} Tensilica Xtensa \\LX6 \end{tabular} & 160 & 512   & 2 MB\\ \hline
         \begin{tabular}{@{}c@{}}Sparkfun Edge\\ Appolo3 Blue \end{tabular} &  \begin{tabular}{@{}c@{}} ARM Cortex\\ M4F \end{tabular}  & 48 & 384  & 1 MB\\ \hline
          \begin{tabular}{@{}c@{}}ST Nucleo\\ Boards \end{tabular}  &  \begin{tabular}{@{}c@{}} ARM Cortex\\ M7 \end{tabular}  & 216   & 320  & 1 MB\\ \hline
          \begin{tabular}{@{}c@{}}Adafruit\\ EdgeBadge \end{tabular}  & ATSAMD51 & 120 & 192  & 512 KB\\
         \hline \hline
\end{tabular}
    
\end{table}
\section{Deploying SF-KWS on TinyML boards: Common challenges \label{sec:tinyml}}
TinyML aims to implement optimized and compact ML models on small, low-power (less than one milliwatt) devices, such as embedded systems and battery-operated microcontrollers. TinyML enables limited power, low latency, and low bandwidth usage. Due to limited resources, most TinyML solutions available today cannot offer on-device model training.  Typically, models are trained on the cloud or on a powerful GPU server before being distributed to the embedded device for running the inference of the trained model on that platform. Hence the major challenge is that the model cannot adapt to new data collected from the on-board or local sensors by fine-tuning a pre-trained model. However, in recently, we have seen few proposals \cite{Gimenez2022} such as TinyOL~\cite{Ren2021} (TinyML with Online-Learning), and Quantization-Aware Scaling or skip-update based an algorithm-system co-design framework which enables incremental on-device training of CNNs on streaming data \cite{Lin2022}, under 256KB SRAM and 1MB Flash. On-device training provides users the benefit from customized AI models without having to transfer the data to the cloud, protecting the privacy. 
\begin{table*}[htp!]
    \centering
    \scriptsize
    \caption{the salient features of various well-known TinyML Frameworks}
    \label{Details of TinyML Frameworks}
    \begin{tabular}{ccccc}
    \hline
    \textbf{Framework} & \textbf{Main Developer} & \textbf{Compatible Platforms} & \textbf{Output Languages} & \textbf{ \begin{tabular}{@{}c@{}}Interoperable External \\Libraries  \end{tabular} } \\
    \hline
    
    TensorFlow Lite\cite{David2021} & Google & ARM Cortex-M & C++ 11 & TensorFlow  \\
    \hline
    ELL\cite{ellmicrosoft}&  Microsoft & \begin{tabular}{@{}c@{}}ARM Cortex-M, ARM Cortex-A\\ Arduino micro:bit  \end{tabular} & C ,C++ & CNTK, Darknet, ONNX \\ 
    \hline
    ARM-NN\cite{armnn} &  ARM & \begin{tabular}{@{}c@{}}ARM Cortex-A \\ARM Mali Graphics, Processors\\ ARM Ethos Processor  \end{tabular} & C & TensorFlow, Caffe, ONNX  \\
    \hline
    CMSIS-NN\cite{cmsisnn} &  ARM & ARM Cortex-M  & C & TensorFlow, Caffe, PyTorch  \\
    \hline
 
    STM 32Cube. AI\cite{STM} &  STMicroelectronics  & STM32 & C & \begin{tabular}{@{}c@{}}Keras, TensorFlow Lite  \end{tabular} \\
    \hline

    AIfES \cite{AIfES} & Fraunhofer IMS & \begin{tabular} {@{}c@{}}Windows (DLL), Raspberry Pi\\ ATMega32U4, ARM Cortex-M4  \end{tabular} & C & \begin{tabular}{@{}c@{}}Keras \\TensorFlow \end{tabular}  \\
    \hline 
   
    uTensor\cite{uTensor} &  Particular developer & mBed boards & C++11 & TensorFlow Lite  \\
    \hline

    TinyMLgen\cite{TinyMLgen} &  Particular developer & \begin{tabular}{@{}c@{}} ARM Cortex-M, ESP32 \end{tabular}  & C++11 & TensorFlow Lite   \\
    \hline

    CMix-NN\cite{Capotondi2020} &  Research group & ARM Cortex-M  & C & Mobilenet \\
    \hline

    Edge Impulse\cite{Edge_Impulse} &  Edge Impulse & \begin{tabular}{@{}c@{}} ARM Cortex-M,ARM Cortex-A\\ Raspberry Pi, ESP32 \\ STMicroelectronics Nucleo  \end{tabular}  & C++ & TensorFlow  \\
    \hline
    \end{tabular}

\end{table*}
Typical MCUs are characterized by limited computational power and restricted memory \cite{Saha2022}, typically less than 2 MB of flash storage and less than 512 KB SRAM as shown in Table \ref{Microcontrollers}. Consequently, enhancing inference latency and optimizing model size emerge as crucial factors in TinyML deployment. Common strategies for reducing latency and enhancing energy efficiency include weight quantization and code optimization. Additionally, pruning and sparsification are well-established techniques for diminishing model size, ensuring that the entire neural network model, comprising its weights, connections, and associated application code (which manages sensor data and responds to model predictions), can fit into the flash memory. Furthermore, the SRAM constraints impose limitations on the temporary memory buffer utilized for storing the model's input and output data. There are a number of available boards that support TinyML. Here are some of the most popular ones(Figure \ref{fig24}):
\begin{figure}[htbp]
\centerline{\includegraphics[scale=0.5]{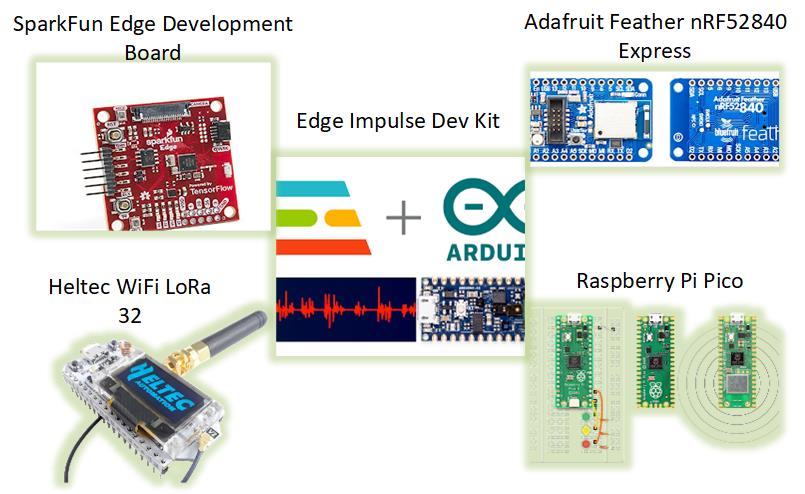}}
\caption{Representative TnyML devices (MCUs) supported by TensorFlow Lite \cite{Saha2022}}
\label{fig24}
\end{figure}
\begin{itemize}
    \item \textbf{Edge Impulse Dev Kit(with Arduino Nano 33 BLE Sense)\footnote{ https://store-usa.arduino.cc/products/arduino-nano-33-ble-sense-with-headers }} is a development board that is specifically designed for TinyML. It features a powerful Neural Decision Processor (NDP) that can run ML models with low power consumption.
    \item \textbf{SparkFun Edge Development Board\footnote{https://www.sparkfun.com/products/15170 }} is another popular option for TinyML. It features a 32-bit microcontroller with 1MB of flash memory and 256KB of RAM.
    \item \textbf{Adafruit TensorFlow Lite for Microcontroller based Kit\footnote{ https://www.adafruit.com/product/4317 }} - small microcontrollers capable of running a scaled-down version of TensorFlow Lite for performing ML computations are available. The EdgeBadge, a diminutive board comparable in size to a credit card, is powered by the ATSAMD51 chip, boasting 512KB of flash and 192KB of RAM. Additionally, 2 MB of QSPI flash is incorporated for file storage, providing a convenient space for TensorFlow Lite files, images, fonts, sounds, or other assets.
    \item \textbf{Himax WE-I Plus EVB Endpoint AI Development Board \footnote{https://www.sparkfun.com/products/17256 }} In collaboration with Google TensorFlow Lite for Microcontrollers framework and Synopsys embARC MLI library, the Himax WE-I Plus EVB delivers a comprehensive development environment for implementing various TensorFlow Lite for Microcontrollers examples. These examples include ``Person detection'', ``Micro speech'', and ``Magic wand'', catering to applications in ``Vision'', ``Voice'' and ``Vibration''.
    \item \textbf{Raspberry Pi Pico\footnote{https://www.raspberrypi.com/products/raspberry-pi-pico}} is a low-cost, high-performance board that is also supported by TinyML frameworks.
\end{itemize}
A TinyML framework is typically used to implement ML models (implemented on well-known ML libraries such as TensorFlow, Scikit-Learn], PyTorch) on edge devices - meaning everything from mobile phones to MCUs.  These frameworks cater to a range of devices, from mobile phones to microcontroller units (MCUs). Various open-source TinyML frameworks  exist, including Google's TensorFlow Lite. This framework provides tools to adapt TensorFlow models for execution on mobile and embedded devices. It comprises two main components: the converter, which transforms TensorFlow models into optimized code for constrained platforms (e.g., ARM Cortex-M series processors like Arduino Nano) and the interpreter, responsible for executing the generated code.  Microsoft has also contributed to the TinyML landscape with its open-source Embedded Learning Library (ELL). This framework facilitates the design and deployment of pre-trained ML models on constrained platforms, supporting ARM Cortex-A and Cortex-M architectures. ARM introduces the CMSIS-NN, a collection of NN kernels optimized for execution on Cortex-M processors. Edge Impulse, a cloud-based solution, simplifies the creation and implementation of ML models for TinyML devices. It involves IoT devices in data collection, FE, model training, and deployment optimization. Supporting TFLM, it uses the EON compiler for model deployment and quantizes models using TensorFlow's Model Optimization Toolkit. Beyond open-source initiatives, some institutions and companies offer licensed products. The Fraunhofer Institute for Microelectronic Circuits and Systems (IMS) presents the Artificial Intelligence for Embedded Systems (AIfES) library [44]. Additionally, several frameworks like MicroMLGen, Weka-porter, EmbML, and m2cgen, interoperable with the widely-used Scikit-learn toolkit, transform traditional ML models (e.g., SVM, decision tree, KNN, Random Forest) for execution on diverse MCUs like Arduino, ESP32, ESP8266, etc. For more insights on TinyML frameworks, refer to Table \ref{Details of TinyML Frameworks}.

Among these, \textbf{Edge Impulse} has emerged as one of the most widely adopted platforms for deploying ML/DL models on IoT devices, such as the Arduino Nano 33 BLE. Below is an overview of the key steps involved in developing and deploying a neural network 
model using Edge Impulse:
\begin{itemize}
	\item \textbf{Initiate Project Setup:} Begin by creating a project on the Edge Impulse dashboard tailored to the deployment target (IoT, edge, or mobile device).
	
	\item \textbf{Integrate Devices with Edge Impulse:} Connect supported hardware platforms (e.g., microcontrollers, smartphones) to Edge Impulse for real-time data acquisition and inference.
	
	\item \textbf{Perform Real-time Data Collection:} Collect live sensor data directly from connected devices, upload datasets manually, or use pre-existing datasets for training.
	
	\item \textbf{Utilize Pre-existing Models:} Access a library of predefined models that can be integrated or used as a baseline for further customization.
	
	\item \textbf{Customize Model and Parameters:} Modify architecture and hyperparameters to balance model accuracy, memory footprint, and latency according to application constraints.
	
	\item \textbf{Conduct Model Training:} Train the model and examine evaluation metrics (e.g., accuracy, loss). If performance is suboptimal, iterate by adjusting model parameters.
	
	\item \textbf{Evaluate Model Performance:} Test the trained model for inference latency, memory requirements, and classification accuracy to ensure it meets deployment criteria.
	
	\item \textbf{Download Deployment Package:} Export the deployment-ready model as a zip archive compatible with the target device.
	
	\item \textbf{Include Necessary Libraries:} For example, when targeting Arduino, integrate the Edge Impulse Arduino library into the Arduino IDE to utilize the trained model.
	
	\item \textbf{Deploy and Infer on Device:} Upload the final model to the device. The system is now ready for real-time inference using live sensor data.
\end{itemize}

\begin{table}[]
    \centering
    \caption{Comparative assessment of different TinyML frameworks applied to various SF-KWS architectures}
    \scriptsize
    \label{Results}
    \begin{tabular}{ccccc}
        \hline \hline
         \multirow{2}{*}{\textbf{Framework}} & \multirow{2}{*}{\textbf{Model}} & \multirow{2}{*}{\textbf{Size(KB)}}  & \textbf{Inference} & \multirow{2}{*}{\textbf{Accuracy}}\\
          & & &  \textbf{Time(ms)}  &\\
         \hline
         \multirow{3}{*}{ Keras } & DNN-S & 999 & 226 & 77.1 \\
         \multirow{3}{*}{ Software}& DNN-L  & 5990 & 874.1 & 86.6 \\
         \multirow{3}{*}{ Models}& CNN-S  & 922.4  & 596.8 & 85.8 \\
                                        & CNN-L  & 5810 & 996.5 & 93.3 \\
                                        & DS-CNN-S  & 480.6 & 603.5 & 90.1 \\
                                        & DS-CNN-L  & 5302.7 & 559 & 91.8 \\
          \hline  
          \multirow{3}{*}{ TensorFlow } & DNN-S & 2147.9 & 2 & 75.3 \\
           \multirow{3}{*}{ Lite }    & DNN-L  & 1985.5 & 169.5 & 82.1 \\
           \multirow{3}{*}{ Models }     & CNN-S  & 280.4  & 81 & 84.9\\
                                        & CNN-L  & 1986 & 272.3 & 92.3 \\
                                        & DS-CNN-S  & 98.7 & 4.8 & 89.5 \\
                                        & DS-CNN-L  & 1652.6 & 8.4 & 91.6 \\
          \hline  
          \multirow{2}{*}{MicroNet}   & MicroNet-S  & 114.5 & 1 & 95.3 \\
                                      & MicroNet-L  & 658.8 & 3.6 & 96.5 \\
          \hline
          \multirow{2}{*}{Edge Impluse}  & mfcc-conv1d(f32) & 14.8 & 98 & 82.4\\
                                    & mfcc-conv1d(i8) & 13 & 97 & 82.6 \\

    \hline   \hline
    \end{tabular}
\end{table}
\section{Experimental study \label{sec:exp}}
\subsection{Case study-I}
\begin{figure}[htbp]
	\centerline{\includegraphics[scale=0.6]{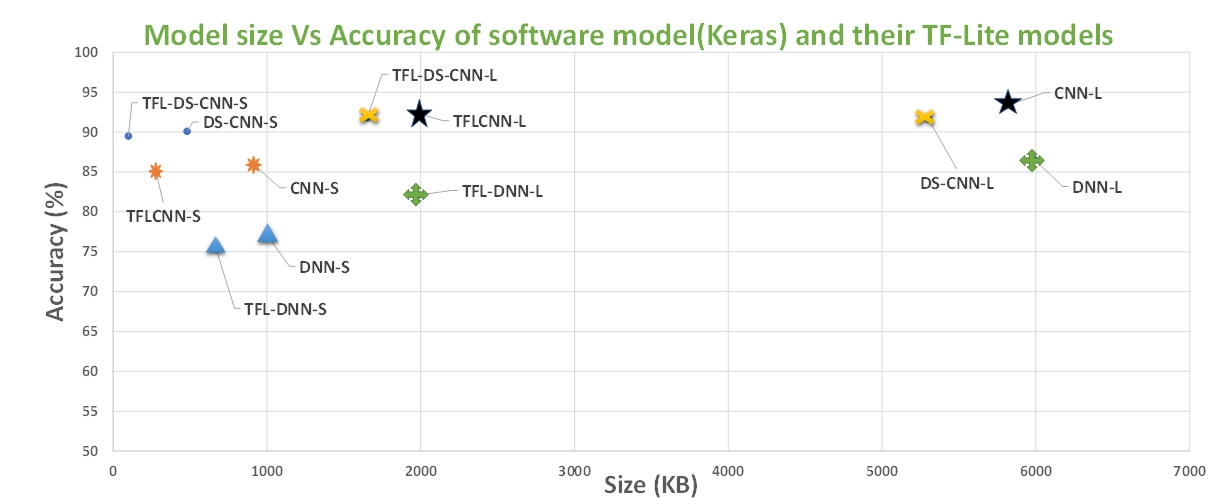}}
	\caption{Comparison of Model Size vs. Accuracy for Software Models (Keras) and their TF-Lite Variants, highlighting variations in model efficiency and performance \cite{Garai2024}.}
	\label{fig25}
\end{figure}

\begin{figure}[htbp]
	\centerline{\includegraphics[scale=0.6]{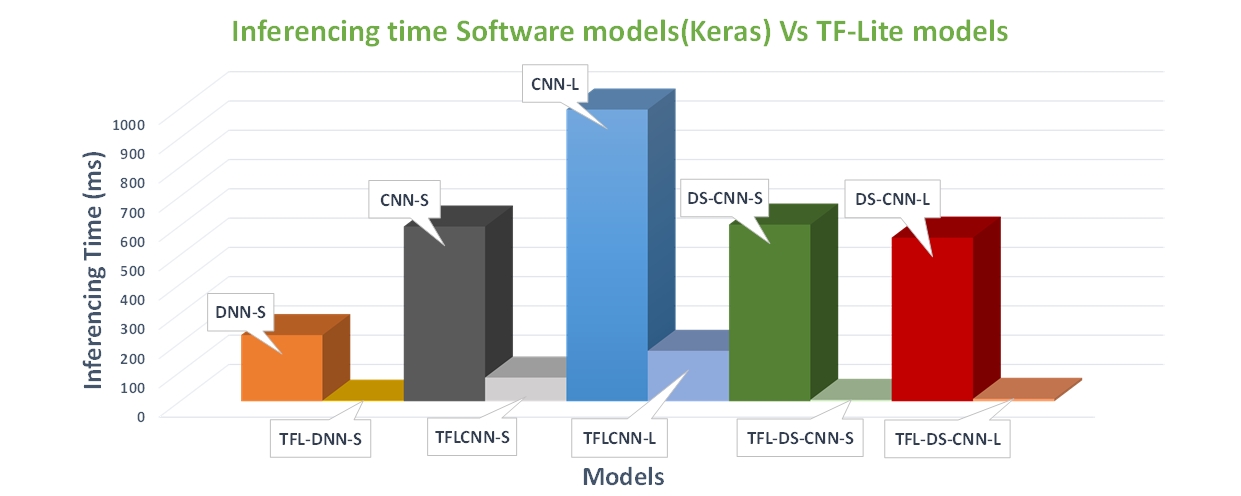}}
	\caption{Inference Time Comparison of Software Models (Keras) vs. TF-Lite Models, showcasing the reduction in execution time for TF-Lite models across different architectures \cite{Garai2024}.}
	\label{fig26}
\end{figure}

We performed a few experiments using TinyML frameworks exclusively on a set of selected model architectures suitable for SF-KWS from the previously discussed categories. The investigation involved a diverse range of architectures sourced from various works, each representing different models with different sizes and configurations to explore this spectrum comprehensively. Our experiment performed on the GSCD aimed at classifying ten distinct classes using diverse architectures. Specifically, we employed DNN, CNN, DS-CNN, and MicroNet models. To further categorize and differentiate these architectures, we classified them based on their size, distinguishing between small (S) and large (L). This resulted in designations such as DNN-S, DNN-L, CNN-S, CNN-L, DS-CNN-S, DS-CNN-L, MicroNet-S, and MicroNet-L~\cite{zhang2017}~\cite{Banbury2021}. This approach allowed us to systematically explore the impact of architecture size on the classification performance within the context of the GSCD. Additionally, for the EdgeImpulse framework, we selected architectures based on MFCC and 1D convolution(mfcc-con1d(f32), mfcc-conv1d(i8)), incorporating both float32 and int8 compression techniques for a comprehensive exploration of model efficiency. Now, we applied different TinyML frameworks on top of this SF-KWS architecture to make them suitable for IoT edge devices. Table \ref{Results} visualises their inferencing time, model size and final deployable at the Cortex-M board.
We utilized Keras software models from the GitHub repository\footnote{https://github.com/ARM-software/ML-zoo}, which were then trained and evaluated as shown in Table \ref{Results}. Additionally, these models were quantized to Int8 using the TensorFlow framework for Tensorflow Lite models, including the MicroNet architectures. In the case of the EdgeImpulse framework, we adopted a model featuring three 1D Conv/Pooling layers with a dropout rate of 0.25 in its network architecture. It is noteworthy that converting Keras software models into quantized Int8 format with the TensorFlow Lite framework led to a remarkable reduction in model size, up to 69\%. This reduction in size did not significantly impact accuracy compared to Keras Software models, and there was also a substantial decrease in inference time. Our findings indicate that TensorFlow Lite CNN-S and DS-CNN-S are particularly well-suited for IoT edge devices, as these models can be easily deployed on the flash memory of targeted edge devices. On the other hand, the Edge Impulse framework significantly reduces the model size, while MicroNet-S involves a trade-off between accuracy and memory size for low-footprint devices. Figure \ref{fig25} visualizes the relationship between accuracy and the number of parameters for all models, including both Keras-based software models and their TensorFlow Lite (TFLite) counterparts. Our analysis indicates that converting Keras models to TFLite reduces model size by approximately 69\% with minimal accuracy loss. Among the TFLite models, MicroNet-S achieved the highest accuracy at 95.3\%. Additionally, Figure \ref{fig26} highlights that TensorFlow Lite not only significantly reduces model size but also improves inference speed by quantizing model weights while preserving performance. 
\subsection{Case study-II}
As mentioned in Section 3.7, we seek Pareto-optimal models using MOO, where we would like to achieve the best possible in two dimensions. Typically, one of these dimensions is Quality-related metrics (accuracy, F1-score, etc.) and the other one is footprint-related metrics (model size, RAM, and so forth).

\begin{figure}[h]
	\centerline{\includegraphics[scale=0.4]{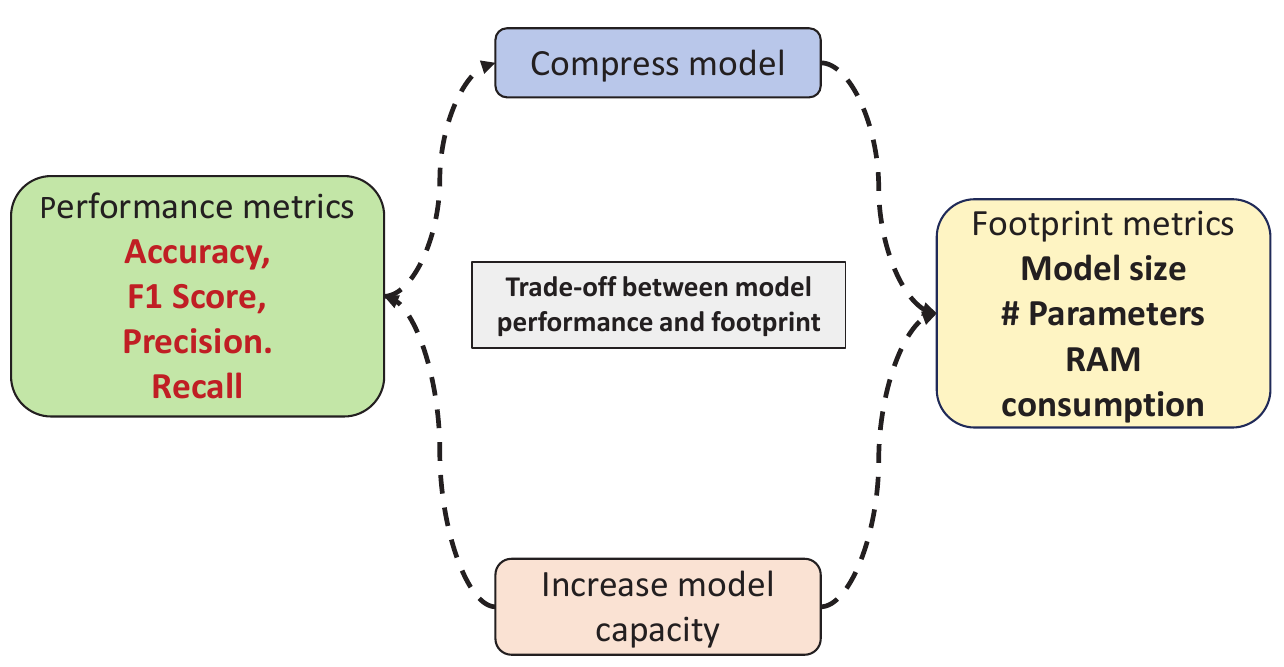}}
	\caption{Tradeoff between model performance and footprint.}
	\label{fig27}
\end{figure}

\begin{figure}[h!]
	\centering
	\includegraphics[width=\textwidth]{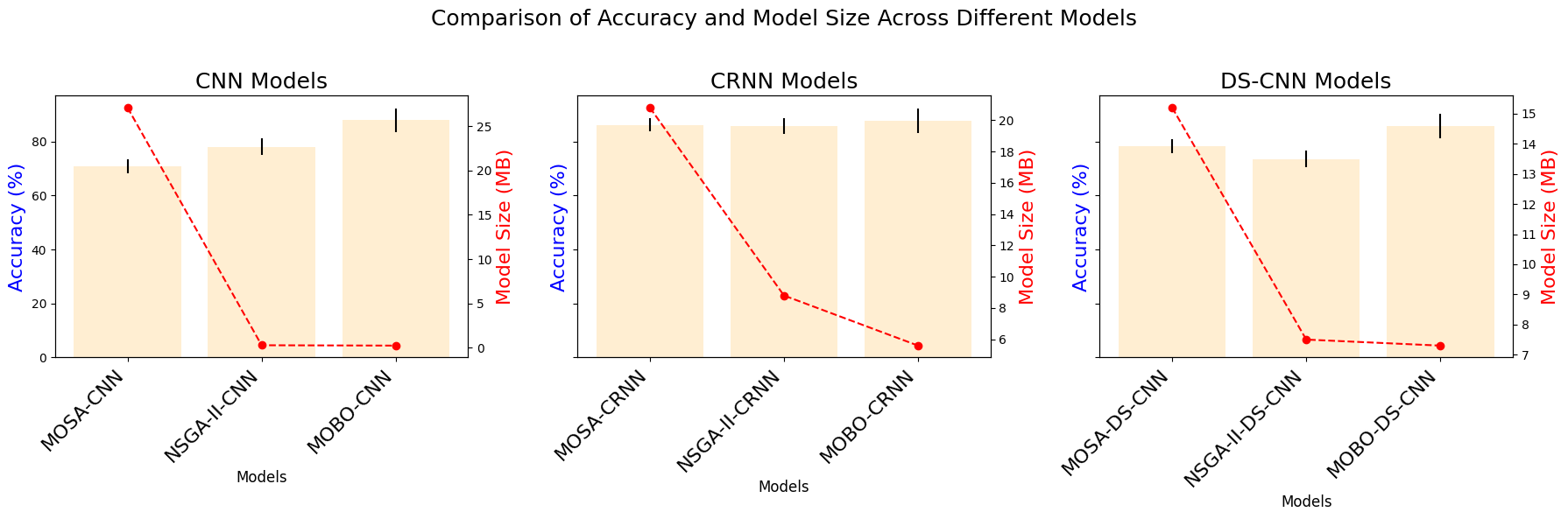}
	\caption{Comparison of Accuracy and Model Size across different CNN, CRNN, and DS-CNN models obtained using three Multi-Objective Optimization  strategies: MOSA, NSGA-II, and MOBO. The bar plots represent the accuracy (left y-axis), while the red dashed lines correspond to the model sizes (right y-axis).}
	\label{fig:moo_comparison}
\end{figure}

There is an inherent trade-off between model quality and footprint metrics. A larger and deeper model generally achieves higher accuracy but comes at the expense of increased model size, latency, and computational cost. Conversely, a smaller model with fewer parameters may be more suitable for deployment but often suffers from reduced accuracy. As shown in Figure \ref{fig27}, one approach is, to begin with a high-quality model and gradually trade some of its accuracies for a smaller footprint by applying compression techniques such as quantization, pruning, or low-rank approximation.
In our experiments, we utilized three Multi-Objective Optimization  techniques which treats accuracy and hardware efficiency as separate objectives, leading to the following formulation \cite{Miettinen1999}:
\begin{equation}
	\max_{\alpha \in A} f_1(\alpha, \mathcal{D}), f_2(\alpha, \mathcal{D}), \dots, f_n(\alpha, \mathcal{D})
\end{equation}
Since multiple objectives often conflict (e.g., reducing model size may lower accuracy), the goal is to find a set of Pareto-optimal solutions rather than a single best solution \cite{Deb2011}. The two main techniques used in MOO are:
\begin{itemize}
	\item \textbf{Scalarization methods:}
This approach transforms the multi-objective problem into a single-objective one by applying a weighted aggregation function:
\begin{equation}
	\max_{\alpha \in A_{sp}} h(f_1(\alpha, \mathcal{D}), f_2(\alpha, \mathcal{D}), \dots, f_n(\alpha, \mathcal{D}))
\end{equation}

where $h$ is typically a weighted sum:
\begin{equation}
	\min_{\alpha \in A_{sp}} w \cdot \text{M}(\alpha, \mathcal{D}) + (1 - w) \cdot (1-ACC(\alpha, \mathcal{D}))
\end{equation}
Here, $\text{ACC}(\alpha, \mathcal{D})$ represents accuracy, $M(\alpha)$ denotes model size, and $w$ is a learnable parameter balancing the two \cite{Deb2011}.
\item \textbf{Heuristic  optimization methods:}
An alternative approach is to use general heuristic-based methods to solve the MOO problem. These methods aim to find a set of Pareto-optimal solutions that balance different objectives, such as accuracy, computational cost, and memory footprint (model size) \cite{Rardin2001}. Common heuristic-based approaches include:
\begin{itemize}
	\item \textbf{SA}: SA is a probabilistic optimization technique inspired by the annealing process in metallurgy. It explores the search space by accepting worse solutions with a certain probability to escape local optima \cite{Bandyopadhyay2008}. SA can be adapted for MOO by incorporating dominance-based acceptance criteria \cite{Guelcue2021}.
	
	\item \textbf{NSGA-II}: NSGA-II is an evolutionary algorithm that ranks solutions based on Pareto dominance and maintains diversity using crowding distance \cite{Deb2002}. It efficiently identifies a diverse set of Pareto-optimal solutions, making it widely used in multi-objective NAS \cite{shaikh2024}.
	
	\item \textbf{Bayesian Optimization}: Bayesian methods use probabilistic models, such as Gaussian Processes or Tree-structured Parzen Estimators (TPE), to approximate the objective functions \cite{Parsa2020}. In MOO, acquisition functions like Expected Hypervolume Improvement (EHVI) help guide the search towards promising regions of the Pareto front \cite{Alibrahim2021}.
\end{itemize}
\end{itemize}
\begin{table}[h]
	\centering
	\scriptsize
	\caption{Hyperparameter search settings for CNN, CRNN, and DS-CNN models}
	\label{tab:hyperparam_settings}
	\begin{tabular}{||l|c|c|c||}
		\hline \hline
		\textbf{Hyperparameter} & \textbf{CNN} & \textbf{CRNN} & \textbf{DS-CNN} \\
		\hline
		No. of conv layers & 2-4 & 1-3 & 3-6 \\
		\hline
		No. of filters & 16-128 & 16-64 & 16-128 \\
		\hline
		Kernel size & (3x3), (5x5), (7x7) & (3x3), (5x5) & (3x3), (5x5) \\
		\hline
		Stride & 1-2 & 1-2 & 1-2 \\
		\hline
		Dropout rate & 0.1-0.5 & 0.1-0.5 & 0.1-0.5 \\
		\hline
		Recurrent layers & - & 1-2 (LSTM) & - \\
		\hline
		LSTM units & - & 32-128 & - \\
		\hline
		Fully connected layers & 1-2 & 1 & 1 \\
		\hline
		Units in dense layer & 64-256 & 64-128 & 64-256 \\
		\hline
		Activation function & ReLU & ReLU & ReLU / Hard-Swish \\
		\hline
		Pooling type & - & - & Max / Average \\
		\hline
		Optimizer & Adam & Adam / RMSprop & Adam \\
		\hline \hline
	\end{tabular}
	\label{table_hyp}
\end{table}
By employing these advanced optimization strategies, hardware-aware NAS enables the discovery of architectures that achieve high accuracy while meeting stringent hardware constraints, making them suitable for deployment in real-world edge and embedded systems \cite{Jin2007}. We have used SA, BO, and NSGA-II to fine-tune the hyperparameters of deep KWS models. Specifically, we optimized three architectures: CNN, CRNN, and DS-CNN, focusing on two conflicting objectives - model accuracy and model size.
The architecture of these models depends on the hyperparameters subject to optimization, such as the number of filters, kernel size, convolutional layers, LSTM layers, fully connected layers, and dropout rate. In Table~\ref{table_hyp}, we provided the ranges of key hyperparameters used for various models. These hyperparameters were optimized using MOO techniques on the GSCD v2 with 10 classes, employing Log-Mel spectrograms with 40 Mel bands as the feature extractor.
Since Pareto-optimality does not yield a single definitive solution, we selected the most suitable models from each MOO method based on domain-specific preferences. When we provide equal preference to both model performance and model size, Bayesian optimization performs  better than NSGA-II as shown in Figure \ref{fig:moo_comparison}. The MOBO-CNN model achieves the highest accuracy with the smallest model size among all CNN variants. Table~\ref{tab:top5models} lists the top 5 models ranked based on a composite performance-efficiency score, highlighting the trade-off between accuracy and model size.

\begin{table}[h!]
	\centering
	\scriptsize
	\caption{Top 5 ranked models based on scalarization score = 0.5 $\times$ Accuracy $-$ 0.5 $\times$ Model Size}
	\begin{tabular}{||l|c|c|c|c|c|c|c|c|c||}
		\hline \hline
		\textbf{Model Name} & \textbf{Conv} & \textbf{LSTM} & \textbf{Filters} & \textbf{Kernel} & \textbf{FC} & \textbf{BN} & \textbf{Dropout} & \textbf{Accuracy (\%)} & \textbf{Model Size (MB)} \\
		\hline
		MOBO\_CNN       & 5 & 0 & 16 & (3,3) & 1 & 1 & 1 & 87.8 & 0.25 \\
		MOBO\_CRNN      & 3 & 1 & 16 & (3,3) & 1 & 0 & 1 & 87.7 & 5.77 \\
		NSGA-II\_CNN    & 5 & 0 & 16 & (3,5) & 1 & 0 & 1 & 78.0 & 0.29 \\
		MOBO\_DS-CNN    & 1 & 0 & 16 & (3,3) & 1 & 1 & 1 & 85.3 & 7.67 \\
		NSGA-II\_CRNN   & 2 & 1 & 16 & (5,3) & 1 & 0 & 1 & 85.4 & 8.19 \\
		\hline \hline
	\end{tabular}
	\label{tab:top5models}
\end{table}

%
%
\section{Conclusion and future scope \label{sec:conclusion}}
The objective of this study has been to furnish a thorough review of the entire pipeline of the SF-KWS system. At the core of this paradigm is a DNN-based acoustic model, which computes posterior probabilities based on extracted speech features. These probabilities are subsequently analyzed to detect the presence of a keyword. While KWS has significantly advanced the field, enabling real-world applications such as voice assistant activation, its potential extends far beyond wake-word detection. In this study, we have systematically explored SF-KWS methodologies, classifying them into seven distinct categories. Furthermore, we have reviewed popular frameworks designed to optimize SF-KWS models for deployment on resource-constrained IoT edge devices with limited memory and power.

The immediate focus for future work will be advancing SF-KWS in two key directions:
\begin{itemize}
	\item Enhancing computational efficiency to ensure SF-KWS models are optimized for TinyML-supported microcontroller devices with minimal power consumption and memory footprint.
	\item Improving real-world performance, including on-device neural network training to enable personalized keyword adaptation and greater robustness in diverse acoustic environments~\cite{Gimenez2022}.
\end{itemize}
Future research will likely focus on the following modular strategies to advance the effectiveness and deployment of SF-KWS systems:
\begin{itemize}
	\item \textbf{Lightweight Architectures for TinyML:} The future research on acoustic models will likely emphasize the development of efficient and lightweight deep learning architectures. Transformer-based models, such as Audio Spectrogram Transformers (AST) \cite{Gong2021}\cite{Samui2023}, have demonstrated superior performance over CNN-based models in various audio classification tasks. However, the complexity of Transformer models due to numerous hyperparameters and high computational requirements makes them challenging to deploy on TinyML devices. To address this, TinyNAS can be employed to optimize KWS networks specifically for microcontrollers by selecting the most efficient combination of layers, kernel sizes, and activation functions \cite{mo20_interspeech}. By jointly optimizing the model structure and inference engine \cite{Liberis2021}, NAS-based KWS solutions achieve superior accuracy with minimal latency and power consumption \cite{Lin2023}. This enables real-time keyword recognition without offloading computations to external servers, ensuring data privacy and reducing energy consumption \cite{Ma2023}.
	\item  \textbf{ Overcoming Data Scarcity with Self-Supervised Learning:} While pure Transformer models lack the spatial inductive biases inherent in CNNs, they require significantly more training data \cite{Vaswani2017}. The success of AST models is largely dependent on supervised pretraining with large labeled datasets, posing a constraint for real-world SF-KWS deployment. Given the high cost of annotating speech data, leveraging self-supervised learning (SSL) approaches such as Self-Supervised AST (SSAST) \cite{Gong2022} can enable models to learn from web-scale unlabeled audio (e.g., from sources like radio or YouTube). This technique helps mitigate data scarcity challenges by extracting meaningful features without extensive labeling \cite{Gao2023}. Furthermore, this approach reduces the dependence on cloud-based training while enabling continuous learning and adaptation to new acoustic environments. For instance, a wearable device with embedded KWS can gradually improve its performance by learning from user-specific variations in speech \cite{Sun2020}. 
	\item \textbf{Sparse Updates for Memory-Efficient Learning:}  Another key advancement is sparse layer and sparse tensor updates \cite{samui2018}, which can significantly reduce the memory footprint of KWS models. TinyML implementations of KWS often rely on deep neural networks with convolutional or recurrent layers. However, storing intermediate activations for backpropagation remains a major bottleneck. Sparse updates selectively modify only essential parts of the network, thereby reducing both memory and computational requirements \cite{Lin2023}. This technique allows KWS models to run efficiently on microcontrollers with less than 256 KB of SRAM, making them viable for battery-operated devices like smartwatches and always-on voice assistants \cite{Fedorov2019}.
	\item \textbf{Joint Optimization:} Traditional approaches sequentially apply optimization techniques, such as NAS, KD, pruning, and quantization, to achieve a compact model. However, this stepwise method often leads to suboptimal results since each optimization stage is applied without considering its impact on previous or subsequent stages. Instead, joint optimization which simultaneously considers NAS, pruning, and quantization has emerged as a superior technique for TinyML based KWS models. Joint optimization significantly increases the search space, as the number of possible hyperparameter combinations grows exponentially \cite{Shafique2021}. To address this challenge, efficient search strategies have been proposed to minimize search time and reduce computational costs. For instance, APQ (Automated Pruning-Quantization) \cite{Wang2020} employs NAS along with accuracy predictors to identify optimal architectures while ensuring effective model compression. This method streamlines the design process, making it particularly suitable for ultra-low-power KWS applications where real-time inference is required on microcontrollers.
\end{itemize}
By integrating these TinyML advancements, small-footprint KWS can achieve real-time, low-power speech recognition while operating entirely on edge devices. This paves the way for more efficient, privacy-preserving, and adaptive voice interfaces in smart homes, wearables, and embedded AI applications.

\section*{Conflict of interest statement}
None declared.

\bibliography{ref_kws_final}

\begin{thebibliography}{100}
\expandafter\ifx\csname url\endcsname\relax
  \def\url#1{\texttt{#1}}\fi
\expandafter\ifx\csname urlprefix\endcsname\relax\def\urlprefix{URL }\fi
\expandafter\ifx\csname href\endcsname\relax
  \def\href#1#2{#2} \def\path#1{#1}\fi

\bibitem{Hoy2018}
M.~B. Hoy, Alexa, siri, cortana, and more: an introduction to voice assistants,
  Medical reference services quarterly 37~(1) (2018) 81--88.

\bibitem{Tristan2020}
S.~Tristan, S.~Sharma, R.~Gonzalez, Alexa/google home forensics, Digital
  Forensic Education: An Experiential Learning Approach (2020) 101--121.

\bibitem{michaely2017keyword}
A.~H. Michaely, X.~Zhang, G.~Simko, C.~Parada, P.~Aleksic, Keyword spotting for
  google assistant using contextual speech recognition, in: 2017 IEEE Automatic
  Speech Recognition and Understanding Workshop (ASRU), IEEE, 2017, pp.
  272--278.

\bibitem{LopezEspejo2021}
I.~L{\'o}pez-Espejo, Z.-H. Tan, J.~H. Hansen, J.~Jensen, Deep spoken keyword
  spotting: An overview, IEEE Access 10 (2021) 4169--4199.

\bibitem{rose1990hidden}
R.~C. Rose, D.~B. Paul, A hidden markov model based keyword recognition system,
  in: International Conference on Acoustics, Speech, and Signal Processing,
  IEEE, 1990, pp. 129--132.

\bibitem{wilpon1991improvements}
J.~G. Wilpon, L.~G. Miller, P.~Modi, Improvements and applications for key word
  recognition using hidden markov modeling techniques, in: [Proceedings] ICASSP
  91: 1991 International Conference on Acoustics, Speech, and Signal
  Processing, IEEE, 1991, pp. 309--312.

\bibitem{Motlicek2012}
P.~Motlicek, F.~Valente, I.~Szoke, Improving acoustic based keyword spotting
  using lvcsr lattices, in: 2012 IEEE International Conference on Acoustics,
  Speech and Signal Processing (ICASSP), IEEE, 2012, pp. 4413--4416.

\bibitem{chen2014}
G.~Chen, C.~Parada, G.~Heigold, Small-footprint keyword spotting using deep
  neural networks, in: 2014 IEEE international conference on acoustics, speech
  and signal processing (ICASSP), IEEE, 2014, pp. 4087--4091.

\bibitem{Saha2022}
S.~S. Saha, S.~S. Sandha, M.~Srivastava, Machine learning for
  microcontroller-class hardware-a review, IEEE Sensors Journal (2022).

\bibitem{kwsprabhakar015}
R.~Prabhavalkar, R.~Alvarez, C.~Parada, P.~Nakkiran, T.~N. Sainath, Automatic
  gain control and multi-style training for robust small-footprint keyword
  spotting with deep neural networks, in: 2015 IEEE International Conference on
  Acoustics, Speech and Signal Processing (ICASSP), IEEE, 2015, pp. 4704--4708.

\bibitem{scott2016suspending}
B.~D. Scott, M.~E. Rafn, Suspending noise cancellation using keyword spotting,
  uS Patent 9,398,367 (Jul.~19 2016).

\bibitem{rybakov20_interspeech}
O.~Rybakov, N.~Kononenko, N.~Subrahmanya, M.~Visontai, S.~Laurenzo, {Streaming
  Keyword Spotting on Mobile Devices}, in: Proc. Interspeech 2020, 2020, pp.
  2277--2281.
\newblock \href {https://doi.org/10.21437/Interspeech.2020-1003}
  {\path{doi:10.21437/Interspeech.2020-1003}}.

\bibitem{Sandler2018}
M.~Sandler, A.~Howard, M.~Zhu, A.~Zhmoginov, L.-C. Chen, Mobilenetv2: Inverted
  residuals and linear bottlenecks, in: Proceedings of the IEEE conference on
  computer vision and pattern recognition, 2018, pp. 4510--4520.

\bibitem{Iandola2016}
F.~N. Iandola, S.~Han, M.~W. Moskewicz, K.~Ashraf, W.~J. Dally, K.~Keutzer,
  Squeezenet: Alexnet-level accuracy with 50x fewer parameters and< 0.5 mb
  model size, arXiv preprint arXiv:1602.07360 (2016).

\bibitem{Warden2019}
P.~Warden, D.~Situnayake, Tinyml: Machine learning with tensorflow lite on
  arduino and ultra-low-power microcontrollers, O'Reilly Media, 2019.

\bibitem{Lin2023}
J.~Lin, L.~Zhu, W.-M. Chen, W.-C. Wang, S.~Han, Tiny machine learning: Progress
  and futures [feature], IEEE Circuits and Systems Magazine 23~(3) (2023)
  8--34.

\bibitem{Shafique2021}
M.~Shafique, T.~Theocharides, V.~J. Reddy, B.~Murmann, Tinyml: Current
  progress, research challenges, and future roadmap, in: 2021 58th ACM/IEEE
  Design Automation Conference (DAC), IEEE, 2021, pp. 1303--1306.

\bibitem{Giraldo2021}
J.~S.~P. Giraldo, M.~Verhelst, Hardware acceleration for embedded keyword
  spotting: Tutorial and survey, ACM Transactions on Embedded Computing Systems
  (TECS) 20~(6) (2021) 1--25.

\bibitem{Tabibian2020}
S.~Tabibian, A survey on structured discriminative spoken keyword spotting,
  Artificial Intelligence Review 53~(4) (2020) 2483--2520.

\bibitem{Garai2024}
S.~Garai, S.~Samui, Exploring tinyml frameworks for small-footprint keyword
  spotting: A concise overview, in: 2024 International Conference on Signal
  Processing and Communications (SPCOM), IEEE, 2024, pp. 1--5.

\bibitem{ChittyVenkata2022}
K.~T. Chitty-Venkata, A.~K. Somani, Neural architecture search survey: A
  hardware perspective, ACM Computing Surveys 55~(4) (2022) 1--36.

\bibitem{Menghani2023}
G.~Menghani, Efficient deep learning: A survey on making deep learning models
  smaller, faster, and better, ACM Computing Surveys 55~(12) (2023) 1--37.

\bibitem{warden2017launching}
P.~Warden, Launching the speech commands dataset, Google Research Blog (2017).

\bibitem{Liang2022}
J.~Liang, X.~Ban, K.~Yu, B.~Qu, K.~Qiao, C.~Yue, K.~Chen, K.~C. Tan, A survey
  on evolutionary constrained multiobjective optimization, IEEE Transactions on
  Evolutionary Computation 27~(2) (2022) 201--221.

\bibitem{Sainath2015conv}
T.~N. Sainath, C.~Parada, {Convolutional neural networks for small-footprint
  keyword spotting}, in: Proc. Interspeech 2015, 2015, pp. 1478--1482.
\newblock \href {https://doi.org/10.21437/Interspeech.2015-352}
  {\path{doi:10.21437/Interspeech.2015-352}}.

\bibitem{Li2022}
M.~Li, A lightweight architecture for query-by-example keyword spotting on
  low-power iot devices, IEEE Transactions on Consumer Electronics 69~(1)
  (2022) 65--75.

\bibitem{Gong2021}
Y.~Gong, Y.-A. Chung, J.~Glass, Ast: Audio spectrogram transformer, Proc.
  Interspeech 2021 (2021).

\bibitem{Berg2021}
A.~Berg, M.~O'Connor, M.~T. Cruz, {Keyword Transformer: A Self-Attention
  Model for Keyword Spotting}, in: Proc. Interspeech 2021, 2021, pp.
  4249--4253.
\newblock \href {https://doi.org/10.21437/Interspeech.2021-1286}
  {\path{doi:10.21437/Interspeech.2021-1286}}.

\bibitem{Samui2023}
S.~Samui, S.~Garai, Time-frequency domain speech enhancement framework using
  audio spectrogram transformer with masked multi-head attention, in: 2023 8th
  International Conference on Computers and Devices for Communication (CODEC),
  2023, pp. 1--2.
\newblock \href {https://doi.org/10.1109/CODEC60112.2023.10465846}
  {\path{doi:10.1109/CODEC60112.2023.10465846}}.

\bibitem{David2021}
R.~David, J.~Duke, A.~Jain, V.~Janapa~Reddi, N.~Jeffries, J.~Li, N.~Kreeger,
  I.~Nappier, M.~Natraj, T.~Wang, et~al., Tensorflow lite micro: Embedded
  machine learning for tinyml systems, Proceedings of Machine Learning and
  Systems 3 (2021) 800--811.

\bibitem{hinton2015distilling}
G.~Hinton, O.~Vinyals, J.~Dean, Distilling the knowledge in a neural network,
  arXiv preprint arXiv:1503.02531 (2015).

\bibitem{Lei2023}
L.~Lei, G.~Yuan, H.~Yu, D.~Kong, Y.~He, Multilingual customized keyword
  spotting using similar-pair contrastive learning, IEEE/ACM Transactions on
  Audio, Speech, and Language Processing 31 (2023) 2437--2447.

\bibitem{Chakravarthi2022}
B.~Chakravarthi, S.-C. Ng, M.~Ezilarasan, M.-F. Leung, Eeg-based emotion
  recognition using hybrid cnn and lstm classification, Frontiers in
  computational neuroscience 16 (2022) 1019776.

\bibitem{Leroy2019}
D.~Leroy, A.~Coucke, T.~Lavril, T.~Gisselbrecht, J.~Dureau, Federated learning
  for keyword spotting, in: ICASSP 2019-2019 IEEE international conference on
  acoustics, speech and signal processing (ICASSP), IEEE, 2019, pp. 6341--6345.

\bibitem{Esch2019}
D.~van Esch, E.~Sarbar, T.~Lucassen, J.~O'Brien, T.~Breiner, M.~Prasad,
  E.~Crew, C.~Nguyen, F.~Beaufays, Writing across the world's languages: Deep
  internationalization for gboard, the google keyboard, arXiv preprint
  arXiv:1912.01218 (2019).

\bibitem{warden2018speech}
P.~Warden, Speech commands: A dataset for limited-vocabulary speech
  recognition, arXiv preprint arXiv:1804.03209 (2018).

\bibitem{chen2014small}
G.~Chen, C.~Parada, G.~Heigold, Small-footprint keyword spotting using deep
  neural networks, in: 2014 IEEE international conference on acoustics, speech
  and signal processing (ICASSP), IEEE, 2014, pp. 4087--4091.

\bibitem{sun2016max}
M.~Sun, A.~Raju, G.~Tucker, S.~Panchapagesan, G.~Fu, A.~Mandal, S.~Matsoukas,
  N.~Strom, S.~Vitaladevuni, Max-pooling loss training of long short-term
  memory networks for small-footprint keyword spotting, in: 2016 IEEE spoken
  language technology workshop (SLT), IEEE, 2016, pp. 474--480.

\bibitem{kumar2018convolutional}
R.~Kumar, V.~Yeruva, S.~Ganapathy, On convolutional lstm modeling for joint
  wake-word detection and text dependent speaker verification., in:
  Interspeech, 2018, pp. 1121--1125.

\bibitem{sorensen2020depthwise}
P.~M. S{\o}rensen, B.~Epp, T.~May, A depthwise separable convolutional neural
  network for keyword spotting on an embedded system, EURASIP Journal on Audio,
  Speech, and Music Processing 2020~(1) (2020) 1--14.

\bibitem{davis1980comparison}
S.~Davis, P.~Mermelstein, Comparison of parametric representations for
  monosyllabic word recognition in continuously spoken sentences, IEEE
  transactions on acoustics, speech, and signal processing 28~(4) (1980)
  357--366.

\bibitem{Vitolo2023}
P.~Vitolo, R.~Liguori, L.~Di~Benedetto, A.~Rubino, G.~D. Licciardo, Automatic
  audio feature extraction for keyword spotting, IEEE Signal Processing Letters
  (2023).

\bibitem{LeCun2015}
Y.~LeCun, Y.~Bengio, G.~Hinton, Deep learning, nature 521~(7553) (2015)
  436--444.

\bibitem{Chen2024}
F.~Chen, S.~Li, J.~Han, F.~Ren, Z.~Yang, Review of lightweight deep
  convolutional neural networks., Archives of Computational Methods in
  Engineering 31~(4) (2024).

\bibitem{howard2017mobilenets}
A.~G. Howard, M.~Zhu, B.~Chen, D.~Kalenichenko, W.~Wang, T.~Weyand,
  M.~Andreetto, H.~Adam, Mobilenets: Efficient convolutional neural networks
  for mobile vision applications, arXiv preprint arXiv:1704.04861 (2017).

\bibitem{Goodfellow2016a}
I.~Goodfellow, Y.~Bengio, A.~Courville, Deep learning, Vol.~1, 2016.

\bibitem{Lebedev2015}
V.~Lebedev, Y.~Ganin, M.~Rakhuba, I.~Oseledets, V.~Lempitsky, Speeding-up
  convolutional neural networks using fine-tuned cp-decomposition, in: 3rd
  International Conference on Learning Representations, ICLR 2015-Conference
  Track Proceedings, 2015.

\bibitem{Shuvo2022}
M.~M.~H. Shuvo, S.~K. Islam, J.~Cheng, B.~I. Morshed, Efficient acceleration of
  deep learning inference on resource-constrained edge devices: A review,
  Proceedings of the IEEE 111~(1) (2022) 42--91.

\bibitem{Sigtia2021}
S.~Sigtia, J.~Bridle, H.~Richards, P.~Clark, E.~Marchi, V.~Garg, Progressive
  voice trigger detection: Accuracy vs latency, in: ICASSP 2021-2021 IEEE
  International Conference on Acoustics, Speech and Signal Processing (ICASSP),
  IEEE, 2021, pp. 6843--6847.

\bibitem{Fu2022}
G.-S. Fu, T.~Senechal, A.~Challenner, T.~Zhang, Unified speculation, detection,
  and verification keyword spotting, in: ICASSP 2022-2022 IEEE International
  Conference on Acoustics, Speech and Signal Processing (ICASSP), IEEE, 2022,
  pp. 7557--7561.

\bibitem{wang2023wekws}
J.~Wang, M.~Xu, J.~Hou, B.~Zhang, X.-L. Zhang, L.~Xie, F.~Pan, Wekws: A
  production first small-footprint end-to-end keyword spotting toolkit, in:
  ICASSP 2023-2023 IEEE International Conference on Acoustics, Speech and
  Signal Processing (ICASSP), IEEE, 2023, pp. 1--5.

\bibitem{Park2024}
E.~Park, D.~Ahn, H.~Kim, Reptor: Re-parameterizable temporal convolution for
  keyword spotting via differentiable kernel search, in: Proc. Interspeech
  2024, 2024, pp. 4518--4522.

\bibitem{Ding2021}
X.~Ding, X.~Zhang, N.~Ma, J.~Han, G.~Ding, J.~Sun, Repvgg: Making vgg-style
  convnets great again, in: Proceedings of the IEEE/CVF conference on computer
  vision and pattern recognition, 2021, pp. 13733--13742.

\bibitem{Yang2022}
H.~Yang, Z.~Yang, L.~Wan, B.~Zhang, Y.~Shi, Y.~Huang, I.~Enchev, L.~Tang,
  R.~Alvarez, M.~Sun, et~al., Lico-net: Linearized convolution network for
  hardware-efficient keyword spotting, arXiv preprint arXiv:2211.04635 (2022).

\bibitem{huang2022}
Y.~Huang, N.~Hou, N.~F. Chen, Progressive continual learning for spoken keyword
  spotting, in: ICASSP 2022-2022 IEEE International Conference on Acoustics,
  Speech and Signal Processing (ICASSP), IEEE, 2022, pp. 7552--7556.

\bibitem{snell2017}
J.~Snell, K.~Swersky, R.~Zemel, Prototypical networks for few-shot learning,
  Advances in neural information processing systems 30 (2017).

\bibitem{Rusci2023}
M.~Rusci, T.~Tuytelaars, On-device customization of tiny deep learning models
  for keyword spotting with few examples, Ieee Micro (2023).

\bibitem{Wang2017}
Y.~Wang, P.~Getreuer, T.~Hughes, R.~F. Lyon, R.~A. Saurous, Trainable frontend
  for robust and far-field keyword spotting, in: 2017 IEEE International
  Conference on Acoustics, Speech and Signal Processing (ICASSP), IEEE, 2017,
  pp. 5670--5674.

\bibitem{Samui2019}
S.~Samui, I.~Chakrabarti, S.~K. Ghosh, Time--frequency masking based supervised
  speech enhancement framework using fuzzy deep belief network, Applied Soft
  Computing 74 (2019) 583--602.

\bibitem{Goodfellow2014}
I.~J. Goodfellow, J.~Shlens, C.~Szegedy, Explaining and harnessing adversarial
  examples, arXiv preprint arXiv:1412.6572 (2014).

\bibitem{Zhang2018}
Z.~Zhang, J.~Geiger, J.~Pohjalainen, A.~E.-D. Mousa, W.~Jin, B.~Schuller, Deep
  learning for environmentally robust speech recognition: An overview of recent
  developments, ACM Transactions on Intelligent Systems and Technology (TIST)
  9~(5) (2018) 1--28.

\bibitem{du2018aishell}
J.~Du, X.~Na, X.~Liu, H.~Bu, Aishell-2: Transforming mandarin asr research into
  industrial scale, arXiv preprint arXiv:1808.10583 (2018).

\bibitem{Mishchenko2019}
Y.~Mishchenko, Y.~Goren, M.~Sun, C.~Beauchene, S.~Matsoukas, O.~Rybakov,
  S.~N.~P. Vitaladevuni, Low-bit quantization and quantization-aware training
  for small-footprint keyword spotting, in: 2019 18th IEEE International
  Conference On Machine Learning And Applications (ICMLA), IEEE, 2019, pp.
  706--711.

\bibitem{higuchi2020stacked}
T.~Higuchi, M.~Ghasemzadeh, K.~You, C.~Dhir, Stacked 1d convolutional networks
  for end-to-end small footprint voice trigger detection, Proc. Interspeech
  2020 (2020).

\bibitem{kim2019query}
B.~Kim, M.~Lee, J.~Lee, Y.~Kim, K.~Hwang, Query-by-example on-device keyword
  spotting, in: 2019 IEEE Automatic Speech Recognition and Understanding
  Workshop (ASRU), IEEE, 2019, pp. 532--538.

\bibitem{hou2019}
J.~Hou, Y.~Shi, M.~Ostendorf, M.~Hwang, L.~Xie,
  \href{https://doi.org/10.1109/LSP.2019.2936282}{Region proposal network based
  small-footprint keyword spotting}, {IEEE} Signal Process. Lett. 26~(10)
  (2019) 1471--1475.
\newline\urlprefix\url{https://doi.org/10.1109/LSP.2019.2936282}

\bibitem{mazumder2021multilingual}
M.~Mazumder, S.~Chitlangia, C.~Banbury, Y.~Kang, J.~M. Ciro, K.~Achorn,
  D.~Galvez, M.~Sabini, P.~Mattson, D.~Kanter, et~al., Multilingual spoken
  words corpus, in: Thirty-fifth Conference on Neural Information Processing
  Systems Datasets and Benchmarks Track (Round 2), 2021.

\bibitem{Qin2019}
X.~Qin, H.~Bu, M.~Li, Hi-mia : A far-field text-dependent speaker verification
  database and the baselines (2019).
\newblock \href {http://arxiv.org/abs/1912.01231} {\path{arXiv:1912.01231}}.

\bibitem{Ghandoura2021}
A.~Ghandoura, F.~Hjabo, O.~{Al Dakkak},
  \href{https://www.sciencedirect.com/science/article/pii/S0952197621001147}{Building
  and benchmarking an arabic speech commands dataset for small-footprint
  keyword spotting}, Engineering Applications of Artificial Intelligence 102
  (2021) 104267.
\newblock \href
  {https://doi.org/https://doi.org/10.1016/j.engappai.2021.104267}
  {\path{doi:https://doi.org/10.1016/j.engappai.2021.104267}}.
\newline\urlprefix\url{https://www.sciencedirect.com/science/article/pii/S0952197621001147}

\bibitem{Arik2017a}
S.~Ã. Arık, M.~Kliegl, R.~Child, J.~Hestness, A.~Gibiansky, C.~Fougner,
  R.~Prenger, A.~Coates, {Convolutional Recurrent Neural Networks for
  Small-Footprint Keyword Spotting}, in: Proc. Interspeech 2017, 2017, pp.
  1606--1610.
\newblock \href {https://doi.org/10.21437/Interspeech.2017-1737}
  {\path{doi:10.21437/Interspeech.2017-1737}}.

\bibitem{tang2018}
R.~Tang, J.~Lin, Deep residual learning for small-footprint keyword spotting,
  in: 2018 IEEE International Conference on Acoustics, Speech and Signal
  Processing (ICASSP), IEEE, 2018, pp. 5484--5488.

\bibitem{choi2019temporal}
S.~Choi, S.~Seo, B.~Shin, H.~Byun, M.~Kersner, B.~Kim, D.~Kim, S.~Ha, Temporal
  convolution for real-time keyword spotting on mobile devices, Proc.
  INTERSPEECH 2019 (2019).

\bibitem{chen2019}
X.~Chen, S.~Yin, D.~Song, P.~Ouyang, L.~Liu, S.~Wei, Small-footprint keyword
  spotting with graph convolutional network, in: 2019 IEEE Automatic Speech
  Recognition and Understanding Workshop (ASRU), IEEE, 2019, pp. 539--546.

\bibitem{Li2020}
X.~Li, X.~Wei, X.~Qin, {Small-Footprint Keyword Spotting with Multi-Scale
  Temporal Convolution}, in: Proc. Interspeech 2020, 2020, pp. 1987--1991.
\newblock \href {https://doi.org/10.21437/Interspeech.2020-3177}
  {\path{doi:10.21437/Interspeech.2020-3177}}.

\bibitem{majumdar2020}
S.~Majumdar, B.~Ginsburg, {MatchboxNet: 1D Time-Channel Separable Convolutional
  Neural Network Architecture for Speech Commands Recognition}, in: Proc.
  Interspeech 2020, 2020, pp. 3356--3360.
\newblock \href {https://doi.org/10.21437/Interspeech.2020-1058}
  {\path{doi:10.21437/Interspeech.2020-1058}}.

\bibitem{kim2021broadcasted}
B.~Kim, S.~Chang, J.~Lee, D.~Sung, Broadcasted residual learning for efficient
  keyword spotting, Proceedings of INTERSPEECH 2021 (2021).

\bibitem{chaudhary2023towards}
A.~Chaudhary, V.~Abrol, Towards on-device keyword spotting using low-footprint
  quaternion neural models, in: 2023 IEEE Workshop on Applications of Signal
  Processing to Audio and Acoustics (WASPAA), IEEE, 2023, pp. 1--5.

\bibitem{akhtar2023}
Z.~Akhtar, M.~O. Khursheed, D.~Du, Y.~Liu, Small-footprint slimmable networks
  for keyword spotting, in: IEEE International Conference on Acoustics, Speech
  and Signal Processing (ICASSP), 2023, pp. 1--5.

\bibitem{tucker2016}
G.~Tucker, M.~Wu, M.~Sun, S.~Panchapagesan, G.~Fu, S.~Vitaladevuni, {Model
  Compression Applied to Small-Footprint Keyword Spotting}, in: Proc.
  Interspeech 2016, 2016, pp. 1878--1882.
\newblock \href {https://doi.org/10.21437/Interspeech.2016-1393}
  {\path{doi:10.21437/Interspeech.2016-1393}}.

\bibitem{Gao2023}
C.~Gao, Y.~Gu, F.~Caliva, Y.~Liu, Self-supervised speech representation
  learning for keyword-spotting with light-weight transformers, in: ICASSP
  2023-2023 IEEE International Conference on Acoustics, Speech and Signal
  Processing (ICASSP), IEEE, 2023, pp. 1--5.

\bibitem{Yang2023}
G.-P. Yang, Y.~Gu, Q.~Tang, D.~Du, Y.~Liu, On-device constrained
  self-supervised speech representation learning for keyword spotting via
  knowledge distillation, in: INTERSPEECH, 2023.

\bibitem{macha2023}
S.~Macha, O.~Oza, A.~Escott, F.~Caliva, R.~Armitano, S.~K. Cheekatmalla,
  S.~H.~K. Parthasarathi, Y.~Liu, Fixed-point quantization aware training for
  on-device keyword-spotting, in: ICASSP 2023-2023 IEEE International
  Conference on Acoustics, Speech and Signal Processing (ICASSP), IEEE, 2023,
  pp. 1--5.

\bibitem{Sun2017a}
M.~Sun, D.~Snyder, Y.~Gao, V.~Nagaraja, M.~Rodehorst, S.~Panchapagesan,
  N.~Strom, S.~Matsoukas, S.~Vitaladevuni, {Compressed Time Delay Neural
  Network for Small-Footprint Keyword Spotting}, in: Proc. Interspeech 2017,
  2017, pp. 3607--3611.
\newblock \href {https://doi.org/10.21437/Interspeech.2017-480}
  {\path{doi:10.21437/Interspeech.2017-480}}.

\bibitem{luo2022}
M.~Luo, D.~Wang, X.~Wang, S.~Qiao, Y.~Zhou, Error-diffusion based speech
  feature quantization for small-footprint keyword spotting, IEEE Signal
  Processing Letters 29 (2022) 1357--1361.

\bibitem{Yang2024}
G.-P. Yang, Y.~Gu, S.~Macha, Q.~Tang, Y.~Liu, On-device constrained
  self-supervised learning for keyword spotting via quantization aware
  pre-training and fine-tuning, in: ICASSP 2024-2024 IEEE International
  Conference on Acoustics, Speech and Signal Processing (ICASSP), IEEE, 2024,
  pp. 10951--10955.

\bibitem{Shan2018}
C.~Shan, J.~Zhang, Y.~Wang, L.~Xie, {Attention-based End-to-End Models for
  Small-Footprint Keyword Spotting}, in: Proc. Interspeech 2018, 2018, pp.
  2037--2041.
\newblock \href {https://doi.org/10.21437/Interspeech.2018-1777}
  {\path{doi:10.21437/Interspeech.2018-1777}}.

\bibitem{bai2019}
Y.~Bai, J.~Yi, J.~Tao, Z.~Wen, Z.~Tian, C.~Zhao, C.~Fan, A time delay neural
  network with shared weight self-attention for small-footprint keyword
  spotting., in: INTERSPEECH, 2019, pp. 2190--2194.

\bibitem{mfslbrahim2019}
E.~A. Ibrahim, J.~Huisken, H.~Fatemi, J.~P. de~Gyvez, Keyword spotting using
  time-domain features in a temporal convolutional network, in: 2019 22nd
  Euromicro Conference on Digital System Design (DSD), IEEE, 2019, pp.
  313--319.

\bibitem{mittermaier2020}
S.~Mittermaier, L.~K{\"u}rzinger, B.~Waschneck, G.~Rigoll, Small-footprint
  keyword spotting on raw audio data with sinc-convolutions, in: ICASSP
  2020-2020 IEEE International Conference on Acoustics, Speech and Signal
  Processing (ICASSP), IEEE, 2020, pp. 7454--7458.

\bibitem{riviello2019binary}
A.~Riviello, J.-P. David, Binary speech features for keyword spotting tasks.,
  in: INTERSPEECH, 2019, pp. 3460--3464.

\bibitem{anderson2019performance}
A.~Anderson, J.~Su, R.~Dahyot, D.~Gregg, Performance-oriented neural
  architecture search, in: 2019 International Conference on High Performance
  Computing \& Simulation (HPCS), IEEE, 2019, pp. 177--184.

\bibitem{veniat2019stochastic}
T.~V{\'e}niat, O.~Schwander, L.~Denoyer, Stochastic adaptive neural
  architecture search for keyword spotting, in: ICASSP 2019-2019 IEEE
  International Conference on Acoustics, Speech and Signal Processing (ICASSP),
  IEEE, 2019, pp. 2842--2846.

\bibitem{mo20_interspeech}
T.~Mo, Y.~Yu, M.~Salameh, D.~Niu, S.~Jui, {Neural Architecture Search for
  Keyword Spotting}, in: Proc. Interspeech 2020, 2020, pp. 1982--1986.
\newblock \href {https://doi.org/10.21437/Interspeech.2020-3132}
  {\path{doi:10.21437/Interspeech.2020-3132}}.

\bibitem{Zhang2021autokws}
B.~Zhang, W.~Li, Q.~Li, W.~Zhuang, X.~Chu, Y.~Wang, Autokws: Keyword spotting
  with differentiable architecture search, in: ICASSP 2021-2021 IEEE
  International Conference on Acoustics, Speech and Signal Processing (ICASSP),
  IEEE, 2021, pp. 2830--2834.

\bibitem{Banbury2021}
C.~Banbury, C.~Zhou, I.~Fedorov, R.~Matas, U.~Thakker, D.~Gope,
  V.~Janapa~Reddi, M.~Mattina, P.~Whatmough, Micronets: Neural network
  architectures for deploying tinyml applications on commodity
  microcontrollers, Proceedings of Machine Learning and Systems 3 (2021)
  517--532.

\bibitem{Busia2022}
P.~Busia, G.~Deriu, L.~Rinelli, C.~Chesta, L.~Raffo, P.~Meloni, Target-aware
  neural architecture search and deployment for keyword spotting, IEEE Access
  10 (2022) 40687--40700.

\bibitem{zhang2017}
Y.~Zhang, N.~Suda, L.~Lai, V.~Chandra, Hello edge: Keyword spotting on
  microcontrollers, arXiv preprint arXiv:1711.07128 (2017).

\bibitem{Peter2022}
D.~Peter, W.~Roth, F.~Pernkopf, End-to-end keyword spotting using neural
  architecture search and quantization, in: ICASSP 2022-2022 IEEE International
  Conference on Acoustics, Speech and Signal Processing (ICASSP), IEEE, 2022,
  pp. 3423--3427.

\bibitem{Toth2014}
L.~T{\'o}th, Combining time-and frequency-domain convolution in convolutional
  neural network-based phone recognition, in: 2014 IEEE International
  Conference on Acoustics, Speech and Signal Processing (ICASSP), IEEE, 2014,
  pp. 190--194.

\bibitem{AbdelHamid2012}
O.~Abdel-Hamid, A.-r. Mohamed, H.~Jiang, G.~Penn, Applying convolutional neural
  networks concepts to hybrid nn-hmm model for speech recognition, in: 2012
  IEEE international conference on Acoustics, speech and signal processing
  (ICASSP), IEEE, 2012, pp. 4277--4280.

\bibitem{song2024ed}
Z.~Song, Q.~Liu, Q.~Yang, Y.~Peng, H.~Li, Ed-skws: Early-decision spiking
  neural networks for rapid, and energy-efficient keyword spotting, in: Proc.
  Interspeech 2024, 2024, pp. 4528--4532.

\bibitem{wang2024global}
S.~Wang, D.~Zhang, K.~Shi, Y.~Wang, W.~Wei, J.~Wu, M.~Zhang, Global-local
  convolution with spiking neural networks for energy-efficient keyword
  spotting, in: Proc. Interspeech 2024, 2024, pp. 4523--4527.

\bibitem{coucke2019efficient}
A.~Coucke, M.~Chlieh, T.~Gisselbrecht, D.~Leroy, M.~Poumeyrol, T.~Lavril,
  Efficient keyword spotting using dilated convolutions and gating, in: ICASSP
  2019-2019 IEEE International Conference on Acoustics, Speech and Signal
  Processing (ICASSP), IEEE, 2019, pp. 6351--6355.

\bibitem{Baevski2020}
A.~Baevski, Y.~Zhou, A.~Mohamed, M.~Auli, wav2vec 2.0: A framework for
  self-supervised learning of speech representations, Advances in neural
  information processing systems 33 (2020) 12449--12460.

\bibitem{Sze2017}
V.~Sze, Y.-H. Chen, T.-J. Yang, J.~S. Emer, Efficient processing of deep neural
  networks: A tutorial and survey, Proceedings of the IEEE 105~(12) (2017)
  2295--2329.

\bibitem{DNNbook2020}
T.-J. Y. J. S.~E. Vivienne~Sze, Yu-Hsin~Chen,
  \href{https://doi.org/10.1007/978-3-031-01766-7}{Efficient Processing of Deep
  Neural Networks}, Synthesis Lectures on Computer Architecture, Springer Cham,
  2020.
\newline\urlprefix\url{https://doi.org/10.1007/978-3-031-01766-7}

\bibitem{LeCun1989}
Y.~LeCun, J.~Denker, S.~Solla, Optimal brain damage, Advances in neural
  information processing systems 2 (1989).

\bibitem{Hassibi1993}
B.~Hassibi, D.~G. Stork, G.~J. Wolff, Optimal brain surgeon and general network
  pruning, in: IEEE international conference on neural networks, IEEE, 1993,
  pp. 293--299.

\bibitem{Jacob2018}
B.~Jacob, S.~Kligys, B.~Chen, M.~Zhu, M.~Tang, A.~Howard, H.~Adam,
  D.~Kalenichenko, Quantization and training of neural networks for efficient
  integer-arithmetic-only inference, in: Proceedings of the IEEE conference on
  computer vision and pattern recognition, 2018, pp. 2704--2713.

\bibitem{Nahshan2021}
Y.~Nahshan, B.~Chmiel, C.~Baskin, E.~Zheltonozhskii, R.~Banner, A.~M.
  Bronstein, A.~Mendelson, Loss aware post-training quantization, Machine
  Learning 110~(11) (2021) 3245--3262.

\bibitem{Gholami2022}
A.~Gholami, S.~Kim, Z.~Dong, Z.~Yao, M.~W. Mahoney, K.~Keutzer, A survey of
  quantization methods for efficient neural network inference, in: Low-power
  computer vision, Chapman and Hall/CRC, 2022, pp. 291--326.

\bibitem{Ostromoukhov2001}
V.~Ostromoukhov, A simple and efficient error-diffusion algorithm, in:
  Proceedings of the 28th annual conference on Computer graphics and
  interactive techniques, 2001, pp. 567--572.

\bibitem{Ding2022}
K.~Ding, M.~Zong, J.~Li, B.~Li, Letr: A lightweight and efficient transformer
  for keyword spotting, in: ICASSP 2022-2022 IEEE International Conference on
  Acoustics, Speech and Signal Processing (ICASSP), IEEE, 2022, pp. 7987--7991.

\bibitem{Niu2021}
Z.~Niu, G.~Zhong, H.~Yu, A review on the attention mechanism of deep learning,
  Neurocomputing 452 (2021) 48--62.

\bibitem{LopezEspejo2022}
I.~López-Espejo, Z.-H. Tan, J.~Jensen, {An Experimental Study on Light Speech
  Features for Small-Footprint Keyword Spotting }, in: Proc. IberSPEECH 2022,
  2022, pp. 131--135.
\newblock \href {https://doi.org/10.21437/IberSPEECH.2022-27}
  {\path{doi:10.21437/IberSPEECH.2022-27}}.

\bibitem{Benmeziane2021}
H.~Benmeziane, K.~El~Maghraoui, H.~Ouarnoughi, S.~Niar, M.~Wistuba, N.~Wang, A
  comprehensive survey on hardware-aware neural architecture search, Ph.D.
  thesis, LAMIH, Universit{\'e} Polytechnique des Hauts-de-France (2021).

\bibitem{tan2019mnasnet}
M.~Tan, B.~Chen, R.~Pang, V.~Vasudevan, M.~Sandler, A.~Howard, Q.~V. Le,
  Mnasnet: Platform-aware neural architecture search for mobile, in:
  Proceedings of the IEEE/CVF conference on computer vision and pattern
  recognition, 2019, pp. 2820--2828.

\bibitem{Ravanelli2018}
M.~Ravanelli, Y.~Bengio, Speaker recognition from raw waveform with sincnet,
  in: 2018 IEEE spoken language technology workshop (SLT), IEEE, 2018, pp.
  1021--1028.

\bibitem{Shrivastava2021}
A.~Shrivastava, A.~Kundu, C.~Dhir, D.~Naik, O.~Tuzel, Optimize what matters:
  Training dnn-hmm keyword spotting model using end metric, in: ICASSP 2021 -
  2021 IEEE International Conference on Acoustics, Speech and Signal Processing
  (ICASSP), 2021, pp. 4000--4004.
\newblock \href {https://doi.org/10.1109/ICASSP39728.2021.9414797}
  {\path{doi:10.1109/ICASSP39728.2021.9414797}}.

\bibitem{Gimenez2022}
N.~L. Giménez, F.~Freitag, J.~Lee, H.~Vandierendonck, Comparison of two
  microcontroller boards for on-device model training in a keyword spotting
  task, in: 2022 11th Mediterranean Conference on Embedded Computing (MECO),
  2022, pp. 1--4.
\newblock \href {https://doi.org/10.1109/MECO55406.2022.9797171}
  {\path{doi:10.1109/MECO55406.2022.9797171}}.

\bibitem{Ren2021}
H.~Ren, D.~Anicic, T.~A. Runkler, Tinyol: Tinyml with online-learning on
  microcontrollers, in: 2021 International Joint Conference on Neural Networks
  (IJCNN), IEEE, 2021, pp. 1--8.

\bibitem{Lin2022}
J.~Lin, L.~Zhu, W.-M. Chen, W.-C. Wang, C.~Gan, S.~Han, On-device training
  under 256kb memory, Advances in Neural Information Processing Systems 35
  (2022) 22941--22954.

\bibitem{ellmicrosoft}
Embedded learning library (ell), \url{https://microsoft.github.io/ELL/},
  accessed: 12 January 2024.

\bibitem{armnn}
{ARM-NN}, \url{https://github.com/ARM-software/armnn}, accessed: 12 January
  2024.

\bibitem{cmsisnn}
{CMSIS-NN}, \url{https://arm-software.github.io/CMSIS_5/NN/html}, accessed: 12
  January 2024.

\bibitem{STM}
{STM32Cube.AI}, \url{https://www.st.com/en/embedded-software/x-cube-ai.html},
  accessed: 12 January 2024.

\bibitem{AIfES}
{AIfES}, \url{https://github.com/Fraunhofer-IMS/AIfES_for_Arduino}, accessed:
  12 January 2024.

\bibitem{uTensor}
{uTensor}, \url{https://github.com/uTensor/uTensor}, accessed: 12 January 2024.

\bibitem{TinyMLgen}
{TinyMLgen}, \url{https://github.com/eloquentarduino/tinymlgen}, accessed: 12
  January 2024.

\bibitem{Capotondi2020}
Cmix-nn, \url{https://github.com/EEESlab/CMix-NN}, accessed: 12 January 2024.

\bibitem{Edge_Impulse}
{Edge Impulse}, \url{https://edgeimpulse.com/}, accessed: 12 January 2024.

\bibitem{Miettinen1999}
K.~Miettinen, Nonlinear multiobjective optimization, Vol.~12, Springer Science
  \& Business Media, 1999.

\bibitem{Deb2011}
K.~Deb, Multi-objective optimisation using evolutionary algorithms: an
  introduction, in: Multi-objective evolutionary optimisation for product
  design and manufacturing, Springer, 2011, pp. 3--34.

\bibitem{Rardin2001}
R.~L. Rardin, R.~Uzsoy, Experimental evaluation of heuristic optimization
  algorithms: A tutorial, Journal of Heuristics 7 (2001) 261--304.

\bibitem{Bandyopadhyay2008}
S.~Bandyopadhyay, S.~Saha, U.~Maulik, K.~Deb, A simulated annealing-based
  multiobjective optimization algorithm: Amosa, IEEE transactions on
  evolutionary computation 12~(3) (2008) 269--283.

\bibitem{Guelcue2021}
A.~G{\"u}lc{\"u}, Z.~Ku{\c{s}}, Multi-objective simulated annealing for
  hyper-parameter optimization in convolutional neural networks, PeerJ Computer
  Science 7 (2021) e338.

\bibitem{Deb2002}
K.~Deb, A.~Pratap, S.~Agarwal, T.~Meyarivan, A fast and elitist multiobjective
  genetic algorithm: Nsga-ii, IEEE transactions on evolutionary computation
  6~(2) (2002) 182--197.

\bibitem{shaikh2024}
A.~A. Shaikh, A.~K. Mukhopadhyay, S.~Poddar, S.~Samui, Toward robust and
  accurate myoelectric controller design based on multiobjective optimization
  using evolutionary computation, IEEE Sensors Journal 24~(5) (2024)
  6418--6429.
\newblock \href {https://doi.org/10.1109/JSEN.2023.3347949}
  {\path{doi:10.1109/JSEN.2023.3347949}}.

\bibitem{Parsa2020}
M.~Parsa, J.~P. Mitchell, C.~D. Schuman, R.~M. Patton, T.~E. Potok, K.~Roy,
  Bayesian multi-objective hyperparameter optimization for accurate, fast, and
  efficient neural network accelerator design, Frontiers in neuroscience 14
  (2020) 667.

\bibitem{Alibrahim2021}
H.~Alibrahim, S.~A. Ludwig, Hyperparameter optimization: Comparing genetic
  algorithm against grid search and bayesian optimization, in: 2021 IEEE
  congress on evolutionary computation (CEC), IEEE, 2021, pp. 1551--1559.

\bibitem{Jin2007}
Y.~Jin, Multi-objective machine learning, Vol.~16, Springer Science \& Business
  Media, 2007.

\bibitem{Liberis2021}
E.~Liberis, {\L}.~Dudziak, N.~D. Lane, $\mu$nas: Constrained neural
  architecture search for microcontrollers, in: Proceedings of the 1st Workshop
  on Machine Learning and Systems, 2021, pp. 70--79.

\bibitem{Ma2023}
L.~Ma, N.~Li, G.~Yu, X.~Geng, S.~Cheng, X.~Wang, M.~Huang, Y.~Jin, Pareto-wise
  ranking classifier for multi-objective evolutionary neural architecture
  search, IEEE Transactions on Evolutionary Computation (2023).

\bibitem{Vaswani2017}
A.~Vaswani, N.~Shazeer, N.~Parmar, J.~Uszkoreit, L.~Jones, A.~N. Gomez,
  {\L}.~Kaiser, I.~Polosukhin, Attention is all you need, Advances in neural
  information processing systems 30 (2017).

\bibitem{Gong2022}
Y.~Gong, C.-I. Lai, Y.-A. Chung, J.~Glass, Ssast: Self-supervised audio
  spectrogram transformer, in: Proceedings of the AAAI Conference on Artificial
  Intelligence, Vol.~36, 2022, pp. 10699--10709.

\bibitem{Sun2020}
Y.~Sun, X.~Wang, Z.~Liu, J.~Miller, A.~Efros, M.~Hardt, Test-time training with
  self-supervision for generalization under distribution shifts, in:
  International conference on machine learning, PMLR, 2020, pp. 9229--9248.

\bibitem{samui2018}
S.~Samui, I.~Chakrabarti, S.~K. Ghosh, Tensor-train long short-term memory for
  monaural speech enhancement, arXiv preprint arXiv:1812.10095 (2018).

\bibitem{Fedorov2019}
I.~Fedorov, R.~P. Adams, M.~Mattina, P.~Whatmough, Sparse: Sparse architecture
  search for cnns on resource-constrained microcontrollers, Advances in Neural
  Information Processing Systems 32 (2019).

\bibitem{Wang2020}
T.~Wang, K.~Wang, H.~Cai, J.~Lin, Z.~Liu, H.~Wang, Y.~Lin, S.~Han, Apq: Joint
  search for network architecture, pruning and quantization policy, in:
  Proceedings of the IEEE/CVF Conference on Computer Vision and Pattern
  Recognition, 2020, pp. 2078--2087.

\end{thebibliography}

\end{document}